\documentclass[useAMS,usenatbib]{mn2e}
\input psfig.tex

\usepackage{latexsym}
\usepackage{amssymb}
\usepackage{amsmath}
\usepackage{graphics}
\usepackage{graphicx}
\usepackage{natbib}
\usepackage{subfigure}
\usepackage{lastpage}
\usepackage{bm} 

\title[Milky Way galaxy model]{Spiral arm kinematics for Milky Way stellar populations}
\author[S. Pasetto et al.]{S. Pasetto $^{1}$\thanks{E-mail:
s.pasetto@ucl.ac.uk or galaxy.model@yahoo.com}, G. Natale $^{2}$, D. Kawata$^1$, C. Chiosi$^{3}$, J. A. S. Hunt$^{1}$, C. Brogliato$^{4}$\\
 $^{1}$University College London, Department of Space \& Climate Physics, Mullard Space Science Laboratory, Holmbury St. Mary, \\ \,\,\,Dorking, Surrey, United Kingdom;\\
 $^{2}$Jeremiah Horrocks Institute, University of Central Lancashire, Preston, PR1 2HE, UK; \\
 $^{3}$Department of Physics \& Astronomy,``Galileo Galilei", University of Padua, Padua, Italy;\\
 $^{4}$Clover-lab, Valene rd, Salo, Brescia, 25087, Italy.  \\
 }

\begin{document}

\date{Accepted for pubblication on MNRAS}


\maketitle

\label{firstpage}

\begin{abstract}
We present a new theoretical population synthesis model (the Galaxy Model) to examine and deal with large amounts of data from surveys of the Milky Way and to decipher the present and past structure and history of our own Galaxy.

We assume the Galaxy to consist of a superposition of many composite stellar populations belonging to the thin and thick disks, the stellar halo and the bulge, and to be surrounded by a single dark matter halo component. A global model for the Milky Way's gravitational potential is built up self-consistently with the density profiles from the Poisson equation. In turn, these density profiles are used to generate synthetic probability distribution functions (PDFs) for the distribution of stars in colour-magnitude diagrams (CMDs). Finally, the gravitational potential is used to constrain the stellar kinematics by means of the moment method on a (perturbed)-distribution function. Spiral arms perturb the axisymmetric disk distribution functions in the linear response framework of density-wave theory where we present an analytical formula of the so-called `reduction factor' using Hypergeometric functions. 

Finally, we consider an analytical non-axisymmetric model of extinction and an algorithm based on the concept of probability distribution function to handle colour magnitude diagrams with a large number of stars. A genetic algorithm is presented to investigate both the photometric and kinematic parameter space.

This galaxy model represents the natural framework to reconstruct the structure of the Milky Way from the heterogeneous data set of surveys such as Gaia-ESO, SEGUE, APOGEE2, RAVE and the Gaia mission.
\end{abstract}


\section{Introduction}\label{Sec1}
The Milky Way (MW) provides and unique environment in which to study the origin and evolution of galaxies on a star-by-star basis, with a precision that is simply impossible to reach for any other galaxy in the Universe. The European Space Agency's cornerstone mission Gaia, together with complementary ground-based spectroscopic follow-ups such as the Gaia-ESO Survey \citep[e.g.,][]{2012Msngr.147...25G}, will map the stellar distribution of the MW with unprecedented accuracy by  providing high-precision phase-space information, physical parameters, and chemical compositions, for roughly one billion of the stars in our Galaxy. The exploitation of this huge amount of  data  cannot be made using the methods and tools   that have been used  for many decades to study much less numerous samples of stars; it requires the development of, and experience with, cutting-edge multi-dimensional data mining tools, as well as sophisticated methodologies to transfer the models from the space of ``simulations''  to the ``plane of observers''.

Star-count techniques are born with the aim to answer a simple astronomical question: why do we see a given distribution of stars in the sky? Since the oldest approach to the star-count equation\citep[e.g.,][]{1953stas.book.....T}, these techniques have represented the most natural way to investigate the closest distribution of stars to us, i.e. the Milky Way. A major advancement of these techniques was achieved by \citet{1984ApJS...55...67B} who applied the concept of stellar populations to the solar neighbourhood \citep[see also][]{1984BAAS...16..733B, 1984ApJ...287..926B} and nowadays, more theoretically sophisticated star-count models are the standard tools to investigate the MW stellar distribution \citep[e.g., the Besan\c{c}on model, ][]{2003A&A...409..523R}. The ultimate step toward the understanding of our Galaxy is thus represented by the extension of the concept of stellar populations to include kinematics, dynamics, photometric and chemical properties together in a global MW modelling approach \citep[e.g.,][]{2000AJ....119..813M, 2006A&A...451..125V}. 

The star-count techniques have the goal to synthetically reproduce the observables obtained from an (unknown)-stellar distribution function (DF), i.e. the number of stars in a given range of, e.g., temperature, velocities, densities, proper motions etc., by considering the data distribution in the space of the observable quantities (e.g., photometry, proper motions, radial velocities, etc.).  To achieve this goal, a number of founding pillars must be assumed to exist on a global scale, e.g., density-profile laws, star-formation histories, age-metallicity relations, age-velocity dispersion relations etc. All these relations will ultimately represent a way of deciphering and constraining the MW history and evolution.

The more independent constraints a model can reproduce, the closer these underlying relations are to the true properties of the system analyzed (the MW in our case). 
The star-count techniques are a Monte-Carlo type solution to a multidimensional integration problem of the star-count equation. Historically, in classical textbooks of statistical astronomy \citep{1953stas.book.....T} the star-count equation is generalized to include the kinematics as follows:
\begin{equation}\label{Eq1}
	\frac{ {d{N_j}} }{ {d{\bm{\gamma}} d{m_{\Delta\lambda}} dC_{\lambda\lambda'}} } = {N_j}{f_j}\left( {\bm{\gamma }} \right),
\end{equation}
where ${N_j}$ is the number of stars for each given stellar population, $j$, with distribution function ${f_j}\left( {\bm{\gamma }} \right)$ in the elemental volume of the phase space $d{\bm{\gamma }} = \left\{ {d{\bm{x}},d{\bm{v}}} \right\} = \left\{ {d\overset{\lower0.5em\hbox{$\smash{\scriptscriptstyle\frown}$}}{\Omega } d{r_{{\rm{hel}}}},d{\bm{v}}} \right\}$. Here  ${r_{{\rm{hel}}}}$ is the heliocentric distance of the stars in an infinitesimal interval of magnitude $d{m_{\Delta\lambda} }$ in the band $\Delta\lambda $  and colour $dC=m_{\Delta\lambda}-m_{\Delta\lambda'}$, $d\overset{\lower0.5em\hbox{$\smash{\scriptscriptstyle\frown}$}}{\Omega }  = dldb \cos b$ (with $l$ and $b$ Galactic longitude and latitude) is the solid angle.  In Section \ref{Sec2} we will review  a generalized framework for Eq.\eqref{Eq1} introduced in \cite{2012A&A...545A..14P}  to recover Eq.\eqref{Eq1} as a special case of a multidimensional marginalization process. 

Two of the major limitations underlying many theoretical works based on analytical expressions for the DF ${f_j}\left( {\bm{\gamma }} \right)$ are the time independence of ${f_j}$  and its axisymmetry properties in the configuration space. Related to the first assumption is the problem of self-consistency: the DFs are not obtained by sampling the phase-space of a system evolved in time under the effect of self-gravity. In this approach, the DF is not numerical but a parametric function. The second assumption of axisymmetry is led by the necessity to keep the treatment of the dynamical evolutions as simple as possible: the corresponding Hamiltonian is cyclic in some variables and hence more suited for analytical manipulation. 

The literature is full of alternatives to overcome these two limitations, e.g., N-body simulations, the Schwarzschild method, the Made-to-Measure method, full theoretical methods \citep[e.g,][]{2013MNRAS.430.1928H,2015MNRAS.450.2132H,2007MNRAS.380..848C,2015A&A...581A.123B,2013A&A...549A..89B} etc. whose review is beyond the goal of this paper. 
In this work, we will relax the axisymmetry assumption for the sole thin disk components by implementing a perturbative approach carried out to the linear order on suitable small parameters to the equation of motion following two different works by \citet{1969ApJ...155..721L} and \citet{1991ApJ...368...79A}. These perturbative linear response frameworks are the only analytical treatment available up to now that can claim observational validation.

The perturbative treatment of \citet{1991ApJ...368...79A} deals with mirror symmetries about the plane of the Galaxy. It has been introduced in the technique we are adopting from Pasetto PhD Thesis 2005 \citep[e.g.,][]{2006A&A...451..125V} where more detail has been given as well as comments about its implementation and observational validation. Nowadays, this work represents a good balance between simplicity and robustness. More recent formulations can be investigated in the future \citep[e.g.,][]{2009A&A...500..781B}. 

The treatment of \citet{1969ApJ...155..721L} is referred as Density Wave theory (DWT) and we will review in what follows the literature that attempts to validate it from the observational point of view. 

The history of the attempts to find an explanation of the spiral features of the MW and external galaxies is long standing and still matter of debate. We recall here (without the presumption to be complete) a few works of observational nature that inspired our star-count implementation of  DWT. The interested reader can look at books such as \citet{1991pagd.book.....S} or \citet{2014dyga.book.....B}. 

\subsection{Observational studies of Density wave theory}\label{Sec1.1}
 %
The existence  of a theory interpreting the spiral arm phenomenon \citep[e.g.,][]{1964SvA.....8..202M, 1964ApJ...140..646L, 1966SvA....10..442M, 1967SvA....10..738M,1968SvA....12..411M,1969ApJ...155..721L,1969SvA....13..411M,1969SvA....13..252M} spurred many research groups to find observational evidence that could either support or deny such a theory. The first attempt to interpret the  mean properties of  observational velocity fields of young stars in terms of the DWT  was by \citet{1973A&A....27..281C}. \citet{1973A&A....27..281C} set up  a  method to interpret the observations in terms of the DWT based on  two  simple ingredients: a multidimensional parametric fit and an asymptotic expansion on small parameters of the basic equations governing the kinematics of the DWT. This seminal study inspired many other studies  in which different results were obtained mainly due to either the adopted multidimensional fitting procedure or the large number of involved parameters  or  the different data sets in usage  and their local/non-local nature in the configuration space. Local models of the velocity space have been considered with asymptotic expansions on different small parameters \citep[e.g.,][]{1977MNRAS.179..663N, 1981A&A....99..311B, 1981MNRAS.196..659B, 1991A&A...241...57C, 1997A&A...323..775M, 1999A&A...341...81M, 2001A&A...372..833F, 2001A&A...379..634G}.

Nowadays this research is still far from being complete \citep[see, e.g.,][]{2013A&A...550A..91J, 2013MNRAS.433.2511G, 2014NewA...29....9G, 2014MNRAS.442.2993V, 2014MNRAS.440.1950R,2016ApJ...821...53V}. Recent studies consider more complex models based on four spiral arms \citep[e.g.,][]{2001ApJ...546..234L,2016AJ....151...55V} and their connections with the pattern of chemical properties of the MW \citep[e.g.,][]{2003ApJ...589..210L, 2004A&A...413..159A} and  the not monotonic features of the MW rotation curve \citep[e.g.,][]{ 2013MNRAS.435.2299B}. Finally, this research field has been recently boosted by numerical simulations. N-body solvers are achieving higher and higher resolution and although they are still missing a complete self-consistent understanding of the spiral arm dynamics, several numerical techniques \citep[e.g., the tree-code,][]{1986Natur.324..446B} allow us to simulate the gravitational interactions among millions of particles with masses of the order of a few thousand solar masses or less \citep[][]{2013ApJ...766...34D, 2012MNRAS.421.1529G, 2012MNRAS.426..167G}. 

The logic flux of the paper is as follows. We first want to present (Section \ref{Sec2}) the concept of stellar population taken from a theory developed in its general form in \citet{2012A&A...545A..14P} and here adapted to the specific case of the MW stellar populations. This will allow us to generalize the previously introduced concept of star-counts in a larger theoretical framework, to set a few assumptions, and to emphasize the goals of this novel Galaxy Model.
We present the normalization of the star-count equation for a field of view of arbitrary size in Section \ref{Sec3} and this allows us to define the density profiles and the consequent MW potential shape (Section \ref{Sec4}). This axysimmetric potential represents the basis for the development of a self-consistent spiral treatment presented in the following section, but as explained above, the formulation adopted in our approach is fully analytical, hence parametric, and so are the density-potential couple introduced in Section \ref{Sec4}. This leaves us with a large number of parameters to deal with in order to model the MW. In section \ref{Sec5} a genetic algorithm is introduced for the study of these parameters which are used to study the MW data surveys. This leads us to the setting of the MW axisymmetric potential (Table 1) that represents the axysimmetric basis used to develop the spiral arms perturbation theory. Hence, in Section \ref{Sec6} the spiral arms formalism is presented with its implication for the density (Section \ref{Sec6.1}) and the CMDs (Section \ref{Sec6.2}) once an ad-hoc extinction model is considered (Section \ref{Sec6.2.1}). The velocity field description is presented in Section \ref{Sec7}. A direct comparison with the most popular Besan\c{c}on model is detailed in Section \ref{Sec8} and the conclusions are presented in Section \ref{Sec9}.

\section{Theory of stellar populations}\label{Sec2}
Robust mathematical foundations for the concept of stellar populations are still missing, but recently \citet{2012A&A...545A..14P} proposed a new formulation  for it. We briefly summarize here the analysis of \citet{2012A&A...545A..14P} because it is the backbone of the population synthesis model we are going to describe here. This  approach extends the classical concepts presented in books as \citet{2005essp.book.....S} or \citet{2011spug.book.....G} to include a phase-space treatment for the stellar populations. These definitions will be crucial for the modelling approach and to formally define our goals. Moreover they will allow us to fix some assumptions we exploited during our work. Hence, we proceed to pin down here the more specific points that in the theory proposed in \citet{2012A&A...545A..14P} are introduced in complete generality.  

\subsection{Theoretical framework: $\mathbb{E_{\rm{MW}}}$}\label{Sec2.1}
\begin{figure}
\includegraphics[width=\columnwidth]{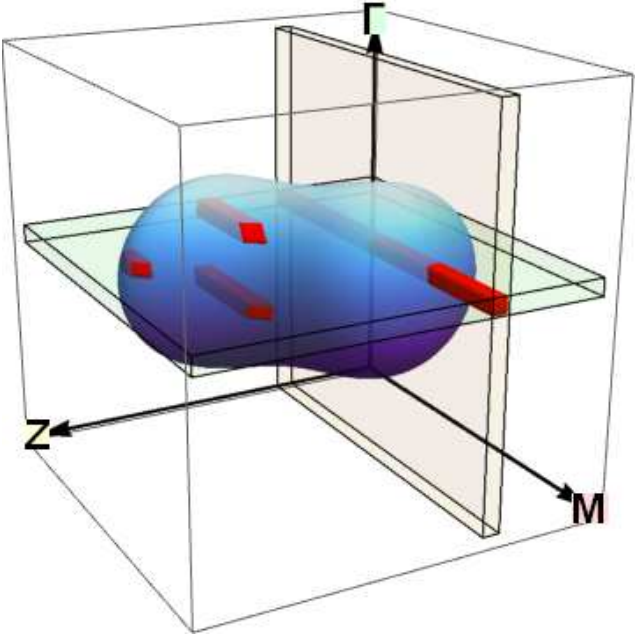}
\caption{Cartoon representing the concept of a foliation of a CSP over SSPs. SSPs result as an intersection (red rectangular area in the Figure) of planes of constant metallicity $dZ$ (light-brown-colour) and constant phase-space $d{\bm{\Gamma }}$ plane (light-green colour) with the CSP (blue). The SSPs are the fundamental ``atoms'' to build up the CSPs. Axes are in arbitrary units.
\label{Fig1} }
\end{figure}

We define every assembly of stars born at different time, positions, with different velocities, masses and chemical composition a composite stellar population (CSP). The space of existence for the Milky Way CSP, $\mathbb{E_{\rm{MW}}}$  is considered as the Cartesian product of the phase-space ${\bm{\Gamma }} = \left( {{x_1},{x_2},...,{x_{3N}},{v_1},{v_2},...,{v_{3N}}} \right)$ ($N$ number of stars of the CSP), the mass space $M$, and the chemical composition space $Z$, $\mathbb{E_{\rm{MW}}} \equiv M \times Z \times {\bm{\Gamma }}$. The inclusion of the time $t$ introduces the ``extended''-existence space $\mathbb{E_{\rm{MW}}} \times \mathbb{R}$. A more formal geometrical definition of this space and its dimensionality for the interested reader is given in \citet{2012A&A...545A..14P}. Because in the extended existence space the MW stars move continuously (losing mass, enriching in metals and travelling orbits in the phase-space) we can safely define a distribution for the CSP in $\mathbb{E_{\rm{MW}}}$, say ${f_{{\rm{CSP}}}^{{\rm{MW}}}} \in {\mathbb{R}^ + }$ real always positive function under the assumption of continuity and differentiability, i.e. ${f_{{\rm{CSP}}}^{{\rm{MW}}}} \in {C^\infty}\left( {{\mathbb{R}^ + }} \right)$. We consider now a sample of identical MW-systems whose initial condition spans a sub-volume of $\mathbb{E_{\rm{MW}}}$, let us refer to it as the ``MW-ensemble''. The number of these systems, $dN$, spanning a mass range, $dM$, a metallicity range, $dZ$, and phase-space interval, $d{\bm{\Gamma }}$, at the instant $t$, is given by
\begin{equation}\label{Eq2}
	dN = N{f_{{\rm{CSP}}}^{{\rm{MW}}}}dMdZd{\bm{\Gamma }},
\end{equation}
and the total number of systems in the ensemble is fixed, finite,  and subject to the important normalization condition
\begin{equation}\label{Eq3}
	\int_{\mathbb{E_{\rm{MW}}}}^{} {{f_{{\rm{CSP}}}^{{\rm{MW}}}}dMdZd{\bm{\Gamma }}}  = 1.
\end{equation}
\cite{2012A&A...545A..14P} proceeded with a foliation of $\mathbb{E_{\rm{MW}}}$ in orthogonal subspaces of metallicity, $dZ$, and phase-space alone, $d{\bm{\Gamma }}$, to define a simple stellar population (SSP) as one of these elemental units. A cartoon of the concept of SSPs is presented in Fig. \ref{Fig1}.

As evident, we can assume that the DF of a CSP can be written as the sum of disjoined DF of SSP,
\begin{equation}\label{Eq4}
	{f_{{\rm{CSP}}}^{{\rm{MW}}}} = \sum\limits_{\rm{SSP} = 1}^n {{f_{{\rm{SSP}}}}},
\end{equation}
where ${f_{{\rm{SSP}}}} = {f_{{\rm{SSP}}}}\left( {M,{Z_0},{{\bm{\Gamma }}_0};{t_0}} \right)$ is the DF of a single stellar population born at time ${t_0}$. In this framework we can give a rigorous geometrical interpretation and definition of CSPs and SSPs \citep[see][]{2012A&A...545A..14P}. This number is related to the granularity of ${f_{{\rm{CSP}}}^{{\rm{MW}}}}$ and hence to the growth of the entropy of the MW as a whole. The study of the number of SSPs is beyond the scope of the present paper and here we will limit $n$ to be a fixed parameter for simplicity. Further considerations on this number are left in Sec. \ref{Sec5} in relation to the Machine Learning approach used to study a given set of observations. As the time passes, the stellar population evolves. According to their masses the stars leave the main sequence and soon after die (supernovae phase) or enter into quiescent stages (white dwarfs phase) injecting chemically processed material into the interstellar medium in form of supernovae remnants or winds. In the same way, the evolution of a SSP in the phase-space obeys the Liouville equation $\frac{{\partial {f_{\bm{\Gamma }}}}}{{\partial t}} =  - \iota \mathcal{L}\left[ {{f_{\bm{\Gamma }}}} \right]$ with $\mathcal{L}\left[ * \right]$  Liouville operator and $\iota $  imaginary units. 

It is of interest for us to recall a few tools which are useful in studying the MW. Within this framework we will make use of the following concepts:
\begin{itemize}
\item 
\textbf{Present-day mass function.}
This is the result of the marginalization of ${f_{{\rm{CSP}}}^{{\rm{MW}}}}$ over the metallicity $Z$  and phase space ${\bm{\Gamma }}$:
\begin{equation}\label{Eq5}
\int_{{\mathbb{R}^{6N}} \times \mathbb{R}}^{} {N{f_{{\rm{CSP}}}^{{\rm{MW}}}}dZd{\bm{\Gamma }}}  = \hat \xi \left( {M;t} \right).
\end{equation}
This can be expressed, e.g., by the approximate relation 
\begin{equation}\label{Eq6}
\hat \xi \left( {M;t} \right) = \left\{ 
{\begin{array}{*{20}{c}}
{\xi \left( M \right)\frac{{{t_{{\rm{MS}}}}}}{\tau }}&{{t_{MS}} < \tau } \\ 
{\xi \left( M \right)}&{{t_{MS}} > \tau ,} 
\end{array}} \right.
\end{equation}
where $\tau  = t - {t_0}$ is the age of the stellar population, ${t_{{\rm{MS}}}} = {t_{{\rm{MS}}}}\left( M \right)$ is the age at which a star exits the main sequence (MS), and $\xi \left( M \right)$ is the initial mass function (IMF) of the MS stars. 
In our model the form of the IMF profiles is limited to multi-segmented power laws and log-normal formula to cover most of the literature. All the IMFs matching a single/multi power-law family of models of the form $\xi \left( M \right)dM = {\xi _0}{\left( {\frac{M}{{{M_ \odot }}}} \right)^{ - \alpha }}\frac{{dM}}{{{M_ \odot }}}$ \citep[e.g.,][]{1955ApJ...121..161S, 1993MNRAS.262..545K, 2001MNRAS.322..231K} are considered as well as the log-normal family of profiles $\xi \left( M \right)dM = {\xi _0}{M^{ - 1}}{e^{ - \frac{1}{{2\sigma _M^2}}{{\left( {\log \frac{M}{{{M_{\min }}}}} \right)}^2}}}\frac{{dM}}{{{M_ \odot }}}$ \citep[e.g.,][]{2003PASP..115..763C} where $\xi _0$, ${{M_{\min }}}$ and ${\sigma _M}$ are free parameters. 
\item 
\textbf{Age-metallicity relation.}
By integration of the DF over the mass, $M$, and the phase-space, ${\bm{\Gamma }}$, we can define the relation,
\begin{equation}\label{Eq7}
	\int_{{\mathbb{R}^{6N}} \times \mathbb{R}}^{} {N{f_{{\rm{CSP}}}^{{\rm{MW}}}}d{\bm{\Gamma }}dM}  = \chi \left( {Z;t} \right),
\end{equation}
which gives the number of stars formed per metallicity interval at the time $t$. Although several studies have been devoted to investigate the age-metallicity relation \citep[][]{2000A&A...358..850R, 2006A&A...453L...9R, 1996A&A...313..792P}, the small volume of the Galaxy covered by the data  does not allow us to apply these age-metallicity relationships  to a global scale model \citep[][]{2014A&A...565A..89B}. The problem   becomes even more puzzling for specific stellar components such as the stellar halo \citep[see, e.g., ][ for the globular cluster case]{2013MNRAS.436..122L}. Even though the age-metallicity relation of \citet{2006A&A...453L...9R} is included  in our model, we will not use it as a standard assumption. 
\item 
\textbf{Phase-space DF and age-velocity dispersion relation.}
By marginalizing ${f_{{\rm{CSP}}}^{{\rm{MW}}}}$ over the mass and metallicity sub-space we can write the formal relation,
\begin{equation}\label{Eq8}
	\int_{{\mathbb{R}^{2}}}^{} {M{f_{{\rm{CSP}}}^{{\rm{MW}}}}dMdZ}  = {e^{ - \iota \mathcal{L}t}}{f_{\bm{\Gamma }}}\left( {{\bm{\Gamma }};{t_0}} \right),
\end{equation}
whose analysis within the framework of a perturbative approach of the DWT will be subject of this paper in the following sections. Here we anticipate only that by taking the moments on the velocities of Eq.\eqref{Eq8} we can obtain the important \textit{age-velocity } dispersion relation implemented in our model:
\begin{equation}\label{Eq9}
\begin{aligned}
			{\sigma _{\bm{v}}}\left( {{\bm{x}};t} \right) &\equiv \int_{{\mathbb{R}^{3N}}}^{} {{d^{3N}}{\bm{v}}{{\left( {{\bm{v}} - {\bm{\bar v}}} \right)}^{ \otimes 2}}\int_{{\mathbb{R}^{2}}}^{} {M{f_{{\rm{CSP}}}^{{\rm{MW}}}}dMdZ} }  \\ 
			&= \int_{{\mathbb{R}^{3N}}}^{} {{{\left( {{\bm{v}} - {\bm{\bar v}}} \right)}^{ \otimes 2}}{e^{ - \iota \mathcal{L}t}}{f_{\bm{\Gamma }}}\left( {{\bm{\Gamma }};{t_0}} \right)} {d^{3N}}{\bm{v}}, 
\end{aligned}
\end{equation}
with ${{\bm{a}}^{ \otimes n}}$ a standard tensor $n$-power of the generic vector ${\bm{a}}$ accounted for its symmetries. A simplified version of this relation for a collisionless stellar system (i.e. where the Liouville operator introduced above is replaced by the Boltzmann operator for collisionless stellar dynamics) are implemented in our model with data interpolated from the values of the work of \citet{2012A&A...547A..71P} and \citet{2004A&A...423..517R} (see Eq.\eqref{Eq44}).
\end{itemize}
Finally, by extension of the previous integral formalism of Eq. \eqref{Eq5}, \eqref{Eq7} and \eqref{Eq8} we introduce the following relations of interest to us:
\begin{itemize}
	\item \textbf{Metallicity/phase-space relationship. }
This relation is formally defined by $\eta \left( {Z,{\bm{\Gamma }};t} \right) \equiv \int_{}^{} {{f_{{\rm{CSP}}}^{{\rm{MW}}}}dM}$ and more interestingly we can project it onto the configuration space:
\begin{equation}\label{Eq10}
	\hat \eta \left( {Z,{\bm{x}};t} \right) = \int_{{\mathbb{R}^{3N}} \times \mathbb{R}}^{} {{f_{{\rm{CSP}}}^{{\rm{MW}}}}{d^{3N}}{\bm{v}}dM}.
\end{equation}
There is indeed observational evidence of the presence of this relation in the chemical radial gradients in the configuration space of the MW thin disk component \citep[see, e.g., ][]{2014A&A...568A..71B, 2013A&A...559A..59B} and it can be eventually implemented on the thick disk \citep[][]{2014ApJ...784L..24C}.
\end{itemize}
For completeness we remind the reader that the stellar-mass/metallicity relation can easily be defined and implemented in our model as presented in \cite{2012A&A...545A..14P} once a larger sample of asteroseismology data becomes available (see references in Section \ref{Sec1}). 

\textit{The goal of our research is to develop a technique to investigate ${\mathbb{E}_{{\rm{MW}}}}$, the existence space of the MW, through  the relations that result from the projection of the unknown ${f_{{\rm{CSP}}}^{\rm{MW}}}$  in the mentioned subspaces of Eqs.\eqref{Eq6}, \eqref{Eq7}, \eqref{Eq8} and \eqref{Eq10}.}

To this aim, we need to relate ${\mathbb{E}_{{\rm{MW}}}}$ to the space of observations. This is made possible by the star-count equation, Eq.\eqref{Eq1}, in combination with  Eq.\eqref{Eq2} in the space of the observational data. To this aim, one has also to solve a crucial point of difficulty with  Eq.\eqref{Eq1}, i.e.  the large fields of view that are often involved.

\section{Star-count equation for large sky coverage}\label{Sec3}
Nowadays and increasingly in the future, we face the challenge of large sky coverage surveys where the gradients of the underlying MW stellar density distributions sensibly vary across the covered survey area. Already large surveys (SDSS, RAVE, SEGUE, etc.) present these characteristics, and the ongoing whole sky survey by Gaia, due to the depth of the magnitude limit and the amplitude of the solid angle considered ($d\overset{\lower0.5em\hbox{$\smash{\scriptscriptstyle\frown}$}}{\Omega }  = 4\pi $), will provide us an enormous amount of data to be considered. 
If a survey spans a large solid angle $d\overset{\lower0.5em\hbox{$\smash{\scriptscriptstyle\frown}$}}{\Omega } $ and has a very deep magnitude limit, then the number of stars per field of view (FOV) becomes large and its realization on a star-by-star basis becomes unpractical. For example, the marginalization of a ${f_{{\rm{CSP}}}^{{\rm{MW}}}}$ over $d{\bm{\Gamma }}$ for large-scale survey data produces a section over $dZ \times dM$, i.e. a Hertzsprung-Russel or a colour magnitude diagram (CMD) that can be over-dense: to realize it graphically we should draw dots-over-dots and count them. This process should be repeated every time we change a single parameter to see the effect of the variation until suitable fitting is achieved. To surpass these CMD realization problems, \citet{2012A&A...545A..14P} presented a novel technique able to substitute the generation of synthetic stars with the computing of a PDF. The convolution of several SSPs along a line-of-sight (l.o.s.), thanks to Eq.\eqref{Eq4}, was then substituting the Monte-Carlo generation of stars for a FOV, de-facto changing the concept of a ``star-count'' model with a probability distribution function (PDF) model.  

We adopted here the same technique to speed up the generation of the ${f_{{\rm{CSP}}}^{{\rm{MW}}}}$, eventually walking back to a star-count type of model by populating the PDF obtained for ${f_{{\rm{CSP}}}^{{\rm{MW}}}}$ only if required. The stellar SSP database used to build the ${f_{{\rm{CSP}}}^{{\rm{MW}}}}$ is the same adopted in \citet{2012A&A...545A..14P}, though any other SSP database can be easily implemented virtually making the modelling approach independent of any particular stellar physics recipes adopted by one or another research group (rotation, overshooting, $\alpha$-enhancement, helium enrichment etc.). More details of this is described for the interested reader in \citet{2012A&A...545A..14P}. 

Nevertheless, this process of populating the PDF for each FOV (that can be as large as the full sky) has to be treated with attention because of the normalization relation Eq.\eqref{Eq3}. In particular, the number of stars generated along the l.o.s. and appearing in the final CMD has to correctly account for the underlying mass fraction of each stellar component $j$ of the Galaxy.

\begin{figure}
\includegraphics[width=\columnwidth]{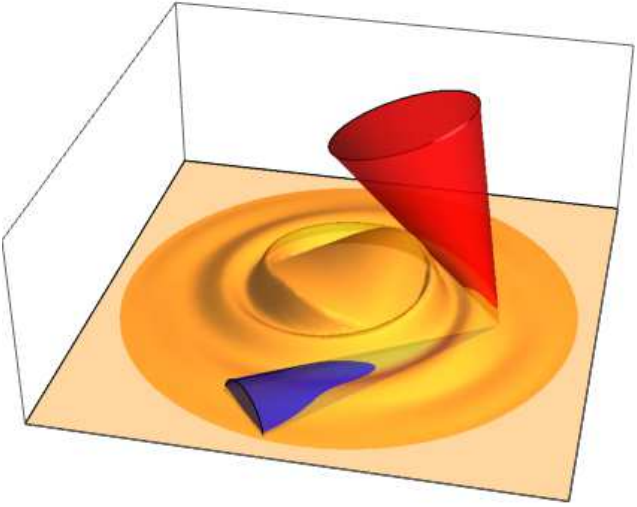}
\caption{Spiral arm stellar isocontour. The intersection of the yellow surface with the FOV (e.g., violet or red) is the integral performed in Eq.\eqref{Eq11}. Two arbitrary solid angles, in blue or red, intersect a single spiral arm SSP over its complicated density profile (orange), from a common solar position slightly outside the plane of the galaxy. The relative contribution to the number of stars in a given l.o.s. is the result of the intersection of the global CSP with the arbitrary cone of the l.o.s..
\label{Fig2} }
\end{figure}

Historically, to deal with Eq.\eqref{Eq1}, or its generalized form in Eq.\eqref{Eq2}, the approach was based on the sum of several close FOV of negotiable opening angle. It was required for $d\overset{\lower0.5em\hbox{$\smash{\scriptscriptstyle\frown}$}}{\Omega } $ to be very small as well as the number of stars per population ${N_j}$. The result of these assumptions was that the underlying density distributions within $d\overset{\lower0.5em\hbox{$\smash{\scriptscriptstyle\frown}$}}{\Omega } $ were to a good approximation constant (if the survey was not too deep in magnitude and hence ${r_{{\rm{hel}}}}$  not too deep). To solve the star-count equation under these approximations was a trivial exercise and in the past decades it has been indeed done by several works in this research area \citep[e.g.,][, and references therein]{2003A&A...409..523R, 2000AJ....119..813M, 2006A&A...451..125V, 2002A&A...392.1129N, 2005A&A...436..895G}.
If the hypothesis of small $d\overset{\lower0.5em\hbox{$\smash{\scriptscriptstyle\frown}$}}{\Omega }  = dldb \cos b$ is to be relaxed, the computing of this number has to be performed numerically as follows: 
\begin{equation}\label{Eq11}
	\begin{aligned}
		N &= \int_{{\mathbb{R}^{3N}}}^{} {{d^{3N}}{\bm{x}}\int_{{\mathbb{R}^{3N}}}^{} {{d^{3N}}{\bm{v}}\int_{{\mathbb{R}^{2}}}^{} {M{f_{{\rm{CSP}}}^{{\rm{MW}}}}\left( {M,Z,{\bm{\Gamma }}} \right)dMdZ} } }  \\ 
		&= \int_{{\mathbb{R}^{3N}}}^{} {{d^{3N}}{\bm{x}}\int_{{\mathbb{R}^{3N}}}^{} {{d^{3N}}{\bm{v}}{e^{ - \iota \mathcal{L}t}}{f_{\bm{\Gamma }}}\left( {\bm{\Gamma }} \right)} }  \\ 
		&= \int_{{\mathbb{R}^{2}}}^{} {d\overset{\lower0.5em\hbox{$\smash{\scriptscriptstyle\frown}$}}{\Omega } } \int_{{\mathbb{R}}}^{} {d{r_{hel}}\mathfrak{J}\rho \left( {{\bm{x}};t} \right),}  
	\end{aligned}
\end{equation}
where $\mathfrak{J} = r_{{\rm{hel}}}^2\left| {\cos b} \right|$ is the Jacobian of the transformation $T$ between the system of galactocentric coordinates $\left( {O,{\bm{x}}} \right)$ to standard galactic coordinates $\left( { \odot ,{r_{{\rm{hel}}}},l,b} \right)$:
\begin{equation}\label{Eq12}
	T:\left\{ \begin{array}{rcl}
		x &=& {R_ \odot } - {r_{{\rm{hel}}}}\cos b\cos l \hfill \\
		y &=& {r_{{\rm{hel}}}}\cos b\sin l \hfill \\
		z &=& {z_ \odot } + {r_{{\rm{hel}}}}\sin b, \hfill \\ 
	\end{array}  \right.
\end{equation}
where ${R_ \odot } = \sqrt {{x_ \odot^2 } + {y_ \odot^2 }} $. After this integral is evaluated, the relative number of stars within a given FOV is obtained as a function of observable quantities (e.g., the galactic coordinates) no matter how large the FOV is (Fig. \ref{Fig2}). Although ${r_{{\rm{hel}}}}$ can be unbounded, in practice it is limited by the survey magnitude limits with Pogson’s law and the dust extinction by taking into account an extinction model (Section \ref{Sec6.2.1}). We point out how the cone-geometry of Fig. \ref{Fig2} for the volume $\int_{}^{} {d\overset{\lower0.5em\hbox{$\smash{\scriptscriptstyle\frown}$}}{\Omega } d{r_{{\rm{hel}}}}} $ is of exemplificative nature. In practice, because every observed star has much larger uncertainty in distance ${r_{{\rm{hel}}}} + \delta {r_{{\rm{hel}}}}$ than in angular position$\left\{ {l \pm \delta l,b \pm \delta b} \right\}$, i.e. $\left| {\frac{{\delta {r_{{\rm{hel}}}}}}{{{r_{{\rm{hel}}}}}}} \right| \gg \left| {\frac{{\delta l}}{l}} \right|$  and $\left| {\frac{{\delta {r_{{\rm{hel}}}}}}{{{r_{{\rm{hel}}}}}}} \right| \gg \left| {\frac{{\delta b}}{b}} \right|$, the mapping of the synthetically generated ${f_{{\rm{CSP}}}^{{\rm{MW}}}}$ of every survey has a different nature. $\left\{ {l,b} \right\}$ are not randomly generated but assumed from the data that we want to analyse without errors while ${r_{{\rm{hel}}}}$ is randomly generated within ${r_{{\rm{hel}}}} \pm \delta {r_{{\rm{hel}}}}$ depending on the particular selection function.

\section{The axisymmetric Milky Way: density distributions, gravitational potentials and kinematics}\label{Sec4}
In the context of the theory of stellar population introduced above we can simplify Eq.\eqref{Eq8} and obtain the mass density $\rho({\bm x})$ as follows:
\begin{equation}\label{Eq13}
	\begin{aligned}
		\rho  &= \int_{{\mathbb{R}^{3N}} \times {\mathbb{R}^2}}^{} {M{f_{{\rm{CSP}}}^{{\rm{MW}}}}dMdZ} {d^{3N}}{\bm{v}} \\ 
		&= \int_{{\mathbb{R}^{3N}} }^{} {{e^{ - \iota \mathcal{L}t}}{f_{\bm{\Gamma }}}\left( {{\bm{\Gamma }};{t_0}} \right){d^{3N}}{\bm{v}}}  \\ 
		&\simeq \int_{{\mathbb{R}^{3}}}^{} {{e^{ - \iota \mathcal{B}t}}{f_{\bm{\gamma }}}\left( {{\bm{\gamma }};{t_0}} \right){d^3}{\bm{v}}}, 
	\end{aligned}
\end{equation}
where in the last row of the equation we reduced the dimensionality of the phase-space by remembering that it is possible to show that the two-body relaxation time ${t_{2b}}$ (considered in the approximation of independent-hyperbolic encounters) is long enough to allow us to treat the Galaxy to a good approximation as a ``collisionless'' system. Hence we can substitute the discrete stellar distribution with a continuous density profile, and the Liouville operator $\mathcal{L}$ can be substituted with the more simple Boltzmann operator $\mathcal{B}\left[ {{f_{\bm{\gamma }}}} \right] \equiv \iota \left\{ {H,{f_{\bm{\gamma }}}} \right\}$, with $H$ one-particle Hamiltonian and $\left\{ {*,*} \right\}$ the Poisson brackets ($\iota$ is the imaginary unit). 

Unfortunately the explicit form of ${f_{{\rm{CSP}}}^{{\rm{MW}}}}$ is unknown (the blue manifold in Fig. \ref{Fig1}) or has to be inferred just from simple theoretical considerations. For this reason, we decided to base our modelling technique on the density distributions of stars and dark matter. From the density profiles the potential and hence the kinematics is computed. Furthermore, from the same density profiles the relative number of stars per bin of colour and magnitude along a l.o.s. in the CMD is computed. This approach is not the only possible way to proceed in analytical modelling, but we are guided by the explicit intention to present a model focused on the interpretation of the data, where the data are the protagonist in leading our understanding of the phenomenon ``Galaxy''. 

Therefore, it is of paramount importance to assign to each  component of the MW a plausible density profile
$\rho({\bm x})$ to derive  a correct global gravitational potential. In the following, we present our treatment of the Poisson equation and hence the global gravitational potential of the MW.
These results will represent the axisymmetric foundations  of our description  the DWT  of spiral arms. The gravitational potential is derived for all components of the MW even if we will focus only on the disk components for which, thanks to their proximity, data of good quality can be acquired and accurate descriptions are possible.

In our model, the location of the Sun is assumed to be  at ${{\bm{x}}_ \odot } = \left\{ {{R_ \odot },{\phi _ \odot },{z_ \odot }} \right\} = \left\{ {8.00,0.00,0.02} \right\}$  kpc in a reference frame centred on the (yet unknown)  mass barycentre of the axisymmetric model of the MW we are going to build up.

\subsection{Axisymmetric SSP models}\label{Axisymmtot}
\subsubsection{Thin and thick disk}\label{AxisymmtotTTD}
As mentioned above in Eq.\eqref{Eq4} we consider a multi-component model of stellar populations. For the ${i^{{\rm{th}}}}$ component of the thin or thick disk, we implemented a double exponential form of the density profiles, that is, with an exponential profile decreasing with Galactocentric radius and vertical distance from the plane. Alternative vertical profiles (power-law and secant-square) are available for investigation but not breaks of the exponential profiles has been implemented \citep[e.g.,][]{2006A&A...454..759P}. Because we are going to develop a kinematics model, no time dependence of the density profiles is assumed. Written in cylindrical coordinates to exploit the $\phi $-symmetry the profile reads:
\begin{equation}\label{Eq14}
	{\rho _D}\left( {R,\phi ,z} \right) = {\rho _ \odot }{e^{ - \frac{{R - {R_ \odot }}}{{{h_R}}} - \frac{{z - {z_ \odot }}}{{{h_z}}}}}.
\end{equation}
This parametric formalism depends on the density at the solar neighbourhood ${\rho _ \odot }$ and two scale parameters: scale length ${h_R}$  and scale height ${h_z}$  for each stellar population considered. It does not contain an explicit dependence on $\phi $. The potential is conveniently expressed as function of one single integral with integrand depending on the Bessel function \citep[e.g.,][]{1987A&A...180...94B} being hence extremely rapid to compute:
\begin{equation}\label{Eq15}
	\begin{aligned}
		{\Phi _D}\left( {R,\phi ,z} \right) &=  - 4\pi G{\rho _0}h_R^{ - 1}\int_0^\infty  {{J_0}\left( {kR} \right){{\left( {h_R^{ - 2} + {k^2}} \right)}^{ - 3/2}}  }  \\ 
		&\times \frac{{h_z^{ - 1}{e^{ - k\left| z \right|}} - k{e^{ - h_z^{ - 1}\left| z \right|}}}}{{h_z^{ - 2} - {k^2}}}dk 
	\end{aligned}
\end{equation}
where ${J_0}$ is the Bessel function of the first kind \citep[e.g.,][]{1972hmfw.book.....A} and the scale parameter for each component ${i^{th}}$ should be taken into account but omitted for the sake of simplicity. 

The kinematic description of these disk populations is in principle obtainable self-consistently from the numerical solution of systems as Eq.(22) in \citet{1992MNRAS.256..166C}. However, aside from the numerical difficulties, this description would require difficult observational validation (e.g., by requiring second order derivatives) that makes it challenging to apply for precise surveys such as the forthcoming Gaia. For this reason we have chosen in favour of an approach based on the Jeans equations. Following \citet{2012A&A...547A..71P} and  \citet{2012A&A...547A..70P} we are not assuming a shape for the ${f_{\mathbf{\gamma }}}\left( {{\mathbf{\gamma }};t} \right)$ but working on the methods-of-moments of the collisionless Boltzmann equation in agreement with the simplification of Eq.\eqref{Eq13}. The mean circular velocity and asymmetric drift can then be studied with the relation:
\begin{equation}
\begin{aligned}
		{{\bar v}_\phi }\left( {R,z} \right) &= \left( \begin{gathered}
		\left| {{{\bar v}_\phi }} \right|\frac{{{r_{{\text{hel}}}}\cos b\cos l}}{R} \hfill \\
		\left| {{{\bar v}_\phi }} \right|\frac{{{R_ \odot } - {r_{{\text{hel}}}}\cos b\cos l}}{R} \hfill \\
		0 \hfill \\ 
	\end{gathered}  \right) \hfill, \\
		\left| {{{\bar v}_\phi }} \right| &= \left( {v_c^2 - \frac{{\partial \ln {\rho _D}}}{{\partial \ln R}}\left( {\sigma _{RR}^2 + \sigma _{Rz}^2} \right) + \left( {\sigma _{RR}^2 + \sigma _{\phi \phi }^2} \right)} \right. \hfill \\
		&{\left. { + R\left( {\frac{{\partial \sigma _{RR}^2}}{{\partial R}} + \frac{{\partial \sigma _{Rz}^2}}{{\partial z}} + \frac{{\partial {\Phi _{{\text{tot}}}}\left( {R,z} \right)}}{{\partial R}}} \right)} \right)^{1/2}} \hfill, \\ 
	\end{aligned} 
\end{equation}
where the dependence of the three non-null diagonal terms $\left\{ {{\sigma _{RR}},{\sigma _{\phi \phi }},{\sigma _{zz}}} \right\}$ on the configuration space will be written in cylindrical coordinates as:
\begin{equation}\label{Eq44}
	{\sigma _{ii,j}}\left( {R,z} \right) = {\nabla _z}{\sigma _{ii,j}}\left( {R,z} \right)\left( {\left| z \right| - {z_{j - 1}}} \right) + {\sigma _{ii,j - 1}}\left( {R,{z_{j - 1}}} \right),
\end{equation}
for $i = \left\{ {R,\phi } \right\}$ and $j = I,II,III$, while the vertical profiles of the thin disk stellar population and the non diagonal term $\sigma _{Rz}$ will be introduced in Section \ref{Sec7}  and where the underlying assumption of $\frac{{\sigma _{RR}^2}}{{\sigma _{zz}^2}} = {\rm{const}}.$ is assumed in agreement with the DWT for spiral arms introduced below. On the plane Eq.\eqref{Eq44} will be forced to match the profiles 
\begin{equation}\label{Eq45}
	\left\{ \begin{array}{rcl}
		\sigma _{RR}^2\left( {R,0} \right) = {\sigma _{RR}}_{, \odot }{e^{ - \frac{{R - {R_ \odot }}}{{{h_R}}}}} \hfill \\
		\sigma _{\phi \phi }^2\left( {R,0} \right) = {\sigma _{zz}}_{, \odot }{e^{ - \frac{{R - {R_ \odot }}}{{{h_R}}}}} \hfill \\
		\sigma _{zz}^2\left( {R,0} \right) = \left( {1 + \frac{{\partial \ln {v_c}}}{{\partial R}}} \right)\frac{{\sigma _{RR}^2\left( {R,0} \right)}}{2} \hfill, \\ 
	\end{array}  \right.
\end{equation}
and the gradients for the three vertical profiles I, II, III are a smooth interpolation of the values in Table \ref{Table2}. Moments of order up to four (obtained directly from cumulants) are evaluated as in Appendix of \cite{2012A&A...547A..70P}.

\begin{table}
\caption{Meridional plane profile of the velocity ellipsoid. For each fixed radius the vertical gradient of the two velocity dispersion tensors is indicated in the fourth and fifth columns. The third column gives the vertical ranges.}
\label{Table2}
\begin{tabular}{llll}
\hline
j 	& ${z_j}$ 												& $\Delta z$  										&  $\frac{{\partial {\sigma _{RR}}}}{{\partial z}}$,$\frac{{\partial {\sigma _{zz}}}}{{\partial z}}$\\
	  & $\left[ {{\rm{kpc}}} \right]$ & $\left[ {{\rm{kpc}}} \right]$ & $\left[ {km\;{s^{ - 1}}\;kp{c^{ - 1}}} \right] $	\\
 & & &  \\	
\hline		
0		& $z_0=0.0$ & $\left| z \right| = 0$ 														& 0.0,0.0 \\
I		& $z_1=0.5$ & $\left| z \right| \in \left] {0.0,0.5} \right]$ 	& 27.2,17.4 \\
II	& $z_2=1.0$ & $\left| z \right| \in \left] {0.5,1.0} \right]$  	& 9.7,5.4 \\
III	&  					& $\left| z \right| > 1.0 $  												& 0.0,0.0 \\
\hline
\end{tabular}
\end{table}

\subsubsection{Stellar halo}\label{AxisymmHalo}
For the ${i^{th}}$ stellar halo component of the MW we follow the model proposed in \citet{2003A&A...409..523R} because it is fine-tuned on the observational constraints, i.e. it is simple in its form but phenomenologically justified. In the original form Robin's profile reads:
\begin{equation}\label{Eq16}
	{\rho _{H*}}\left( r \right) = \frac{{{\rho _{0,H*}}}}{{{r_ \odot }}}\left\{ {\begin{array}{*{20}{c}}
  {{r^\alpha }}&{r > {h_{rH*}}} \\ 
  {h_{r{H^*}}^\alpha }&{r \leqslant {h_{rH*}},} 
\end{array}} \right.
\end{equation}
where ${\rho _{0,H*}}$ is the central stellar halo density, and ${h_{rH*}}$ the scale length parameter. Because we are interested in the potential formalism of this density model, we compute its corresponding potential solving Poisson’s equation in spherical coordinates and guaranteeing continuity (but not differentiability) to the formulation as follows:
\begin{equation}\label{Eq17}
{\Phi _{H*}}\left( r \right) = \left\{ \begin{gathered}
  4\pi G{\rho _{0,H*}}\frac{{r_ \odot ^{ - \alpha }}}{r}\frac{{\left( {\alpha  + 2} \right)h_{r,H*}^{\alpha  + 3} + {r^{\alpha  + 3}}}}{{\left( {\alpha  + 2} \right)\left( {\alpha  + 3} \right)}} \wedge r > {h_{r,H*}} \hfill \\
   - 2\pi G{\rho _{0,H*}}\frac{{3h_{r,H*}^2 - {r^2}}}{3}{\left( {\frac{{{h_{r,H*}}}}{{{r_ \odot }}}} \right)^\alpha } \wedge r \leqslant {h_{r,H*}}, \hfill \\ 
\end{gathered}  \right.
\end{equation}
where the scale parameters dependence of the ${i^{th}}$-component of stellar halo is omitted. 

For the kinematics description of the stellar halo several models for $f_{{\text{SSP}}}^{H*}$ are available in the literature with different flavours of parameters (multi-scale parameters, anisotropy, etc.). Nevertheless, self-consistent models rely on a description of the halo as a dynamically relaxed population. This is clearly an non-physical assumption for a stellar population as the halo which is composed by old stars but dynamically-young, non phase-mixed, rich in substructures as stellar streams \citep[see e.g.][]{2006ApJ...642L.137B}.

\subsubsection{Dark matter and hot-coronal gas}\label{AxisymmtotDM}
The only component that we can start from the potential shape is the dark matter component, because its presence is indirectly manifest but it is not directly observed. We select a simple balance between gravity and centrifugal forces for circular orbits to obtain the logarithmic potential:
\begin{equation}\label{Eq18}
	{\Phi _{DM}}\left( {R,\phi ,z} \right) \equiv \frac{{v_{0,DM}^2}}{2}\log \left( {h_{R,DM}^2 + {R^2} + {q^{ - 2}}{z^2}} \right),
\end{equation}
where ${v_0}$ is the scale velocity, $h_{R,DM}^{}$ the scale length and $q$ the flattening factor. The density profile can again be obtained by use of Poisson’s equation as:
\begin{equation}\label{Eq19}
	{\rho _{DM}} = \frac{{v_0^2}}{{4\pi G}}\frac{{{q^2}\left( {h_{r,DM}^2\left( {2{q^2} + 1} \right) + {R^2} + 2{z^2}} \right) - {z^2}}}{{{{\left( {{q^2}\left( {h_{r,DM}^2 + {R^2}} \right) + {z^2}} \right)}^2}}}.
\end{equation}
No compelling reasons exist so far to split  dark matter in more components. and the model does not consider sub-structures of the dark matter component \citep[e.g.,][]{2011ApJ...731...58Y}.  We expect not to detect granularity in the dark matter distribution from  the kinematic anomalies of the closest stars kinematics. In view of this,  the presence of granularity in  the dark matter  distribution, mimicking dark matter streams, are neglected in our model.

We complete the review by mentioning that optionally we can include axisymmetric components adding a hot coronal gas. This does not influence the closest stellar dynamics of the MW stars but in mass it is thought to contribute up to $ \sim 5 \times {10^{10}}{M_ \odot }$ within $ \sim 200$ kpc from the MW galaxy centre \citep[see, e.g.,][ and reference therein for a model including it]{2012A&A...542A..17P}. 

\subsubsection{Bulge}\label{AxisymmtotBul}
A separate work is in preparation on the kinematical treatment of the central part of the Galaxy which is of course very important. Unfortunately, up to now the modelling of the bulge is still imprecise and a subject of debate. A recent finding of \citet{2015ApJ...812L..29D} shows an example of the ongoing research and constantly changing knowledge that we have about the central regions of the MW. Nevertheless, in the total potential a bulge component has to be accounted for and we adopt the following spherical density-potential ``couple'' \citep[][]{1990ApJ...356..359H} from which kinematics is implemented too:
\begin{equation}\label{Eq20}
	\left\{ \begin{aligned}
		{\rho _B}\left( r \right) &= \frac{{{M_B}{h_{r,B}}}}{{2\pi r{{\left( {r + {h_{r,B}}} \right)}^3}}} \hfill \\
		{\Phi _B}\left( r \right) &=  - \frac{{G{M_B}}}{{r + {h_{r,B}}}}, \hfill  
	\end{aligned}  \right.
\end{equation}
where again the dependence of the scale parameters $\left\{ {{M_B},{h_{r,B}}} \right\}$, bulge mass and scale radius respectively, from the stellar bulge component is understood even though the subscript is omitted.

This series of equations represent the basic potential in axisimmetric approximation. The chosen density parameters that we are going to assume for these profiles are presented in Table \ref{Table1} as results of the technology that we are going to introduce in Sec.\ref{Sec5}.

These parameters are chosen in such a way that they nicely reproduce some important observational constraints (see also Appendix A of \citet{Pasetto2005}, and \citet{2012A&A...547A..70P}):

\subsection{Axisymmetric SSP constraints}
\subsubsection{Circular velocity}\label{Sec4.1}
To date several studies have covered the most important dynamical constraints on the MW potential, i.e. its rotation curve, from several data sets and with different techniques, both for the total stellar rotation, for gas rotation or for single MW stellar populations \citep[e.g.,][ to quote a few]{2012MNRAS.424L..44D, 2013RAA....13..849X, 2013MNRAS.432.2402F, 2014ApJ...785...63B, 2014A&A...563A.128L, 2011MNRAS.411.1480D, 2008ApJ...679.1288L, 2008ApJ...684.1143X}. From the potential adopted here we obtained the rotation curve analytically as ${v_c} = \sqrt {r\frac{{\partial {\Phi _{  }}}}{{\partial r}}} $, where the individual components are not difficult to evaluate. Using Eq.\eqref{Eq18} we get for the dark matter component:
\begin{equation}\label{Eq21}
	v_{c,DM}^2 = \frac{{{R^2}{v_0}^2}}{{{R^2} + h_{r,{\rm{DM}}}^2}};
\end{equation}
from Eq.\eqref{Eq17} for the stellar halo profile we have  
\begin{equation}\label{Eq22}
v_{c,{H^*}}^2 = \left\{ {\begin{array}{*{20}{c}}
  {\frac{{4\pi G}}{{{r^2}\left( {\alpha  + 3} \right)}}\frac{r}{{r_ \odot ^\alpha }}\sum\limits_{{H^*}}^{} {{\rho _{0,{H^*}}}\left( {{r^{\alpha  + 3}} - h_{r,{H^*}}^{\alpha  + 3}} \right)} }&{{h_{r,{H^*}}} < r} \\ 
  {\frac{{4\pi G}}{3}\frac{r}{{r_ \odot ^\alpha }}\sum\limits_{{H^*}}^{} {{\rho _{0,{H^*}}}h_{r,{H^*}}^\alpha } }&{r < {h_{r,{H^*}}},} 
\end{array}} \right.
\end{equation}
where we assumed the same $\alpha \forall {H^*}$, ${H^*} \in \mathbb{N}$ indexing the stellar populations, i.e. with a simple abuse of notation we wrote $v_{c,{H^*}}^2 = \sum\limits_{{H^*} = 1}^{{N_{{H^*}}}} {v_{c,{H^*}}^2} $ with ${N_{{H^*}}}$ number of  stellar halo populations implemented in Tab.\ref{Table1} (one in this case). For the bulge population from Eq.\eqref{Eq20}  we get
\begin{equation}\label{Eq23}
	v_{c,B}^2 = GR\sum\limits_B^{} {\frac{{{M_B}}}{{{{\left( {R + {h_{r,B}}} \right)}^2}}}},
\end{equation}
 with $B \in \mathbb{N}$  indexing the populations as above. The disk components are only slightly more complicated by the presence of the Bessel function that can be nevertheless handled numerically \citep[e.g.,][]{1972hmfw.book.....A} from Eq.\eqref{Eq15} in the form:
\begin{equation}\label{Eq24}
	v_{c,D}^2 = 4\pi GR\sum\limits_D^{} {{\rho _{0,D}}\int_0^\infty  {dk\frac{1}{{{h_{R,D}}{{\left( {h_{R,D}^{ - 2} + {k^2}} \right)}^{3/2}}}}\frac{{k{J_1}(kR)}}{{h_{z,D}^{ - 1} + k}}} },
\end{equation}
 	where $D \in \mathbb{N}$ indexes the disk populations.

\begin{figure}
\includegraphics[width=\columnwidth]{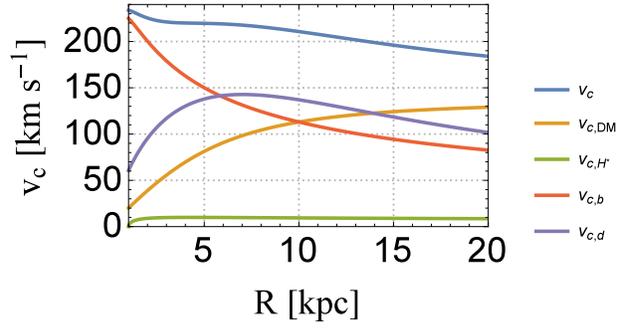}
\caption{Circular velocities as a function of the radius and contribution from each stellar component of the model of the galaxy. See text for the definitions of the equation for the rotation curve of each sub-population.
\label{Fig3} }
\end{figure}

Finally in Figure \ref{Fig3}, we present the velocity curves of the various components of the Galaxy according to the corresponding density profiles already discussed above and with parameters summarized in Table \ref{Table1} as a result of the technique presented in Sec.\ref{Sec5}.

\subsubsection{Oort functions: ${O^{\pm} }$}\label{Sec4.2}
The slope of the rotation curve, locally related to the Oort’s constants, has long been known to depend on the local gas content, which does not monotonically vary with the radius and contributes significantly to the local gradient of the rotation curve \citep[][]{1998MNRAS.297..943O, 2007MNRAS.377.1163M, 2003ApJ...599..275O}. The profile of these functions outside the solar neighborhood is what we refer to as ``Oort functions''. We will present in the next section a map distribution of the gas content in relation to dust distribution and extinction (Fig.\ref{Fig8}). In the future the estimation of the Oort function will represent a challenge for large kinematic surveys such as Gaia. The Oort functions are defined as  ${O^\pm}\left( R \right) \equiv  \pm \frac{1}{2}\left( {\frac{{{v_c}}}{R} \mp \frac{{d{v_c}}}{{dR}}} \right)$. Perhaps the greatest difficulty in estimating the Oort functions derives from the presence of the derivatives in their definition. Unfortunately, current observations of the rotational motion of the Milky Way are not good enough to allow a calculation of the derivatives in ${O^ \pm }\left( R \right)$ directly from the data \citep[][]{1986MNRAS.221.1023K,1987AJ.....94..409H}. 
 It is also possible to determine ${O^ + } - {O^ - } = \frac{{{v_c}}}{R}$ in an independent way from the individual values of ${O^ + }$  and ${O^ - }$ from proper motion surveys in the direction $l = 90^\circ $ or $l = 270^\circ $. Because along these directions the stars have a small dependence on the Galactocentric radius so the estimations are less affected by the radial dependence of the Oort functions. Finally, the combination $ - \frac{{{O^ - }}}{{{O^ + } - {O^ - }}}$ can be estimated from the velocity ellipsoid of random stellar motions. For the first function ${O^ + } \equiv O^ +\left( R \right)$ a compact formulation can be obtained as follows. For the dark matter component, 
\begin{equation}\label{Eq25}
	{O^+_{DM}}\left( R \right) = \frac{1}{{4{v_{c}}}}\frac{{2{R^3}v_0^2}}{{{{\left( {{R^2} + h_{r,DM}^2} \right)}^2}}},
\end{equation}
for the stellar halo components of Robin's density profiles 
\begin{equation}\label{Eq26}
{O^+_{{H^*}}}\left( R \right) =  - \frac{{\pi G}}{{{v_{c}}}}\frac{{r_ \odot ^{ - \alpha }}}{{\left( {\alpha  + 3} \right){r^2}}}\sum\limits_{{H^*}}^{} {{\rho _{0,{H^*}}}\left( {\alpha {r^{\alpha  + 3}} + 3h_{r,{H^*}}^{\alpha  + 3}} \right)} ,
\end{equation}
 for ${h_{r,{H^*}}} < r$, while it is clearly null inside the scale radius. For the bulge components’ contribution we can write 
\begin{equation}\label{Eq27}
	{O^+_B}\left( R \right) = \frac{{G{M_B}}}{{4{v_{c}}}}\sum\limits_B^{} {\frac{{3R + {h_{r,B}}}}{{{{\left( {R + {h_{r,B}}} \right)}^3}}}},
\end{equation}
 and finally for the stellar disks contribution we can write:
\begin{equation}\label{Eq28}
	{O^+_D}\left( R \right) = \frac{{\pi GR}}{{{v_{c}}}}\sum\limits_D^{} {\frac{{{\rho _{0,D}}}}{{{h_{R,D}}}}\int_{\mathbb{R}}^{} {dk\frac{{{k^2}{J_2}(kR)}}{{h_{z,D}^{ - 1} + k}}\frac{1}{{{{\left( {h_{R,D}^{ - 2} + {k^2}} \right)}^{3/2}}}}} }.
\end{equation}
Analogously for the ${O^ - }$ function we can write:  for the dark matter component
\begin{equation}\label{Eq29}
	{O^- _{DM}}\left( R \right) = \frac{{v_0^2}}{{2{v_{c}}}}\frac{{{R^3} + 2Rh_{r,DM}^2}}{{{{\left( {{R^2} + h_{r,DM}^2} \right)}^2}}},
\end{equation}
for the stellar halo it reads 
\begin{equation}\label{Eq30}
{O^- _{{H^*}}}\left( R \right) = \frac{{\pi G}}{{{v_{c}}}}\frac{{r_ \odot ^{ - \alpha }}}{{(\alpha  + 3){r^2}}}\sum\limits_{{H^*}}^{} {{\rho _{0,{H^*}}}\left( {(\alpha  + 4){r^{\alpha  + 3}} - h_{r,{H^*}}^{\alpha  + 3}} \right)} 
\end{equation}
Differently from ${O^ + }$ the contribution from the stellar halo in the central zones for Robin's profile is not null, but 
\begin{equation}\label{Eq31}
{O^- _{{H^*}}}\left( R \right) = \frac{{16\pi Gr}}{3}\sum\limits_{{H^*}}^{} {{\rho _{0,{H^*}}}{{\left( {\frac{{{h_{r,{H^*}}}}}{{{r_ \odot }}}} \right)}^\alpha }},
\end{equation}
for $r \leqslant {h_{r,{H^*}}}$. The level of this contribution is nevertheless extremely weak and added here only for completeness. Effectively it is null compared with the  dominant contribution of the bulge component:
\begin{equation}\label{Eq32}
	{O^- _B}\left( R \right) = \frac{1}{{4{v_{c}}}}\sum\limits_B^{} {G{M_B}\frac{{R + 3{h_{r,B}}}}{{{{\left( {R + {h_{r,B}}} \right)}^3}}}} .
\end{equation}
Finally, the most significant contributions that account for star and the local gas distributions are given by:
\begin{equation}\label{Eq33}
	{O^- _D}\left( R \right) = 4\pi G\sum\limits_D^{} {\frac{{{\rho _{0,D}}}}{{{h_{R,D}}}}\int_{\mathbb{R}}^{} {dk\frac{k}{{h_{z,D}^{ - 1} + k}}\frac{{kR{J_0}\left( {kR} \right) + 2{J_1}\left( {kR} \right)}}{{{{\left( {h_{R,D}^{ - 2} + {k^2}} \right)}^{3/2}}}}} } .
\end{equation}
With this equation and the parameters of Table 1 we obtain the following values at the solar position: ${O^ + } = 15.1\,{\rm{km}}\;{{\rm{s}}^{{\rm{ - 1}}}}\;{\rm{kp}}{{\rm{c}}^{{\rm{ - 1}}}}$  and ${O^ - } =  - 13.1\,{\rm{km}}\;{{\rm{s}}^{{\rm{ - 1}}}}\;{\rm{kp}}{{\rm{c}}^{{\rm{ - 1}}}}$. For comparison the study of Hipparcos proper motions by \citep[][]{1997MNRAS.291..683F,1997ESASP.402..625F} yield ${O^ + }\left( {{R_ \odot }} \right) = 14.8 \pm 0.8\;{\rm{km}}\;{{\rm{s}}^{{\rm{-1}}}}\;{\rm{kp}}{{\rm{c}}^{{\rm{-1}}}}$ and ${O^ + }\left( {{R_ \odot }} \right) - {O^ - }\left( {{R_ \odot }} \right) = 27.2 \pm 0.9$ ${\rm{km}}\;{{\rm{s}}^{{\rm{ - 1}}}}\;{\rm{kp}}{{\rm{c}}^{{\rm{ - 1}}}}$ and by \citet[][]{1998MNRAS.294..429D} yield ${O^ + }\left( {{R_ \odot }} \right) = 14.5 \pm 1.5\;{\rm{km}}\;{{\rm{s}}^{{\rm{-1}}}}\;{\rm{kp}}{{\rm{c}}^{{\rm{-1}}}}$ and ${O^ + }\left( {{R_ \odot }} \right) - {O^ - }\left( {{R_ \odot }} \right) = 27.20 \pm 1.5$ ${\rm{km}}\;{{\rm{s}}^{{\rm{ - 1}}}}\;{\rm{kp}}{{\rm{c}}^{{\rm{ - 1}}}}$.

\subsubsection{Vertical force}\label{Sec4.3}
The last significant constraint that we consider in the determination of the MW potential is the force acting vertically on the plane. This constraint is of paramount importance to tune the vertical profiles of the disk and the vertical epicyclic oscillations of the orbits, thus several studies have investigated the vertical structure of the Milky way on the basis of different observations \citep[][]{2014A&A...566A..87J, 2001ApJ...553..184C, 1997A&A...320..428H, 2012ApJ...755..115B, 2008ApJ...679.1288L, 2014A&A...566A..87J, 2003A&A...398..141S}. We determine the vertical force at any location within the galaxy as follows. 

For the dark matter component, we evaluate the vertical gradient of the potential at any radial and vertical location as
\begin{equation}\label{Eq34}
	{F_{z,DM}} = \frac{{{v_0}^2z}}{{{q^2}\left( {{R^2} + h_{r,DM}^2} \right) + {z^2}}},
\end{equation}
that retains information of the flattening parameter $q$ of the DM halo. Unfortunately, the alignment of the DM halo component with the principal axis of symmetry of the gravitational potential is far from clear. The triaxiality and the directions of the eigenvectors of the inertia tensor of the DM mass distribution is at present unknown and the problem of the stability of rotating disks inside triaxial halos is weakly understood from the theoretical point of view and still a matter of debate \citep[e.g.,][ and reference therein]{2013MNRAS.434.2971D}. We will take $q$ into account only for completeness and eventually add a flattening of the DM profiles while moving inward in the Galaxy. For the stellar halo components, the same computation yields:
\begin{equation}\label{Eq35}
{F_{z,{H^*}}}\left( {R,z} \right) = \frac{{4\pi Gzr_ \odot ^{ - \alpha }}}{{\alpha  + 3}}\sum\limits_{{H^*}}^{} {{\rho _{0,{H^*}}}\left( {{r^\alpha } - \frac{{h_{r,{H^*}}^{\alpha  + 3}}}{{{r^3}}}} \right)} ,
\end{equation}
for $\sqrt {{R^2} + {z^2}}  = r > {h_{r,{H^*}}}$, and 
\begin{equation}\label{Eq36}
{F_{z,{H^*}}}\left( {R,z} \right) = 4\pi G\sum\limits_{{H^*}}^{} {\frac{{{\rho _{0,{H^*}}}z}}{3}r_ \odot ^{ - \alpha }h_{r,{H^*}}^\alpha } ,
\end{equation}
otherwise. Analogously for the bulge components we get
\begin{equation}\label{Eq37}
	{F_{z,B}}\left( {R,z} \right) = \frac{z}{{\sqrt {{R^2} + {z^2}} }}\sum\limits_B^{} {\frac{{G{M_B}}}{{{{\left( {\sqrt {{R^2} + {z^2}}  + {h_{r,B}}} \right)}^2}}}} .
\end{equation}
Finally, for the disk component we get:
\begin{equation}\label{Eq38}
	\begin{aligned}
		{F_{z,D}}\left( {R,z} \right) &= 4\pi G\sum\limits_D^{} {\frac{{{\rho _{0,D}}}}{{{h_{R,D}}{h_{z,D}}}} \times }  \\ 
		&\times \int_{\mathbb{R}}^{} {dk\frac{{{e^{k\left| z \right|}} - {e^{h_{z,D}^{ - 1}\left| z \right|}}}}{{{k^2} - h_{z,D}^{ - 2}}}\frac{{k{J_0}(kR){e^{ - \left| z \right|\left( {h_{z,D}^{ - 1} + k} \right)}}}}{{{{\left( {h_{R,D}^{ - 2} + {k^2}} \right)}^{3/2}}}}} .  
	\end{aligned} 
\end{equation}
Our MW potential model with the values of Table \ref{Table1} presents a value of $\frac{{\left| {{F_z}\left( {1.1{\rm{kpc}}} \right)} \right|}}{{2\pi G}} = 70.0$ for the total vertical force on the plane that match exactly the standard literature values of \cite{1989MNRAS.239..605K} \citep[see also][]{1989MNRAS.239..571K,1989MNRAS.239..651K} and $\frac{{\left| {{F_z}\left( {2.0{\rm{kpc}}} \right)} \right|}}{{2\pi G}} = 87.9$ at the solar potion $R_\odot = 8.0$ and $\phi_\odot = 0.0$ \citep[e.g., see][for compatible values at $R_\odot = 8.5$]{2014A&A...571A..92B}. 
We consider these as the major contributors to  the shape of the underlying MW potential. Adding other  constraints will not significantly change the distribution of the stars in the CMDs and their kinematics. 

\subsubsection{Further constraints}\label{Sec4.4}
By integrating the density profiles of Eqs.\eqref{Eq14},  \eqref{Eq16},  \eqref{Eq19} and \eqref{Eq20} we obtain the total mass as a direct sum of the mass of the components. For the dark matter this integration reads:
\begin{equation}\label{Eq39}
	{M_{DM}} = \frac{{{v_0}^2}}{G}\frac{{r_{\max }^3}}{{r_{\max }^2 + h_{r,DM}^2}},
\end{equation}
for the stellar halo components
\begin{equation}\label{Eq40}
	{M_{H*}} = \frac{{4\pi {r_ \odot }^{ - \alpha }}}{{3\left( {\alpha  + 3} \right)}}\sum\limits_{{H^*}}^{} {\frac{{{\rho _{0,{H^*}}}}}{{{d_{0,{H^*}}}}}\left( {3r_{\max }^{\alpha  + 3} + \alpha h_{r,{H^*}}^{\alpha  + 3}} \right)},
\end{equation}
for the bulge
\begin{equation}\label{Eq41}
	{M_B} = 4\pi Gr_{\max }^2\sum\limits_B^{} {\frac{{{M_B}}}{{{{\left( {{h_{r,B}} + {r_{\max }}} \right)}^2}}}},
\end{equation}
and for the disks, by proceeding arbitrarily with an integration in cylindrical coordinates, we can write
\begin{equation}\label{Eq42}
	{M_D} = 4\pi \sum\limits_D^{} {\frac{{{\rho _{0,D}}{e^{ - {R_{\max }}h_{R,D}^{ - 1}}}}}{{h_{R,D}^{ - 2}h_{z,D}^{ - 1}}}\left( { {e^{{R_{\max }}h_{R,D}^{ - 1}}} - {R_{\max }}h_{R,D}^{ - 1} - 1} \right)}.
\end{equation}
With the parameters in Table 1 we obtain a total mass of ${M_{  }} = 1.12 \times {10^{12}}{M_ \odot }$ for $r_{\rm{max}}=100$ kpc \citep[e.g.,][]{2016arXiv160401216H, 2015ApJ...806...96L, 2014ApJ...794...59K,2014ApJ...785...63B, 2014A&A...562A.134B, 2002ApJ...573..597K, 2008ApJ...679.1239W, 2008ApJ...684.1143X, 2002ApJ...573..597K,2011MNRAS.414.2446M,2002SSRv..100..129G,2001MNRAS.326..164O}. This constraint has several implications on the orbits of the dwarf galaxy satellites of the MW, and several studies have focused on total mass determination and on the escape speed from the MW \citep[e.g.,][]{2014ApJ...785...63B, 2007MNRAS.379..755S, 2013AAS...22125411L, 1998MNRAS.294..429D, 2008ApJ...684.1143X, 1999MNRAS.310..645W, 2011MNRAS.414.2446M}. 

Finally, the determination of the local surface mass density is tightly related to the integration presented for the total mass. This is computed in our modelling only for the disk components ${\Sigma _D} = 2\sum\limits_D^{} {\frac{{{\rho _{c,D}}}}{{{h_{z,D}^{ - 1}}}}{e^{ - R{h_{R,D}^{ - 1}}}}} $. This is a relevant constraint especially in relation to the disk modelling of the spiral arms that we are going to present here. With the parameters of Table 1 we estimate a value of ${\Sigma _D} = 41\;{{\rm{M}}_ \odot }{\rm{p}}{{\rm{c}}^{{\rm{ - 2}}}}$ at $R=R_\odot$

We do not consider here  a few other issues of minor importance  that a standard axisymmetric model should take into account such as, e.g.,  the terminal velocities for the inner Galaxy \citep{2006A&A...451..125V}. Although it may
provide a better constraint than Oort's constants for an axisymmetric galaxy,  it it severely affected by  non-circular motions of the ISM. Its interpretation needs much  more precise mapping of the galactic gas distribution \citep[see also][]{2015A&A...578A..14C, 2013IAUS..292..101G}. In this respect, Section \ref{Sec6.2.1} we will present our new non axisymmetric distribution of gas.


\begin{table*}
\caption{Kinematic and dynamical properties of the MW components. The first two thin disk stellar components implement the spiral arm treatment described in the text. Because a map of the metallicity gradients ${\nabla _{\bm{x}}}\left[ {\frac{{Fe}}{H}} \right]$ is still uncertain, no standard default values are assumed and they are used as free parameters.}
\label{Table1}
\begin{tabular}{lllcc}
\hline
Components 	& Scale parameters 	& $\Delta t$  &  $\left[ {\frac{{Fe}}{H}} \right]$  & ${{\bm{\sigma }}_{ii}}_ \odot $\\
	& & $[Gyr]$ & [dex] & [${\rm{km}}\;{{\rm{s}}^{ - 1}}$] \\
\hline
											& $ \{{{\rho _D},{h_R},{h_z}}\}_\odot $		&  &  &   \\
											& $ \left[ {{{\rm{M}}_ \odot }\;{\rm{kp}}{{\rm{c}}^{{\rm{ - 3}}}}{\rm{,kpc}}{\rm{,kpc}}} \right] $ & & & \\								
Thin disk pop 1 (sp) 	& $1.29\times 10^7,2.57,0.06$ & [0.1, 0.5[ 	& [-0.70, 0.05[ & 27.0,15.0,10.0 \\
Thin disk pop 2 (sp) 	& $1.93\times 10^7,2.59,0.06$ & [0.5, 0.9[ 	& [-0.70, 0.05[ & 30.0,19.0,13.0 \\
Thin disk pop 3 			& $4.96\times 10^7,2.96,0.07$ & [0.9, 3.0[ 	& [-0.70, 0.05[ & 41.0,24.0,22.0 \\
Thin disk pop 4 			& $3.38\times 10^7,2.99,0.09$ & [3.0, 7.5[ 	& [-0.70, 0.05[ & 48.0,25.0,22.0 \\
Thin disk pop 5 			& $3.34\times 10^7,3.41,0.25$ & [7.5, 10.0[ & [-0.70, 0.05[ & 52.0,32.0,23.0 \\
Thick disk      			& $2.40\times 10^6,2.23,1.35$ & [10.0,12.0[ & [-1.90,-0.60[ & 51.0,36.0,30.0 \\
ISM             			& $2.26\times 10^7,4.51,0.20$ & & & \\
& & & & \\
											& $\{{{\rho _{0,H*}},{d_{0,H*}},h_{r{H^*}},\alpha }\} $ &  & & \\		
											& $ \left[ {{{\rm{M}}_ \odot }\;{\rm{kp}}{{\rm{c}}^{{\rm{ - 3}}}}{\rm{,kpc}}{\rm{,kpc}}} \right] $ & & & \\								
Stellar halo pop 1		& $2.18 \times {10^4},1.00,1.00, - 2.44$ & [12.0,13.0[ &  $<-1.90$  & 151.0,116.0,95.0 \\
& & & & \\
											& $\{{M_B},{h_{r,B}}\}$ &  & & \\		
											& $\left[ {{{\rm{M}}_ \odot }\;{\rm{,kpc}}} \right]$ & & & \\								
	Bulge	pop 1					& $3.4 \times {10^{10}},0.7$ & [6.0,12.0[ & [-0.40,+0.30[ & \\
& & & & \\
								& $\{ {{v_0},h_{r,DM},q} \}$& & & \\
							  & $\left[ {{\rm{km}}\;{{\rm{s}}^{-1}}{\rm{,kpc}}} \right]$& & & \\
	Dark matter		& $139.04,6.70,0.89 $& & & \\							
\hline
\end{tabular}
\end{table*}

\section{Machine learning}\label{Sec5}
To obtain values representative of the MW stellar, gas and DM components (Table \ref{Table1}) we tune the free parameters of the density-potential couples formulated in the previous section on theoretical and observational constraints.
The choice to represent a given set of data with a fixed number $n$ of SSP is one of the major underlying constraints that we have adopted in our formalism. This forced us to a statistical interpretation of \textit{parametric}-nature and hence a \textit{supervised}-machine learning approach. Unsupervised learning in the framework of Neural Networks will be explored in a future investigation (Pasetto et al 2016, in preparation). 

Within this parametric approach, among the most sophisticate and robust techniques available to date are the genetic algorithms. A genetic algorithm is an adaptive stochastic optimization algorithm involving search and optimization, and it was   first introduced by \citet{1975anas.book.....H}. Holland created an electronic organism as a binary string (``chromosome''), and then used the genetic and evolutionary principles of fitness-proportionate selection for reproduction, random crossover, and mutation to explore the space of solutions. The so-called ``genetic programming languages'' apply the same principles using an expression tree instead of a bit string as a ``chromosome''. In astronomy, the Pikaia genetic algorithm has been already considered in the galactic kinematics in \citep[see, 2005 Pasetto, PhD thesis,][]{1995ApJS..101..309C, 2003JCoPh.185..176M}. We consider the following quite generic task to model a given dataset with a set of adjustable parameters. This task consists of finding the single parameter set that minimizes the difference between the model's predictions and the data. A ``top-level'' view of the canonical genetic algorithm for this task can be read as follows: we start by generating a set (``population'') of trial solutions, usually by choosing random values for all model parameters; then evaluate the goodness of fit (``fitness'') of each member of the current population (e.g., through a chi-square measure with the data). Then the algorithm selects pairs of solutions (``parents'') from the current populations, with the probability of a given solution being selected made proportional to that solution's fitness. It breeds the two solutions selected and produces two new solutions (``off-spring''). It repeats the selection of the population and its progeny until the number of off-springs equals the number of individuals in the current population by replacing the new population of off-springs over the old one. It then repeats the whole sequence until some termination criterion is satisfied (i.e. the best solution of the current population reaches a fit goodness exceeding some pre-set value). 

A genetic-algorithm based approach to a given optimization task, as defined above, resembles a kind of forward-modelling: no derivatives of the fit function goodness with respect to model parameters is needed to be computed. Nothing in the procedure outlined above depends critically on using a least-squares statistical estimator; any other robust estimator could be used, with little or no change to the overall procedure. In the kinematical applications, the model needs to be evaluated (i.e., given a parameter set, compute a synthetic dataset and the associated goodness of fit). 

In the process of CMD fitting the genetic algorithm has a long history in the Padua group starting from the works of \citet{2002A&A...392.1129N} and has been implemented in the kinematic fitting of observational data in \citet{2006A&A...451..125V}. The algorithm has been run on true data to reproduce radial velocities \citep[][]{2002ApJ...574L..39G}, the GSC-II proper motion catalogue \citet{2006A&A...451..125V} and the RAVE dataset equipped with 2MASS proper motions in \citet{2012A&A...547A..71P}, and \citet{2012A&A...547A..70P}. The detailed study of the MW potential is beyond the goal of the present paper (and maybe meaningless at the sunrise of the Gaia-era), but we limit ourselves to present in Table 1 the guest parameters for the MW potential just introduced and achieved so far. They will represent the starting values of the founding potential that we are going to perturb in the next section to obtain the spiral arms description which represents the core of this work.


\begin{figure*}
\centering
\resizebox{\hsize}{!}{\includegraphics{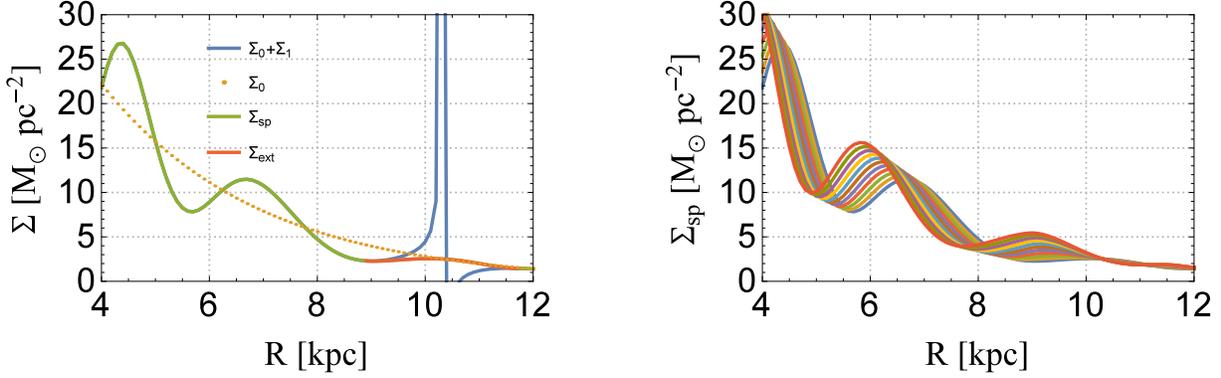}}
\caption{(left panel) On-the-plane section of the perturbed density profile from Eq.\eqref{Eq55} at $\phi  = 0$. The blue line is the profile with singularity at the resonance location $R = {R_{{\rm{res}}}}$, the red line is the analytical continuation from ${R_{{\rm{res}}}} \pm {\varepsilon _R}$ with ${\varepsilon _{{\rm{res}}}} = 1.4{\rm{kpc}}$ and the underlying green profile is the overall profile continued over the resonance (see text for details). (Right panel) Same green perturbed density profile of left panel but for a random set of varying $\phi $ angles (and random colour). The purpose is just to show how smooth the passage is between one line and another at different $\phi $-s.
\label{Fig4} }
\end{figure*}

\section{Density description of non-axisymmetric features}\label{Sec6}
The axisymmetric potential that we have introduced above and summarized in \ref{Table1} represents the starting point for the perturbative approach that we introduce hereafter.

As previously anticipated, the first framework that we are introducing is the density wave theory (DWT). It deals in its original form with the description of the in-plane motion of the stars in a spiral galaxy. It is a linear response theory for an unperturbed generalized Schwarzschild distribution function (SDF):
\begin{equation}\label{Eq43}
\begin{gathered}
  Q \equiv {\left( {{\bm{v}} - {\bm{\bar v}}} \right)^T}{{\bm{\sigma }}_{\bm{v}}^{-1}}\left( {{\bm{x}};t} \right)\left( {{\bm{v}} - {\bm{\bar v}}} \right), \hfill \\ 
  {f^{{\rm{Sch}}}} \equiv {e^{ - \frac{1}{2}Q\left( {\bm{x}} \right) + \eta }} \hfill \\
\end{gathered}
\end{equation}
where $Q$  is a quadratic positive definite form, ${{\bm{\sigma }}_{\bm{v}}}\left( {{\bm{x}};t} \right)$ a second rank symmetric tensor defined in Eq.\eqref{Eq9}, $\eta \left( {\bm{x}} \right)$ a continuous and differentiable scalar function and with the superscript $(*)^T$ we refer to the transpose of an array and with  $(*)^{-1}$ to the inverse element of an array (not the inverse matrix). It is normalized accordingly with ${\left( {2\pi } \right)^{ - 3/2}}{\left| {\bm{\sigma }} \right|^{ - 1/2}}{e^{ - \eta /2}} \equiv {\left( {2\pi } \right)^{ - 3/2}}{\left| {\bm{\sigma }} \right|^{ - 1/2}}{\Sigma _0}\left( R \right)$. We will recall in what follows the basis of this theoretical framework of the DWT without explicit proof, but we will present a new hypergeometric form for the expression of the first moments of the perturbed DF that were previously known only in an integral form. We will highlight the advantages of our formulation. 

The second perturbative framework adopted here has been developed by \citet{1991ApJ...368...79A} and it has been previously adopted in our modelling technique by Pasetto, PhD thesis 2005 in \citet{2006A&A...451..125V}. Here we will only recall the theoretical basis of this second perturbative framework dealing with the vertical behaviour of the kinematics above and below the disk plane, and we will compare it with the DWT. 
In the axisymmetric case, the matrix ${{\bm{\sigma }}_{\bm{v}}}\left( {{\bm{x}};t} \right)$ acquires an especially simple diagonal form and the dependence of the three non-null diagonal terms has been already introduced in Sec. \ref{AxisymmtotTTD}.

\subsection{Linear response theory to a spiral perturbation pattern }\label{Sec6.1}
The most popular self-consistent fully analytical treatment available in literature to study the spiral arms is based so far on the DWT proposed by \citet{1969ApJ...155..721L}, \citet{1969SvA....13..411M}, \citet{1969SvA....13..252M}. In these works, a sinusoidal perturbation to the axisymmetric potential for a discoidal stellar distribution is considered. This description stands on two theoretical pillars of the stellar dynamics: the epicycles approximation theory and linear response theory to Boltzmann collisionless equation (see also Sec.\ref{Sec2}). Here we limit ourselves to introduce the basic functions as definitions, without proofs, forwarding the reader to specialized text on stellar dynamics for a coherent exposition of these topics \citep[e.g.,][]{2014dyga.book.....B}. If we perturb an axisymmetric potential with a sinusoidal wave, we need to search for the self-consistent condition for a potential of a spiral drawn by a shape-function $\psi \left( R \right) \equiv  - 2\cot \left( p \right)\log \left( {\frac{R}{{{R_0}}}} \right)$ where $p$ is the pitch angle $p \sim 8^\circ $, $R_0$ is the starting radius of the spiral perturbation, ${R_0} \sim 2.6{\rm{kpc}}$, and $m = 2$ is the number of spiral arms that we assume. The values adopted here are examples for the MW case but they have not been deduced through data analysis of Sec.\ref{Sec5}. A large literature review has been presented in Sec. \ref{Sec1.1} from where we extracted the adapted values for the exercise presented below. The variable ${\Omega _p}$ is the rotation pattern of the spiral structure; the theory so far is developed for a constant ${\Omega _p}$, even though no strong observational constraints are available to justify this assumption. Recent N-body simulation studies suggest ${\Omega _p} = {\Omega _c}\left( R \right)$, where ${\Omega _c}\left( R \right)$ is the angular speed at R \citep[e.g.,][]{2011ApJ...735....1W, 2012MNRAS.421.1529G}, while the other studies interpret the spiral arms in N-body simulations as overlapping multiple-density waves covering different radial ranges, with different pattern speed, $\sum\limits_i^{} {{\Omega _{p,i}}} $. and slower pattern speeds in the outer region \citep[][]{2012MNRAS.426.2089R,2014ApJ...785..137S}. Nevertheless, no strong observational evidence is available to date to formalize ${\Omega _p}={\Omega _p}(R,\phi,z;t)$. 

In what follows we will simply assume ${\Omega _p} \sim 35.0\;{\rm{km}}\;{{\rm{s}}^{ - 1}}\;{\rm{kp}}{{\rm{c}}^{ - 1}}$. We define ${\Phi ^a}\left( R \right) \equiv  - \Phi _0^aR\;{e^{ - \frac{R}{{{h_s}}}}}$ to be amplitude of the spiral arm potential profile, where an indicative scale length ${h_s} \sim 2.5$ kpc is assumed and $\Phi _0^a \sim 887.0\;{\rm{k}}{{\rm{m}}^2}{{\rm{s}}^{ - 2}}{\rm{kp}}{{\rm{c}}^{ - 1}}$ from \citet[][]{2014MNRAS.440.1950R}. These numerical values are taken form the literature reviewed in Sec. \ref{Sec1.1} and are of illustrative nature to the present capability of our modelling approach alone. They are not meant to be best fitting values through the technology explained in Sec.5 to any particular survey. 

The DWT is developed in epicycle approximation. The epicycle approximation is probably the weakest of the assumptions adopted in our model.  In the next section, we will review some observational evidence of the failure of this approximation in the solar neighbourhood. Here we proceed simply by adopting the modification that the perturbative linear approach induces on this approximation, without presenting a critical review, even though improved tools are already available \citep[][]{1999AJ....118.1190D}. The main role of this approximation is to decouple in Eq.\eqref{Eq43} the radial/azimuthal from the vertical direction thus simplifying them. From the potential introduced in the previous sections we can define the rotation frequency as $\Omega \left( R \right) \equiv \frac{{{v_c}}}{R}$ together with its derivative $\frac{{\partial \Omega }}{{\partial R}} = \frac{1}{R}\frac{{\partial {v_c}}}{{\partial R}} - \frac{{{v_c}}}{{{R^2}}}$. The radial epicycle frequency is then given by $\kappa  = 2\Omega \sqrt {1 + \frac{R}{{2\Omega }}\frac{{\partial \Omega }}{{\partial R}}} $. Finally, recalling that the wave-number is the derivative of the shape function introduced above, $k = \frac{{\partial \psi }}{{\partial R}}$, we can compute Toomre’s number as $X \equiv \frac{k}{\kappa }{\sigma _{RR}}.$ In our approach, Toomre’s number can eventually acquire a vertical dependence trough the velocity dispersion profiles introduced above (Eq.\eqref{Eq44}). Because a self-consistent theory for the vertical motion of the stars in the presence of spiral arms is missing, a large freedom is left to the researcher to investigate different approaches.

After the introduction of these quantities, we are in the position to make use of the results of \citet{1969ApJ...155..721L}. A solution of the evolution equation (i.e. the linearized Boltzmann equation) $\iota \frac{{\partial {f_1}}}{{\partial t}} - {\mathcal{B}_0}\left[ {{f_1}} \right] - {\mathcal{B}_1}\left[ {{\Phi _1}} \right] = 0$ is considered in the form:
\begin{equation}\label{Eq46}
	{f_1}\left( {{\bm{x}},{\bm{v}};t} \right) = \int_{ - \infty }^t {\left\langle {{\nabla _{{\bm{x'}}}}{\Phi _1},\frac{{\partial {f^{{\rm{Sch}}}}\left( {{\bm{x'}},{\bm{v'}}} \right)}}{{\partial {\bm{v'}}}}} \right\rangle dt'}, 
\end{equation}
with natural boundary conditions ${f_1} \to 0$  as $t \to  + \infty $. Here $\left\langle {*,*} \right\rangle $ represents the standard inner product. Under the assumption that the perturbations take the form of spiral waves ${\Phi _1}\left( {{\bm{x}};t} \right) = {\Phi ^a}\left( R \right){e^{\iota \left( {m\phi  - \omega t + \int_{}^R {kdR} } \right)}}$, in a ``tightly wound'' approximation, i.e. $\left| {kR} \right| \gg 1$, we get rapidly to the form for the perturbed DF on the plane as:
\begin{equation}\label{Eq47}
	{f_1} =  - \frac{{{\Phi _1}}}{{\sigma _{RR}^2}}{f^{{\rm{Sch}}}}\left( \begin{gathered}
		1 - \rm{sinc}^{-1}\left( {\nu \pi } \right) \times  \hfill \\
		\times \frac{1}{{2\pi }}\int_{ - \pi }^\pi  {{e^{\iota \left( {\nu \tau  + X\left( {u\sin \tau  + v\left( {1 + \cos \tau } \right)} \right)} \right)}}d\tau }  \hfill \\ 
	\end{gathered}  \right),
\end{equation}
with ${\rm{sin}}{{\rm{c}}^{ - 1}}\left( {\nu \pi } \right) \equiv \frac{{\sin \left( {\nu \pi } \right)}}{{\nu \pi }}$ the ``sinc'' function, where we set here for simplicity the frequency ratio $\nu  \equiv \frac{{\omega  - m\Omega_p }}{\kappa }$, and $\frac{1}{{{\gamma ^2}}} \equiv \frac{{\sigma _{\phi \phi }^2}}{{\sigma _{RR}^2}} = \frac{{{\kappa ^2}}}{{{{\left( {2\Omega } \right)}^2}}}$ in agreement with the hypothesis underlying Eq.\eqref{Eq45}. Finally we simplified the notation writing the peculiar velocities as  ${\bm{v}} - {\bm{\bar v}} = \left\{ {\frac{{{v_R}}}{{{\sigma _{RR}}}},\gamma \frac{{{v_\phi } - {v_c}}}{{{\sigma _{RR}}}},\frac{{{v_z}}}{{{\sigma _{zz}}}}} \right\} = \left\{ {u,v,w} \right\}$.

We are ready now to proceed to compute the first order moments of this DF that we adopted in our kinematic model.
 The moment of order zero and one was already carried out in numerical form by the authors in Appendix A of \citet{1969ApJ...155..721L} to the first order, and the second order central moments were recently proposed by \citet{2014MNRAS.440.1950R} in a work focused on the vertex deviation and the bracketing of the resonances. Nevertheless, in the original work by \citet{1969ApJ...155..721L} and in the work \citet{2014MNRAS.440.1950R} the numerical integral was passed over in favour of a more compact analytical formalism, and the divergences due to the resonances were not considered. 

We present here a different solution for these moments in the form of Hypergeometrical functions instead of numerical integrals. We will underline later the advantages of our formulation in the context of the present modelling approach. We will also offer a necessary solution to cover the resonances and to make the model suitable for the star-count approach that we are developing here.
\begin{figure*}
\centering
\resizebox{\hsize}{!}{\includegraphics{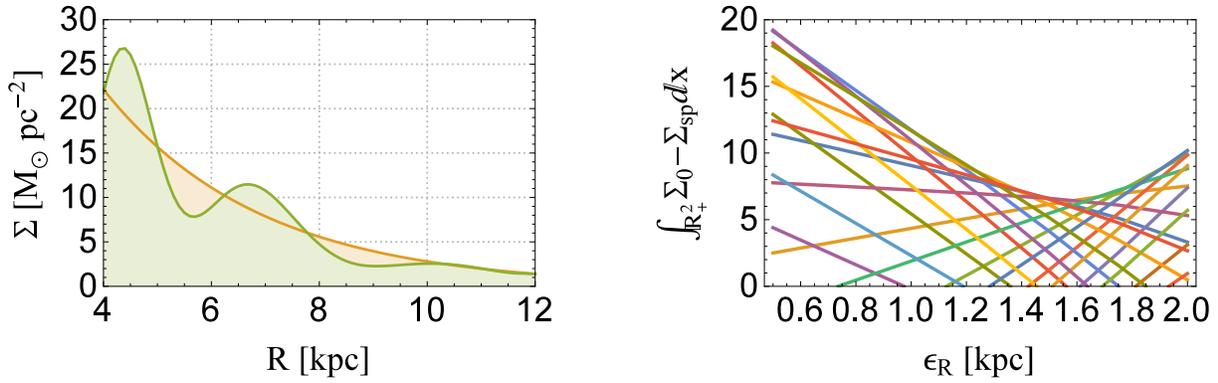}}
\caption{(left panel): the integral of the continued density profile ${\Sigma _{{\rm{sp}}}}$ (green) and the unperturbed axisymmetric (orange). (right panel): Integration profile at different azimuthal directions $\phi $, each for each colour, versus ${\varepsilon _R}$.
\label{Fig5} }
\end{figure*}
\subsubsection{Zero order moments of the perturbed DF}\label{Sec6.1.1}
The family of the perturbed density profiles result as the zero order moment of the total DFs given by $f = {f^{{\rm{Sch}}}} + {f_1}$ with ${f^{{\rm{Sch}}}}$ defined by Eq.\eqref{Eq43} and ${f_1}$ by Eq.\eqref{Eq47}. We write:
\begin{equation}\label{Eq48}
	\Sigma \left( {R,\phi ;t} \right) \equiv \int_{{\mathbb{R}^3}}^{} {fd{v_R}d{v_\phi }d{v_z}}  = \int_{{\mathbb{R}^3}}^{} {\left( {{f^{{\rm{Sch}}}} + {f_1}} \right)d{v_R}d{v_\phi }d{v_z}} .
\end{equation}
By exploiting the notation introduced above, we can write
\begin{equation}\label{Eq49}
	\begin{aligned}
		\frac{{{\Sigma _1}}}{{{\Sigma _0}}} &= \frac{{\int_{{\mathbb{R}^3}}^{} {{f_1}d{v_R}d{v_\phi }d{v_z}} }}{{\int_{{\mathbb{R}^3}}^{} {{f^{{\rm{Sch}}}}d{v_R}d{v_\phi }d{v_z}} }} \hfill \\
		&=  - \frac{{{\Phi _1}}}{{\sigma _{RR}^2}}\frac{1}{{{\Sigma _0}}}\int_{{\mathbb{R}^2}}^{} {\left( {1 - \rm{sinc}^{-1}\left( {\nu \pi } \right) \times } \right.}  \hfill \\
		&\times \left. {\frac{1}{{2\pi }}\int_{ - \pi }^\pi  {{e^{\iota \left( {\nu \tau  + X\left( {u\sin \tau  + v\left( {1 + \cos \tau } \right)} \right)} \right)}}d\tau } } \right){f^{{\rm{Sch}}}}du dv dw, \hfill 
	\end{aligned}
\end{equation}
and in particular, we reach the form
\begin{equation}\label{Eq50}
\begin{aligned}
		\Sigma  &= {\Sigma _0} + {\Sigma _1} \\ 
		&= {\Sigma _0} - {\Sigma _0}\frac{{{\Phi _1}}}{{\sigma _{RR}^2}}\left( {\frac{1}{{{\Sigma _0}}}\frac{1}{{2\pi }}\int_{{\mathbb{R}^3}}^{} {\left( {1 - \rm{sinc}^{-1}(\nu \pi) \times } \right.} } \right. \\ 
		&\times \left. {\left. {  \int_{ - \pi }^\pi  {{e^{\iota \left( {\nu \tau  + X\left( {u\sin \tau  + v\left( {1 + \cos \tau } \right)} \right)} \right)}}d\tau } } \right){f^{{\rm{Sch}}}}du dv dw} \right),  
	\end{aligned} 
\end{equation}
which is the obvious generalization of the work of \citet{1969ApJ...155..721L} to the case of vertical velocity DFs. If we remember that $\int\limits_{{\mathbb{R}^3}}^{} {\frac{1}{{{{\left( {2\pi } \right)}^{3/2}}}}{e^{ - \left( {{u^2} + {v^2} + {w^2}} \right)}}} du dv dw = 1$, the terms inside the external brackets reads simply
\begin{equation}\label{Eq51}
	\begin{aligned}
			&= 1 - \frac{\rm{sinc}^{-1}(\nu \pi)}{{2\pi }} \int_{{\mathbb{R}^3}}^{} {dudvdw{e^{ - \frac{{{u^2} + {v^2} + {w^2}}}{2}}} \times }  \hfill \\
			&\times \frac{1}{{{{\left( {2\pi } \right)}^{3/2}}}}\int_{ - \pi }^\pi  {{e^{\iota \left( {\nu \tau  + X\left( {u\sin \tau  + v\left( {1 + \cos \tau } \right)} \right)} \right)}}d\tau } . \hfill  
	\end{aligned}
\end{equation}
At this point, by changing the integration order, we can obtain
\begin{equation}\label{Eq52}
	\begin{aligned}
		&=  - \frac{{\rm{sinc}^{-1}\left( {\nu \pi } \right)}}{{2\pi }}\frac{1}{{{{\left( {2\pi } \right)}^{3/2}}}}\int_{ - \pi }^\pi  {d\tau }  \times  \\ 
		&\times \int_{{\mathbb{R}^3}}^{} {dudvdw{e^{ - \frac{{{u^2} + {v^2} + {w^2}}}{2}}}{e^{\iota \left( {s\tau  + X\left( {u\sin \tau  + v\left( {1 + \cos \tau } \right)} \right)} \right)}}}  \\ 
		&=  - \frac{{\rm{sinc}^{-1}\left( {\nu \pi } \right)}}{{2\pi }}\int_{ - \pi }^\pi  {d\tau } \left( {{{\rm{e}}^{\iota \nu \tau  - {X^2}\left( {1 + \cos \tau } \right)}}} \right) \\ 
		&=  - \frac{{\rm{sinc}^{-1}\left( {\nu \pi } \right)}}{{2\pi }}\int_{ - \pi }^\pi  {d\tau } {e^{ - {X^2}\left( {1 + \cos \tau } \right)}}\left( {\cos \left( {\nu \tau } \right) + \iota \sin \left( {\nu \tau } \right)} \right) \\ 
		&=  - \frac{{\rm{sinc}^{-1}\left( {\nu \pi } \right)}}{{2\pi }}\int_{ - \pi }^\pi  {d\tau } {e^{ - {X^2}\left( {1 + \cos \tau } \right)}}\cos \left( {\nu \tau } \right) \\ 
		&= { - _{\left( {\frac{1}{2},1} \right)}}{{\tilde F}_{\left( {1 - \nu ,1 + \nu } \right)}}\left( { - 2{X^2}} \right) \\ 
		&\equiv { - _{\left( {\frac{1}{2},1} \right)}}{{\hat F}_{\left( {1 - \nu ,1 + \nu } \right).}} 
	\end{aligned}
	\end{equation}
Here we introduced the generalized Hypergeometric function:
\begin{equation}\label{Eq53}
	\begin{aligned}
		_2{{\tilde F}_2}\left( {{a_1},{a_2};{b_1},{b_2};z} \right) &\equiv \frac{{_2{F_2}\left( {{a_1},{a_2};{b_1},{b_2};z} \right)}}{{\Gamma \left( {{b_1}} \right)\Gamma \left( {{b_2}} \right)}} \\ 
		&= \frac{1}{{\Gamma \left( {{b_1}} \right)\Gamma \left( {{b_2}} \right)}}\sum\limits_{k = 0}^\infty  {\frac{{{{\left( {{a_1}} \right)}_k}{{\left( {{a_2}} \right)}_k}}}{{{{\left( {{b_1}} \right)}_k}{{\left( {{b_2}} \right)}_k}}}} \frac{{{z^k}}}{{k!}},
	\end{aligned} 
\end{equation}
	with ${\left( a \right)_n} \equiv a(a + 1) \ldots (a + n - 1) = \frac{{\Gamma (a + n)}}{{\Gamma (a)}}$ the Pochhammer symbol and $\Gamma $ the Eulero Gamma function. In particular, we advance the notation of the Hypergeometric function to 
\begin{equation}\label{Eq54}
		_{\left( {{a_1},{a_2}} \right)}\hat F_{\left( {{b_1},{b_2}} \right)}^{} \equiv {\;_2}{\tilde F_2}\left( {{a_1},{a_2};{b_1},{b_2}; - 2{X^2}} \right).
\end{equation}
We can then recollect the terms to express the density in a compact way as:
\begin{equation}\label{Eq55}
	\begin{aligned}
		\Sigma  &= {\Sigma _0} + {\Sigma _1} \\ 
		&= {\Sigma _0}\left( {1 - \frac{{{\Phi _1}}}{{\sigma _{RR}^2}}\left( {1{ - _{\left( {\frac{1}{2},1} \right)}}{{\hat F}_{\left( {1 - \nu ,1 + \nu } \right)}}} \right)} \right) \\ 
		&= {\Sigma _0}\left( {1 - \frac{{{\Phi _1}}}{{\sigma _{RR}^2}}\frac{{{X^2}}}{{1 - \nu }}\Re } \right). 
	\end{aligned}
\end{equation}
This represents the formula for the density profile perturbed by the spiral arms that we are going to implement.

As a corollary of this result, it is evident that we are able for the first time to propose a form for the ``reduction-factor'' $\Re $. This was historically introduced in \citet{1969ApJ...155..721L} as the factor to which we have to reduce the response of a stellar disk below the value of a cold disk (this is presented by direct integration in Appendix A too). This compact formulation of the density perturbation due to spiral perturbations presents extremely rapid computation benefices because of the presence of the hypergeometric function $_2{F_2}$. This will turn out to be especially useful for a technique that wants to be able to realize mock catalogues, where these integrals have to be computed a larger number of times to span a huge parameter space or to realize a high number of stars by populating PDFs.
The plot of the density profiles for the values of the potential of Table \ref{Table1} and the parameters assumed above are in Fig. \ref{Fig4}.
As evident from the plot, the previous equation Eq.\eqref{Eq55} presents a singularity at the resonances that we are going to treat in the next section.
\begin{figure*}
\centering
\resizebox{\hsize}{!}{\includegraphics{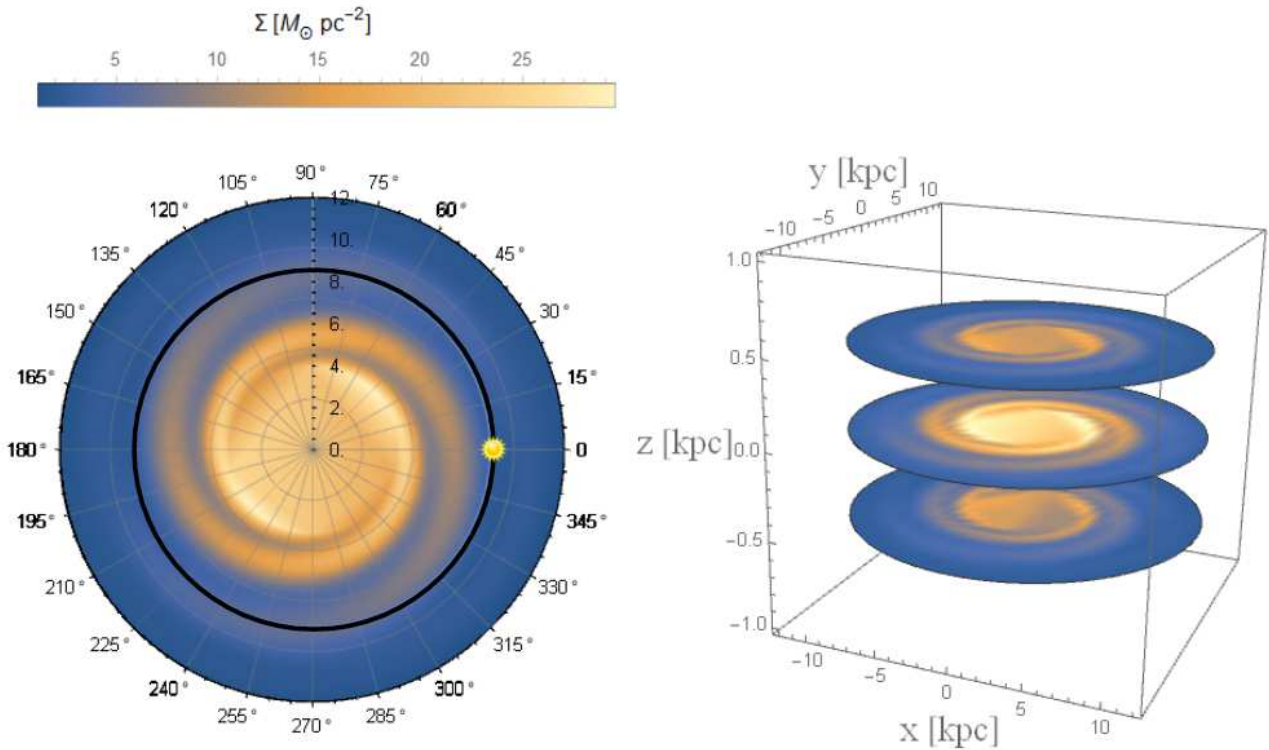}}
\caption{(left panel) Density profile in the plane of the Milky Way. The Sun is located at at ${\left\{ {R,\phi ,z} \right\}_ \odot } = \left\{ {8.0,0.0} \right\}$[kpc]. A black circle suggests the solar radius, despite a fixed position for $\phi_\odot$ not being investigated here, it is assumed $\phi_\odot=0$ [deg] for simplicity. The figure shows the plane at $z = 0$. (right panel) out of plane density distribution for two slices symmetric below and above the plane.
\label{Fig6} }
\end{figure*}
\subsubsection{Interpolation schemes over the resonances}\label{Sec6.1.2}
As evident from Fig. \ref{Fig4}, at the radius where the resonances are located (i.e. wherever $1 - \nu  = 0$) a divergence in the density profile of $\Sigma $ is present. To satisfy the normalization equation Eq.\eqref{Eq3} we need to cover this divergence. From Fig. \ref{Fig4} it is evident how the closure required by the stellar population theory in Eq.\eqref{Eq3} (in the special case of Eq.\eqref{Eq13} and \eqref{Eq43}) leads to a failure of the normalization condition. The PDF generated by Eq.\eqref{Eq55} cannot be populated in a star-count dot-by-dot fashion because of the infinite number of stars necessary to fill the locality of the radius $R = {R_{{\rm{res}}}}$. 

Two are the options immediately available at this point that we tested:
\begin{enumerate}
	\item  We can apply a bilinear interpolation in cylindrical coordinates. We solve the condition of continuity ${\left. \Sigma  \right|_{{p_1}}} = {a_{10}}$ and differentiability ${\left\{ {\frac{{\partial \Sigma }}{{\partial R}},\frac{{\partial \Sigma }}{{\partial \phi }},\frac{{\partial \Sigma }}{{\partial R\partial \phi }}} \right\}_{{p_1}}} = \left\{ {{a_{11}},{a_{12}},{a_{13}}} \right\}$ in each of the 4 points of the grid where the potential scheme introduced above has been valued. The linear matrix for the system of 16 equations in 16 unknowns is invertible and can be solved for two radii, one internal to the Lindblad resonance ${R_{{\rm{res}}}} - {\varepsilon _R}$, and one external to it, at ${R_{{\rm{res}}}} + {\varepsilon _R}$. Finally the function Eq.\eqref{Eq55} is extended (as the red curve in Fig. \ref{Fig4}).
	\item We can develop the function Eq.\eqref{Eq55} on a orthogonal set of basis in cylindrical coordinates (e.g., the Bessel function ${J_\alpha }$ introduced in Sec. \ref{Sec4}). Then we can mimic the behaviour of the DF $\Sigma \left( R \right) \sim \sum\limits_{n = 1}^\infty  {{c_n}{J_{\alpha ,n}}\left( R \right)} $ where ${J_{\alpha ,n}}\left( R \right) \equiv {J_\alpha }\left( {{z_{\alpha ,n}}\frac{R}{{{R_{\max }}}}} \right)$ and ${z_{\alpha ,n}}$ is the zero of the Bessel function ${J_a}$, with coefficients ${c_n} = \frac{{\left\langle {\Sigma ,{J_{\alpha ,n}}} \right\rangle }}{{\left\langle {{J_{\alpha ,n}},{J_{\alpha ,n}}} \right\rangle }}$. This approach passes through a long computing of inner products and hence is very slow, it does not respect precisely the values of the original $\Sigma $ and, while it can be worked out efficiently once eigenfunctions of the Laplacian operator are considered, it loses efficiency when the purpose is to cover the resonances on the velocity space.
\end{enumerate}
 
To bypass these difficulties encountered, we developed here a general scheme that works rapidly both for the treatment of resonances on the densities (i.e. first order moments of ${f^{{\rm{Sch}}}}$) as well as for resonances for the moments of higher order (mean, dispersion etc.).

To achieve this goal we proceed to investigate here a method that only extends a fiven profile function along the radial direction $R$ with a polynomial $P$ of degree $\deg \left( P \right) = 4$, i.e. $P = \sum\limits_{i = 0}^4 {{c_i}{x^i}} $. The methodology can of course work equally well with $\deg \left( P \right) = 3$ to match the number of constraints at the points $P\left( {{R_{{\rm{res}}}} - {\varepsilon _R}} \right)$ and $P\left( {{R_{{\rm{res}}}} + {\varepsilon _R}} \right)$ where it is valued together with its derivative. Nevertheless, the same scheme with $\deg \left( P \right) = 4$, allows to impose to the first and second order moments closer values to the corresponding unperturbed functions, thus allowing us to gently reduce the perturbations to the density and velocity fields to zero if desired. 

The reason for this polynomial solution to work is the azimuthal symmetry of the underlying unperturbed model. As evident from the condition of resonances, $1 - \frac{{m\left( {{\Omega _p} - \Omega \left( R \right)} \right)}}{{\kappa \left( R \right)}} = 0$, the divergences have no dependence on the azimuthal angle $\phi $. In the framework of the DWT we can individuate the resonances location only by analysis of the radial direction $R$. The results of this interpolation scheme are presented in Fig. 4 (right panel), where a very close azimuthal spanning is operated to check the validity of the continuation polynomial scheme presented, with evidently satisfactory results. 

The choice of the exact value that the scheme induces at the resonance is by itself a free parameter that we investigate here below.

\subsubsection{The choice of the interpolating radius}\label{Sec6.1.3}
The only parameter left unspecified in this interpolation scheme is the radius at which the scheme has to take over the DWT predictions. This is a single parameter, one condition is sufficient to fix it and the most natural one is based on the continuity equation. We require that the difference in mass between the continued ${\Sigma _{{\rm{sp}}}}$ and the unperturbed ${\Sigma _0}$  axisymmetric density distributions are the same (see Fig. \ref{Fig5})
\begin{equation}\label{Eq56}
	\Theta \left( {{\varepsilon _R}} \right) = \int_{{\mathbb{R}_ + } \times \left[ {0,2\pi } \right[} {\left( {{\Sigma _0} - {\Sigma _{{\rm{sp}}}}\left( {{\varepsilon _R}} \right)} \right)RdR d\phi} .
\end{equation}
This condition is equivalent to minimize the impact of the arbitrary shape that we chose to use to cover the resonances. If we convolve the integrals over all the angular directions we obtain Fig. 5 (right panel). As is evident in the figure, the minimal difference between the integrated mass predicted by ${\Sigma _0}$ and ${\Sigma _{{\rm{sp}}}}$ is achieved for ${\varepsilon _R} \sim 1.5$ kpc.
Finally in Fig. \ref{Fig6} the plot of the density profiles in the plane and above and below the plane are shown. 

A black line marks the solar radius: the solar location is \textit{assumed} to be at $\phi_\odot = 0$ but it is not a result of an investigation of any dataset. So far all the values obtained in Table \ref{Table1} are the results from studies in the axisymmetric formalism of Sec. \ref{Sec4}. A non-axisymmetric investigation of the solar position in the MW plane is within the DWT framework is, to our knowledge, not available (and beyond the goal of this paper). 

\begin{figure}
\includegraphics[width=\columnwidth]{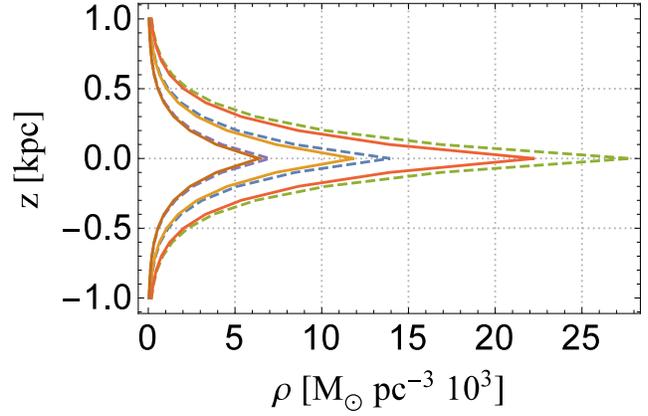}
\caption{Vertical profile of the spiral arm component at $R=6$, $R=8$  and $R=10$ kpc (solid lines). The unperturbed exponential is added for comparison with a dashed line.
\label{Fig7} }
\end{figure}

The vertical density profile of the spiral arms is not directly obtained from the DWT, which is developed only in the plane. Here we are not searching for  a self-consistent determination of the density profile, instead we assume decoupling of the vertical and radial profile in the configuration space assigning the axisimmetric density profile of the disk stellar population to the spiral arms profile too (Fig. \ref{Fig7}).
As evidenced in the Figure, the effect of the spiral arms is a tiny contraction of the vertical profile with respect to the corresponding unperturbed one. This is in response to the dependence of the density profiles to the velocity dispersions. Because of Eq.\eqref{Eq9}, i.e. the so-called ``age-velocity dispersion'' relation evident in the MW, the older the SSP of the spiral arm is, the smaller is this contraction.  

\subsection{Colour magnitude diagrams of spiral features}\label{Sec6.2}
Once the density is computed,  we know the relative contribution of all the stellar populations that we want to implement in our model (Table 1). At a given distance we compute the synthetic photometry of an observed FOV by distributing the SSPs, or the stars, along the density profiles according to their relative contribution. The new approach presented in Sec. \ref{Sec2} allows us to use  virtually any database available in literature (and this part of the software is freely available upon request to the authors). 
If we want to include the treatment of the spiral arm density distribution on the photometry, the major problem is the extinction along the l.o.s. It has to be accounted for accordingly with the spiral arm distribution of the stars and the gas. In particular, if we want to populate a PDF representative of a CMD for the stellar density distribution computed with the density profiles introduced above, we need an extinction model accounting for a gas distribution following the spiral arms distribution too. 

To account for this extinction, we developed a model of gas distribution based on the spiral density profiles introduced above, but for cold disks ($\Re  = 1$ in Eq.\eqref{Eq55}).
\begin{figure}
\includegraphics[width=\columnwidth]{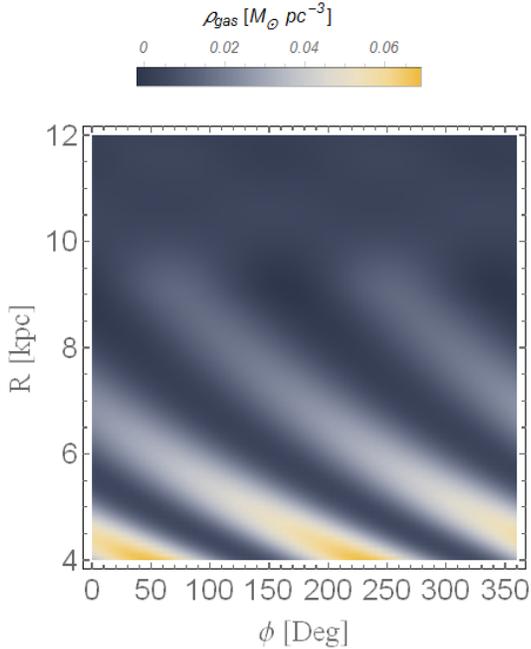}
\caption{ISM gas distribution from Eq.\ref{Eq55}. Note that this kind of density plots map a linear (i.e. $R$) and an angular (i.e. $\phi$) quantity over square. This is causing strong distortions over the range of $R>8$ kpc and careful attention has to be paid in its interpretation.
\label{Fig8} }
\end{figure}
\subsubsection{Extinction model}\label{Sec6.2.1}
While propagating throughout a galaxy, the intensity of the star light decreases because of absorption and scattering due to the presence of interstellar dust. The combined effect, called extinction, has to be taken into account in order to derive the stars intrinsic luminosity from its observed flux. In order to predict the effect of interstellar dust on the observed CMDs, we calculated the extinction towards each SSP or each star in our model galaxy as follows.

We assumed that the dust is traced by the gas in our galaxy model and that its density, relative to the gas density (shown in Fig. \ref{Fig8}), as well as its optical properties, are well described by the dust model of \citet{2007ApJ...657..810D}. This dust model has been calibrated for the dust extinction curve, metal abundance depletion and dust emission measurements in the local Milky Way. From this dust model we consider the extinction coefficient ${k_{\lambda ,\operatorname{ext} }}$ per unit gas mass. From ${k_{\lambda ,\operatorname{ext} }}$ and the gas density distribution, ${\rho _{{\rm{ISM}}}}$, we derive the optical depth crossed by the star light along the path between each star and the observer (located at the sun position): ${\tau _\lambda } =  \int_ \odot ^* {{k_{\lambda ,\operatorname{ext} }}{\rho _{{\rm{ISM}}}}d{r_{{\rm{hel}}}}} $. Then, the extinction in magnitudes is derived as ${A_\lambda } = 2.5{\tau _\lambda }\log e$.

The determination of the predicted observed flux of a star taking into account dust extinction in a galaxy model is affected by several caveats. First, the optical properties of the dust are known to change substantially for different l.o.s. within the Milky Way \citep[e.g., ][]{1999PASP..111...63F}. Therefore, any Milky Way dust model can only be interpreted as an average model for many directions within the Galaxy. Furthermore, the amount of obscuration due to the dust is known to change significantly between the ``diffuse'' and ``dense'' ISM. In particular for very young stars, still embedded in their parent molecular cloud, our approach is surely underestimating their extinction (since the \citet{2007ApJ...657..810D} model is calibrated for the diffuse ISM and molecular clouds are not resolved in our model for the gas distribution).
In this work, we assumed that the dust follows the gas distribution within our galaxy model and that the dust optical properties are uniform. Although quite simple, this approach is sufficient to show the general effect of the presence of spiral arms on the predicted CMD (Fig. \ref{Fig10}).
\begin{figure*}
\centering
\resizebox{\hsize}{!}{\includegraphics{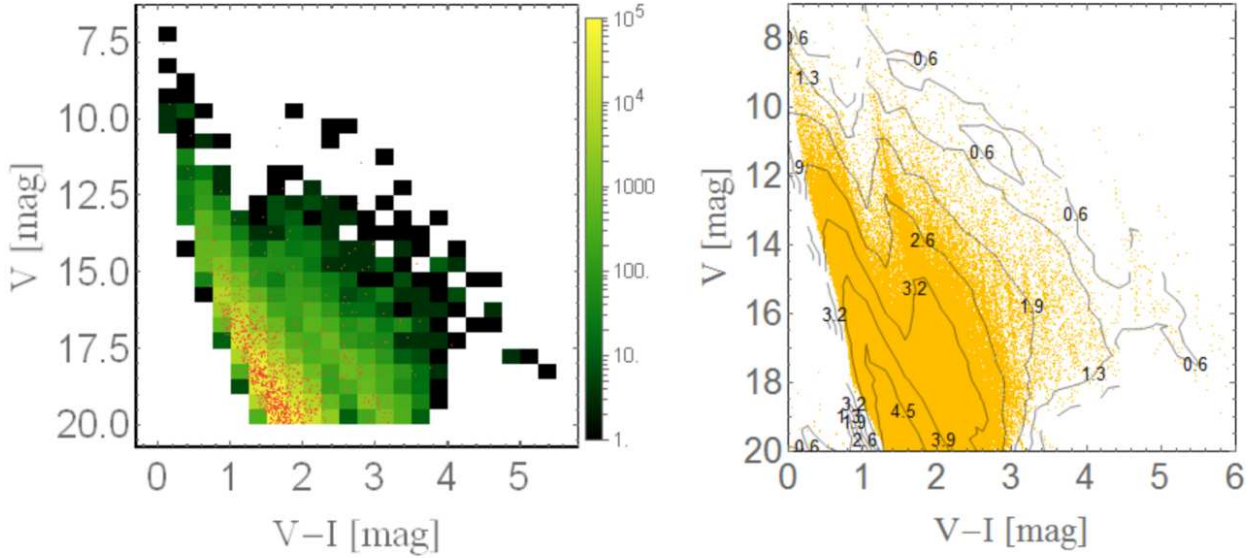}}  
\caption{CMD in V and I band for a field $l \in \left[ {88,92} \right]^\circ $ $b \in \left[ { - 2,2} \right]^\circ $ and magnitude limit $V < 20$ mag for the present model (left panel) and for the Besan\c{c}on model (right panel, see Sec. \ref{Sec8} for details). In our model (left panel) field stars are represented with PDF and only spiral arms star population is visualized in a scatter-type CMD. The logarithmic scale used in this plot is mandatory to interpret the plot and computed as explained in Sec.\ref{Sec3}. For comparison, the Besan\c{c}on model on the right panel is presenting the classical scatter-type CMD where we overlapped the isocontour for the number of stars.
\label{Fig9} }
\end{figure*}
\subsubsection{Sources of stellar tracks, isochrones, SSPs in different photometric systems}\label{Sec6.2.1bis}

We adopt the  stellar models and companion isochrones and SSPs with magnitudes and colours in various photometric systems  of the Padua data-base   because they have been  widely tested and used  over the years in many areas of observational stellar astrophysics  going from the CMDs of stellar clusters, to populations synthesis either star-by-star or integral photometry (magnitudes and colours) or spectral energy distributions, and others.

\begin{itemize}
	\item \textbf{Stellar tracks.} We will not review the physics of these stellar tracks here but we just mention that over the years, these models were calculated including semi-convection in massive stars
\citep[e.g., ][]{1970Ap&SS...8..478C}, ballistic-convective overshooting from the core  \citep{1981A&A...102...25B}, overshooting from the the bottom of the convective envelope \citep{1991A&A...244...95A}, turbulent diffusion from the convective  core and convective shells \citet{1996A&A...313..145D,1996A&A...313..159D,1999A&A...342..131S}, plus several additional improvements and revisions
\citep[][]{1993A&AS...97..851A, 1994A&AS..105...39F, 1994A&AS..106..275B,1994A&AS..105...29F, 1994A&AS..105...29F, 2001AJ....121.1013B,2003AJ....125..770B,2008A&A...484..815B, 2009A&A...508..355B}.
The stellar models in use are those by \citet{2008A&A...484..815B,2009A&A...508..355B}, which cover a wide grid of helium $Y$, metallicity $Z$, and enrichment ratio $\Delta Y / \Delta Z$. The associated isochrones include the effect of mass loss by stellar wind and the thermally pulsing AGB phase according to the models calculated by \citet{2007A&A...469..239M}.

	\item \textbf{The database of SSPs.} We briefly report here on the data base of isochrones and SSPs that has been calculated for the purposes of this study. The code in use is the last version of YZVAR developed over the years by the Padova group and already used in many studies \citep[for instance][]{1981A&A....98..336C,1986sfdg.conf..449C,1989A&A...219..167C,1995A&A...301..381B,1995A&A...295..655N,1996ApJ...469L..97A,2001AJ....121.1013B,
2003AJ....125..770B} and recently extended to   obtain isochrones and
SSPs in a large region of the $Z-Y$ plane. The details on the interpolation scheme at given $\Delta Y/ \Delta Z$ are given in \citet{2008A&A...484..815B,2009A&A...508..355B}.
The present isochrones and SSPs are in the Johnson-Cousins-Glass system as defined by \citet{1990PASP..102.1181B}  and \citet{1988PASP..100.1134B}.  The formalism adopted to derive the bolometric corrections  is described in \citet{2002A&A...391..195G}, whereas the definition and values of the zero-points are  as in
\citet{2007A&A...469..239M} and \citet{2007A&A...468..657G} and will not be repeated here.
Suffice it to recall that the bolometric corrections  stand on an updated and extended library of stellar spectral fluxes. The core of the library now consists of the ԏDFNEWԠATLAS9 spectral fluxes from \citet{2003IAUS..210P.A20C}, for ${T_{{\rm{eff}}}} \in \left[ {3500,50000} \right]$ K, ${\log _{10}}g \in \left[ { - 2,5} \right]$ (with g the surface gravity), and scaled-solar metallicities ${\rm{[M/H]}} \in \left[ { - 2.5, + 0.5} \right]$. This library is extended at the intervals of high $T_{\rm{eff}}$ with pure black-body spectra. For lower $T_{\rm{eff}}$, the library is completed with the spectral fluxes for M, L and T dwarfs from \citet{2000ASPC..212..127A}, M giants from \citet{1994A&AS..105..311F}, and finally the C star spectra from \citet{2001A&A...371.1065L}. Details about the implementation of this library, and in particular about the C star spectra, are provided in \citet{2007A&A...469..239M}. It is also worth mentioning that in the isochrones we apply the bolometric corrections derived from this library without making any correction for the enhanced He content which has been proved by \citet{2007A&A...468..657G} to be small in most common cases.

The database of SSP cover the photometric projection of any reasonable $\mathbb{E_{\rm{MW}}}$. The number of ages ${N_\tau }$  of the SSPs are sampled according to a law of the type $\tau  = i \times {10^j}$ for $i = 1,...,9$ and $j = 7,...,9$, and for $N_Z$ metallicities are $Z = \left\{ {0.0001,0.0004,0.0040,0.0080,0.0200,0.0300,0.0400} \right\}$. The helium content associated to each choice of metallicity is according to the enrichment law $\Delta Y / \Delta Z =2.5$.  Each SSP has been calculated allowing a small age range around the current value of age given by $\Delta\tau = 0.002\times 10^j $ with $j=7,...,9$. No $\alpha$-enhanced or He-enhanced tracks are in use in this example (e.g., $\alpha$-enhanced tracks can easily be taken from an external database, e.g., \citet[][]{2006ApJ...642..797P} or He-enhanced from \citet{2008A&A...484..815B}), and interpolated as in the previous scheme, although this extra-dimension SSP interpolation is beyond the goals of the present paper focused on the kinematics of the spiral arms populations. In total, the data base  contains
 ${N_\tau } \times {N_Z} \cong 150$ SSP. This grid is fully sufficient for our purposes. For future practical application of it, finer grids of SSPs can be calculated and made available.
To calculate SSPs one needs the initial mass function (see comments after Eq.\eqref{Eq6}) of stars of which the are many formulations in the literature. Care must paid that the IMF of the SSPs is the same of the Galaxy model to guarantee self-consistency of the results.
By construction, the IMF contains a normalization factor which depends on the IMF itself and the type of constraint one is using, e.g. the total number of stars in a certain volume, the total mass of stars in a certain  galaxy component etc. In the case of a SSP, the normalization constant is usually defined imposing the total mass of the SSP to be  $M_{\rm{SSP}} = 1 M_\odot$, so that it can immediately be used to find the total luminosity (magnitude) of a stellar assembly with a certain total mass (See section \ref{Sec3} and Eq.\eqref{Eq11}).
Needless to say that other libraries of stellar models and isochrones can be used to generate the database of SSPs, the building blocks of our method. The same is true about the code generating the  SSPs: we have adopted our code YZVAR, of course other similar codes in literature can be used provided they may reach the same level of performance. A code generating SSPs from any database of stellar populations is available upon request to the authors of \citet{2012A&A...545A..14P}. Equally for the photometric systems. So the matrix method for generating the  DFs for  starting SSPs does not depend on a particular choice for the data base of stellar tracks, isochrones, and photometric system.

	\item \textbf{Simulation of photometric errors and completeness.}
Real data on the magnitudes (and colours) of the stars are affected by photometric errors, whose amplitude in general increases at decreasing luminosities (increasing magnitude). The photometric errors come together with the data itself provided they are suitably reduced and calibrated. Photometric errors can be easily simulated in  theoretical CMDs.
The procedure is simple and straightforward \citep[see for instance][for all details]{2012A&A...545A..14P}.
To compare data acquired along a given line of sight with theory one has to know the completeness of the former as a function of the magnitudes and pass-band  \citep{1988AJ.....96..909S,1995AJ....110.2105A}. This is long known problem, and tabulations of the completeness factors must be supplied in advance. The only thing to mention here is that correcting for completeness will alter the DF of stars in the cells of the observational CMD we want to analyze. These tabulations of completeness factors must be  supplied by the user of our method.

	\item \textbf{CMDs rasterisation.} Modern, large surveys of stellar populations easily generate CMDs containing millions of stars or more,   of different age, chemical composition, position in the host galaxy, suffering different reddening and extinction etc. so that even plotting the CMD can be a problem not to speak about deciphering it for the underlying star formation and chemical enrichment  histories, mass and spatial distribution of the stellar component under examination. We have already introduced the concept PDF for a stellar population, we want now to particularize it to the case of the CMD and introduce the concept of a tesselated CMD. Given two photometric pass-bands $\delta \lambda$ and  $\delta \lambda'$ and associated magnitudes and colors, one can soon build two CMDs  $m_{\delta\lambda}$ vs
$m_{\delta\lambda'}$ - $m_{\delta\lambda}$  and $m_{\delta\lambda'}$ vs [$m_{\delta\lambda'}$-$m_{\delta\lambda}$] and divide this in elementary cells of size $\Delta m_{\delta\lambda}$ and $\Delta [m_{\delta\lambda}$-$m_{\delta\lambda}]$.
To the population of each cell contribute stars from all SSPs whose evolutionary path crosses the cell. The regions occupied by stars in the main sequence, red giant, red clump, and asymptotic giant phase this latter stretching to very low effective temperatures (red colors)  are well evident and long lived phases display a higher number of stars compared to the short lived ones. As explained some blurring of the CMD can be  caused by varying extinction across the galaxy under consideration. In the case of deep FOV, the effect of different distances for the stars is included, adding a further dimension to the problem treated with Eq.\eqref{Eq11}.
\end{itemize}

\subsubsection{PDF vs Color-Magnitude Diagrams}\label{Sec6.2.2}
In Fig.\ref{Fig9} (left panel) we present the PDF of the CMD in V and I pass-bands for an ideal  field ${\left\{ {l,b} \right\}_{{\rm{cen}}}} \in \left\{ {90^\circ ,0^\circ } \right\}$ with opening angle $ \sim 2^\circ $ and a limiting magnitude of $V < 20$ mag. We note that the PDF is  not normalized to $\left[ {0,1} \right]$  but by means of the integral Eq.\eqref{Eq11} is normalized to the number of stars effectively predicted by the IMF, star formation history and density profiles as introduced above. The PDF of spiral arms alone has been populated by orange dots to reproduce a scatter diagram (i.e., the classical CMD) for the stellar population and superposed to the PDF distribution of the field stars. The stars in the spiral arm almost overlap  with the main sequence stars of the field. The faint evolved sequence is visible at the right of the main sequence. The right panel shows the distribution of stars in the analogous field generated by the Besan\c{c}on model\footnote{on-line ver. of the July 5, 2013, 9:46 CEST} where standard scatter plots are used for the CMDs. A detailed comparison of the kinematics of the MW stellar populations in our model and in the Besan\c{c}on model is left for Sec.\ref{Sec8}. Here we want to point out how our technique \citep[presented in][]{2012A&A...545A..14P} is particularly fast because of the use of PDFs as opposite to the scatter diagrams used in the Besan\c{c}on model. This is a key feature of our model and it is particularly suited to deal with the upcoming era of large surveys where the realization of several CMDs per second is necessary in order to explore large parameter spaces. Vice versa this exploration seems to be a practical impossibility in models such as the Besan\c{c}on (if not by involving heavy parallel calculus machines) where the techniques requires the realization of scatter plots with dots-over-dots plotted in each colour-magnitude square (cfr. explanation in Sec\ref{Sec3}). Clearly from the Besan\c{c}on model we can reach the same PDF distribution presented in our technique (e.g., by binning the scatter plot in the right panel of Fig. \ref{Fig9}) but the two models work fundamentally in the opposite direction. Our modelling approach works by generating convolved PDFs and then (if necessary) by populating them to obtain a scatter plot, vice versa the Besan\c{c} model works by obtaining first the scatter plots diagrams and then (if necessary) by binning them to have PDF.

\begin{figure*}
\centering
\resizebox{\hsize}{!}{\includegraphics{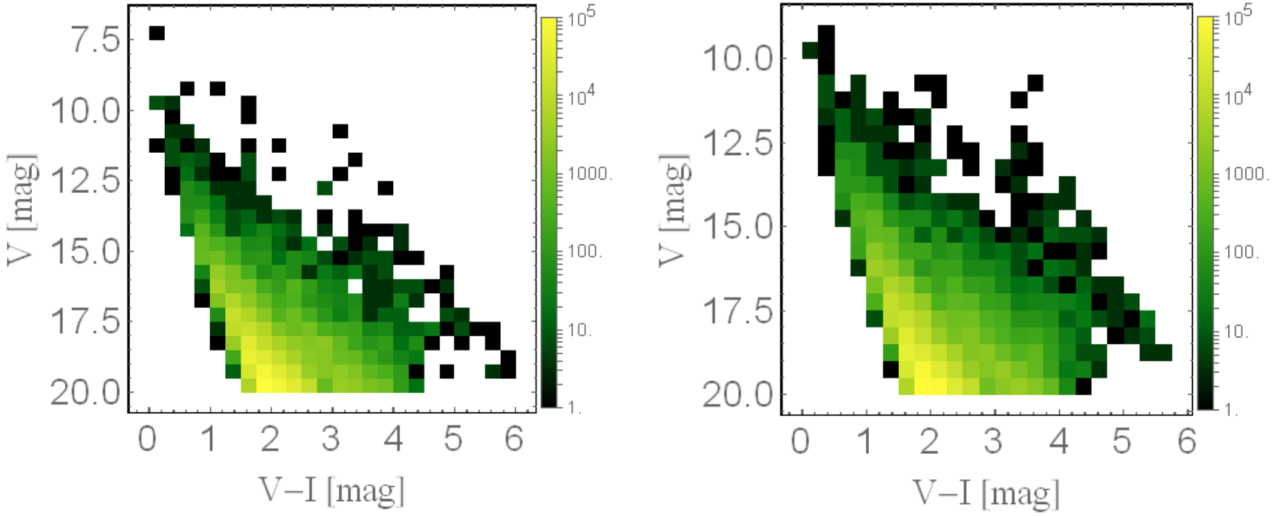}}
\caption{(left panel). CMD for the sole field stellar population derived by using the extinction model introduced in Sec. \ref{Sec6.2.1}. (right panel) CMD realized assuming a double exponential extinction profile on the sole field stars.
\label{Fig10} }
\end{figure*}

Finally, in Fig. \ref{Fig10} we show two examples of tesselated CMDs for for a population of field stars observed in the V and I pass-bands and in which two different extinction profile have been adopted. The left panel shows the case with extinction represented by the asymmetric gas spiral distribution of Fig.\ref{Fig8}, whereas in the right panel two exponential laws  are used for the extinction. The logarithmic colour code is proportional to the value of the PDF in each cell.
Once the stars are distributed on the CMD, we can test the effect of the  new extinction model.
The large  impact of the gas distribution along the spiral arms in the shaping  the composite CMD is soon evident. Analysing this particular field in more detail  is beyond the goal of the present study. However, it is worth mentioning that the combined effects of the large viewing angle   and the density gradient in mass density due to  spiral arms have combined to  stretch the distribution of stars to the red side of the CMD. This is  because of the more concentrated gas/dust distribution rising up rapidly at about 1.3 kpc from the solar position along the l.o.s.
In a near future, surveys like  Gaia will be able to provide great insight on the distribution of gas and dust across the   MW. In this context, the technique we have developed  can soon be adapted to a star-by-star approach to directly determine the extinction.

\section{Velocity field description of non-axisymmetric features: azimuthal and vertical tilt of the velocity ellipsoid}\label{Sec7}
The tilt of the velocity ellipsoid with respect to the configuration space axis generates   non-null diagonal terms in the matrix ${{\bm{\sigma }}_{\bm{v}}}\left( {\bm{x}} \right)$ introduced  in Eq.\eqref{Eq43}. We derive here  these
non-null terms in the context of two independent theories.

Following what was done for the zero-order moment  of ${f^{{\rm{Sch}}}}$, we rely on the DWT to derive the moments of order one or more for the velocity field on the plane, and on the study by \citet{1991ApJ...368...79A} for the velocity moments on the meridional plane. 

Finally once the moments of the velocity DF are obtained, the velocity of the field is derived from the diagonalization of ${{\bm{\sigma }}_{\bm{v}}}\left( {\bm{x}} \right)$ by simply solving the eigen-system:
\begin{equation}\label{Eq57}
	\det \left( {{\bm{\sigma^{-1} }} - \lambda {\bm{I}}} \right) = 0,
\end{equation}
with ${\bm{I}}$ the unit matrix and $\lambda $  the eigenvalues,  and populating the corresponding tilted PDF.

\subsection{Radial-azimuthal velocity field}\label{Sec7.1}

\subsubsection{Radial mean stream velocity}\label{Sec7.1.1}
\begin{figure*}
\centering
\resizebox{\hsize}{!}{\includegraphics{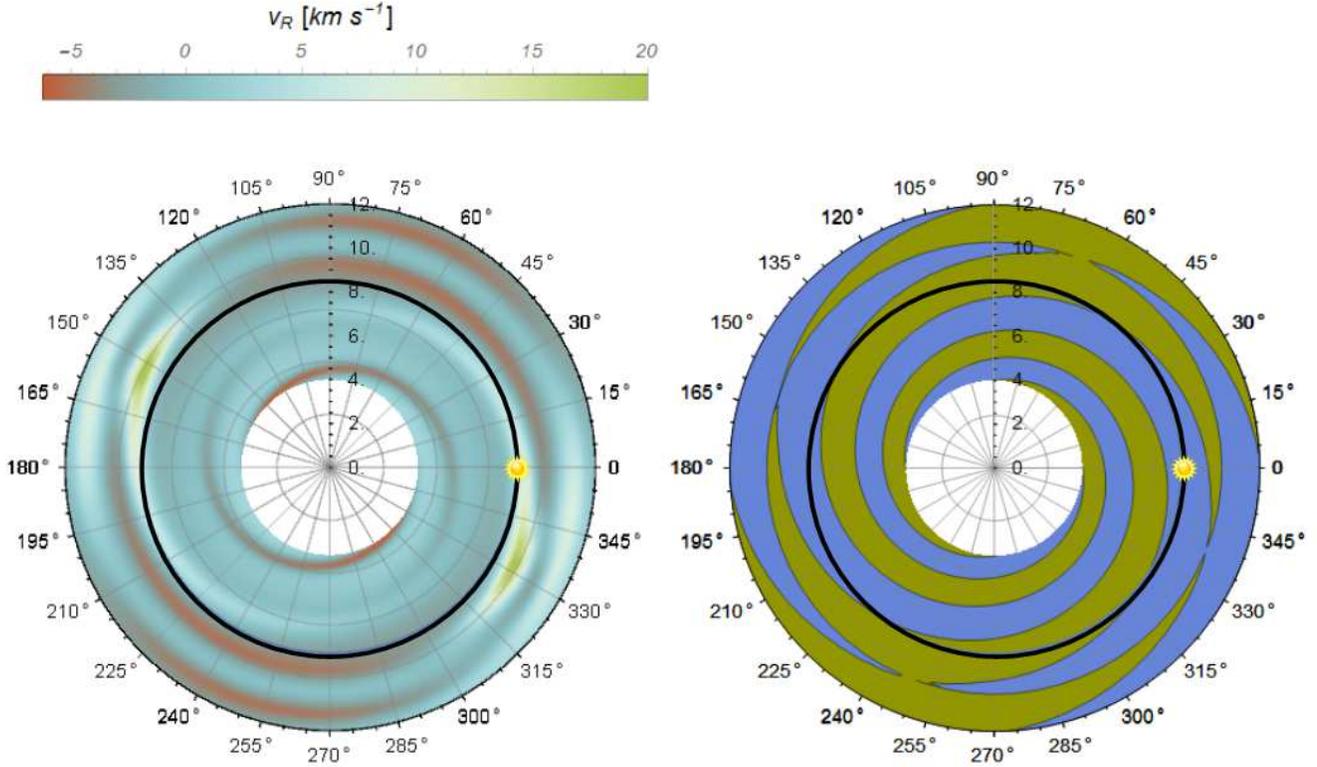}}
\caption{(left pane) Mean radial velocity ${\bar v_R}$ for a spiral stellar population. The Sun is arbitrarly located at ${\left\{ {R,\phi ,z} \right\}_ \odot } = \left\{ {8.0,0.0,0.02} \right\}$[kpc, deg, kpc] and its circle radius shown in black. The Fig. is on the plane at $z = 0$. (right panel) Density contrast ${\Sigma _{{\rm{sp}}}} - {\Sigma _0}$ at one contour level. Density contrast is negative in blue-colour.
\label{Fig11} }
\end{figure*}

The computation of the velocity field proceeds exactly as above for the moment of order zero. We start with the radial moment defined as:
\begin{equation}\label{Eq58}
	\begin{aligned}
		\bar v_R  &= \frac{1}{\Sigma }\int_{{\mathbb{R}^3}}^{} {{f }{v_R}d{v_R}d{v_\phi }d{v_z}}  \\ 
		&= \frac{{{\Sigma _0}}}{\Sigma }{{\bar v}_{0,R}} + \frac{{{\Sigma _0}}}{\Sigma }\frac{1}{{{\Sigma _0}}}\int_{{\mathbb{R}^3}}^{} {{f_1}{v_R}d{v_R}d{v_\phi }d{v_z}}  \\ 
		&= \frac{1}{\Sigma }\int_{{\mathbb{R}^3}}^{} {{f_1}{v_R}d{v_R}d{v_\phi }d{v_z}} , \\ 
	\end{aligned}
\end{equation}
where ${\bar v_{0,R}} = \frac{1}{{{\Sigma _0}}}\int_{}^{} {{f^{{\rm{Sch}}}}{v_R}d{v_R}d{v_\phi }d{v_z}}  = 0$ and remembering Eq.\eqref{Eq47} we evidently need to compute the following
\begin{equation}\label{Eq59}
\begin{gathered}
  \bar v_R  =  - \frac{{{\Sigma _0}}}{\Sigma }\frac{{{\Phi _1}}}{{\sigma _{RR}^2}}\frac{1}{{{\Sigma _0}}}\int_{{\mathbb{R}^3}}^{} {{f^{{\rm{Sch}}}}\left( {1 - \frac{{\rm{sinc}^{-1}\left( {\nu \pi } \right)}}{{2\pi }}} \right. \times }  \hfill \\
  \left. { \times \int_{ - \pi }^\pi  {{e^{\iota \left( {\nu \tau  + X\left( {u\sin \tau  + v\left( {1 + \cos \tau } \right)} \right)} \right)}}d\tau } } \right)u\frac{{\sigma _{RR}^3}}{\gamma }{\sigma _{zz}}dudvdw. \hfill \\ 
\end{gathered} 
\end{equation}
But the RHS of the previous equation reads 
\begin{equation}\label{Eq60}
\begin{gathered}
  =  - \frac{{{\Sigma _0}}}{\Sigma }\frac{{{\Phi _1}}}{{\sigma _{RR}^2}}\frac{1}{{{\Sigma _0}}}\frac{{\sigma _{RR}^3}}{\gamma }{\sigma _{zz}}\left( {\int_{{\mathbb{R}^3}}^{} {{f^{\rm{Sch}}}ududvdw}  - \int_{{\mathbb{R}^2}}^{} {{f^{\rm{Sch}}} \times } } \right. \hfill \\
  \left. { \times \frac{{\rm{sinc}^{-1}\left( {\nu \pi } \right)}}{{2\pi }}\int_{ - \pi }^\pi  {{e^{\iota \left( {\nu \tau  + X\left( {u\sin \tau  + v\left( {1 + \cos \tau } \right)} \right)} \right)}}d\tau } } \right)ududvdw \hfill, \\ 
\end{gathered} 
\end{equation}
where the first term in the round brackets on the is identically null because no average radial motion is expected on an axisymmetric disk. We simplify further Eq.\eqref{Eq59} by introducing explicitly Eq. \eqref{Eq43} as
\begin{equation}\label{Eq61}
\begin{gathered}
  \bar v_R  = \frac{{{\Sigma ^{{\rm{Sch}}}}}}{\Sigma }\frac{{{\Phi _1}}}{{{\sigma _{RR}}}}\frac{{\rm{sinc}^{-1}\left( {\nu \pi } \right)}}{{{{\left( {2\pi } \right)}^2}}}\int_{ - \pi }^\pi  {d\tau  \times }  \hfill \\
   \times \int_{{\mathbb{R}^2}}^{} {dudvdw{e^{\iota \left( {\nu \tau  + X\left( {u\sin \tau  + v\left( {1 + \cos \tau } \right)} \right)} \right)}}u{e^{ - \frac{{{u^2} + {v^2} + {w^2}}}{2}}}} , \hfill \\ 
\end{gathered} 
\end{equation}
so that carrying out explicitly the integral on the bottom row of the previous equation we get
\begin{equation}\label{Eq62}
		\bar v_R  = \frac{{{\Sigma _0}}}{\Sigma }\frac{{{\Phi _1}}}{{{\sigma _{RR}}}}\frac{X}{{2\pi }}\rm{sinc}^{-1}\left( {\nu \pi } \right)\int_{ - \pi }^\pi  {d\tau \iota \sin \tau {e^{\iota \nu \tau  - {X^2}\left( {1 + \cos \tau } \right)}}} ,
\end{equation}
and finally, making use of the Eq.(53), we are able to write the first order moment in the radial direction as:
\begin{equation}\label{Eq63}
	\begin{aligned}
		\bar v_R  &= \frac{{{\Sigma _0}}}{\Sigma }\frac{{{\Phi _1}}}{{{\sigma _{RR}}}}\frac{X}{2} \rm{sinc}^{-1}\left( {\nu \pi } \right)\times  \hfill \\
		&\times \left( {{}_{\left( {\frac{1}{2},1} \right)}{{\hat F}_{\left( { - \nu ,2+\nu  } \right)}} - {\kern 1pt} {}_{\left( {\frac{1}{2},1} \right)}{{\hat F}_{\left( {\nu ,2 - \nu } \right)}}} \right). \hfill \\ 
	\end{aligned} 
\end{equation}
We plot an example of the computing of Eq.\eqref{Eq63} in Fig. \ref{Fig11}. As evidenced in the moment of order zero, the amplitude of the response in the velocity field at the resonances grows beyond the limits permitted by the linear response theory of \citet{1969ApJ...155..721L} and the theory breaks down. Analytical continuation is applied also in this case within the same framework developed for the first order moment above. 

For a spiral pattern with trailing spiral arms, $i > 0$, it is easy to prove that the wave number $k = \frac{{\partial \psi }}{{\partial R}} < 0$ and so $\psi $ is a  decreasing function of $R$. Hence, for example, if we are considering the regions where ${\Omega _p} > \Omega \left( R \right)$ (see also Fig. \ref{Fig6}) we have that $\nu  > 0$. Inside the spiral arm it is ${\Sigma _1} > 0$ so that we must have ${\Phi _1} < 0$ . Hence, with the centre of the spiral being given by the phase $\varphi  = 0$, we recover the results expected by the density spiral wave theory that presents mean radial motion toward the galactic centre inside the spiral arm and a motion outward in the inter-arm regions $\varphi  =  \pm \pi $ depending on the location where ${\Omega _p} = \Omega \left( R \right)$ as shown in Fig.\ref{Fig11} just above the $\sim 6.5$ kpc \citep[][]{2014MNRAS.440.2564F}. As evidenced by the one contour style of this Fig. where ${\Sigma _{{\rm{sp}}}} - {\Sigma _0}$  has been plotted, the mean radial velocity field of Fig. \ref{Fig11} is in-phase with the density as expected from DWT. 
This test is not only performed to graphically validate the computation of the velocity moments through hypergeometric functions, but more importantly, to evidence the goodness of the continuation scheme of Sec. \ref{Sec6.2.1} over the resonance on the velocity space. As evident from the plot the expectation holds even above and very close to the resonances, $R \in \left[ {{R_{{\rm{res}}}} - {\varepsilon _{{\rm{res}}}},{R_{{\rm{res}}}} + {\varepsilon _{{\rm{res}}}}} \right]$, a result which is not obvious to prove analytically.

\subsubsection{Azimuthal mean velocity}
\begin{figure*}
\centering
\resizebox{\hsize}{!}{\includegraphics{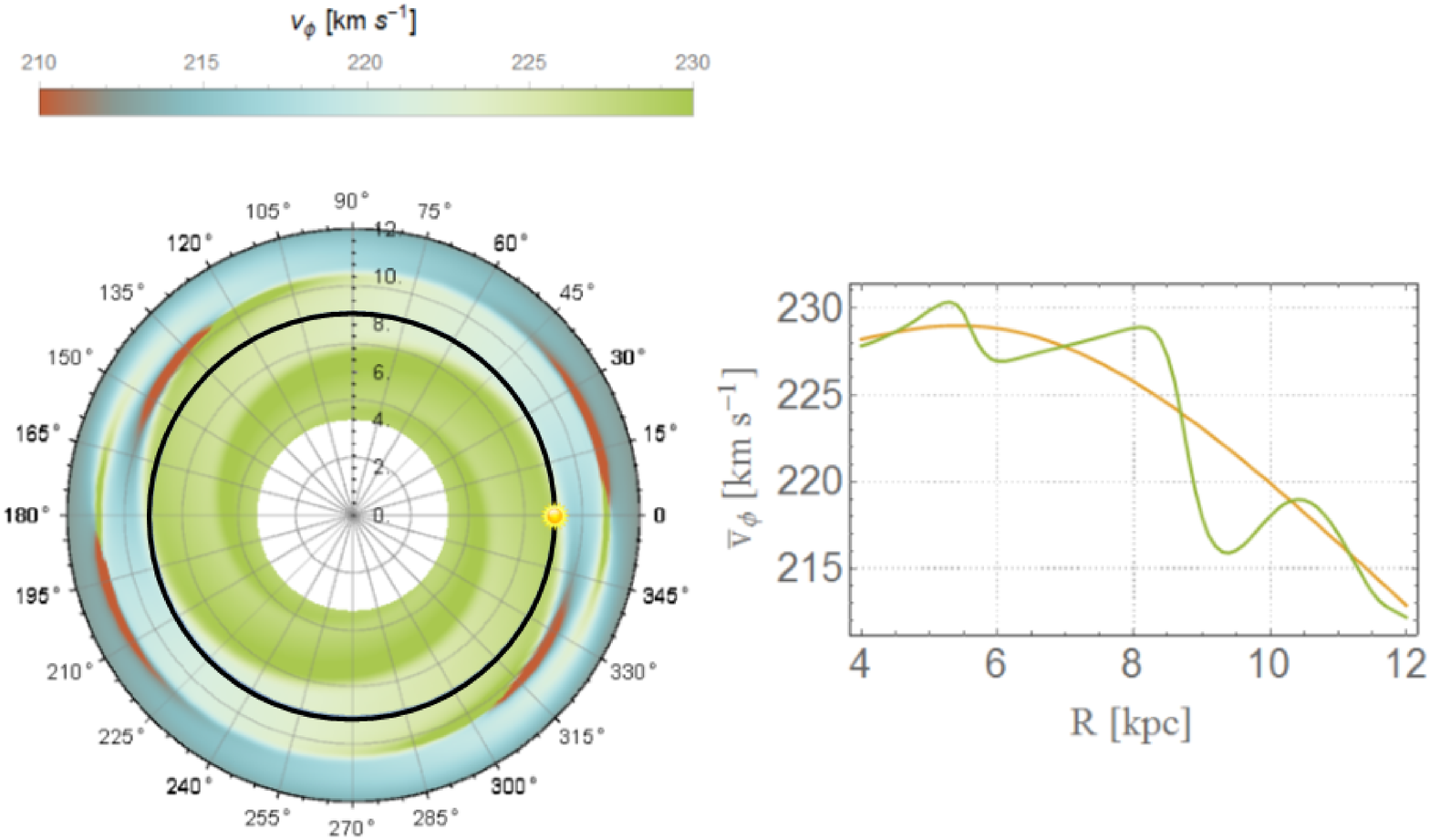}}
\caption{(left panel), Mean stream azimuthal velocity over all the disk. (right panel) Mean stream azimuthal velocity for spiral arm distributed SSP (green). It’s evident the strong influence of the spiral arms in the measure circular velocity (see text for detail). For comparison the rotation curve of Fig. \ref{Fig3} is added in the range of interest of the spiral arms.
\label{Fig12} }
\end{figure*}

\label{{Sec7.1.2}}
It is even more interesting to describe the influence of the spiral arms in the mean azimuthal velocity for its implications regarding the location of the Sun relative to the Local Standard of Rest. Because of the similarity of the integrations performed previously in the radial direction, we report here simply the results. We obtain for the azimuthal direction 
\begin{equation}\label{Eq64}
	\bar v_\phi   = \frac{{{\Sigma ^{{\rm{Sch}}}}}}{\Sigma }{v_c} - \frac{{{\Sigma ^{{\rm{Sch}}}}}}{\Sigma }{\Phi _1}\frac{{{v_c}}}{{\sigma _{RR}^2}}\left( {1 - \frac{{\nu \pi }}{{\sin \left( {\nu \pi } \right)}}{\;_{\left( {\frac{1}{2},1} \right)}}{{\hat F}_{\left( {1 - \nu ,1 + \nu } \right)}}} \right),
\end{equation}
that we plot on Fig. \ref{Fig12}.
As shown in this figure, the average azimuthal perturbation on the circular velocity is of the order of 5 to 10 ${\rm{km}}\;{{\rm{s}}^{ - 1}}$, i.e. compatible with the motion of the sun relative to the LSR. This result is extremely interesting no matter where the resonance is located. At every radius, the spiral arm presence affects the mean motion and can severely bias the works aimed to determine the motion of the Sun with respect to the circular velocity (the Local Standard of Rest). This is in line with what is already evidenced by numerical simulations \citep[e.g., ][]{2005AJ....130..576Q,2014MNRAS.440.2564F,2014MNRAS.443.2757K}. Unfortunately, up to now the result has only a theoretical value because it is affected by the uncertainties on the resonances' locations, on the validity of the DWT, and on the uncertainty of the Sun's location.

\subsubsection{Dispersion velocity tensor}\label{Sec7.1.3}
The moments of order two can be calculated by direct integration, but the procedure is more cumbersome and the integrals not easily tractable analytically. A simpler and fully algebraic procedure is followed here bypassing the direct integration in favour of the second order not central moments. The desired results will be then achieved with the help of the previously obtained Eq. \eqref{Eq63} and \eqref{Eq64}. We present here the computation for the first of these moments; the results for the following orders are obtainable following a similar procedure. 
From the definition of non-central radial moment of order two we write 
\begin{equation}\label{Eq65}
	\begin{aligned}
		{\overline {v_R^2}  } &= \frac{1}{\Sigma }\int_{{\mathbb{R}^3}}^{} {{f }v_R^2d{v_R}d{v_\phi }d{v_z}}  \\ 
		&= \frac{1}{\Sigma }\int_{{\mathbb{R}^3}}^{} {\left( {{f^{{\rm{Sch}}}} + {f_1}} \right)v_R^2d{v_R}d{v_\phi }d{v_z}}  \\ 
		&= \frac{{{\Sigma _0}}}{\Sigma }\sigma _{RR}^2 + \frac{1}{\Sigma }\int_{{\mathbb{R}^3}}^{} {{f_1}v_R^2d{v_R}d{v_\phi }d{v_z}} , 
	\end{aligned} 
\end{equation}
and remembering the definition of ${f_1}$ from Eq.\eqref{Eq47} we obtain:
\begin{equation}\label{Eq66}
\begin{gathered}
  {\overline {v_R^2}  } = \frac{{{\Sigma _0}}}{\Sigma }\sigma _{RR}^2 - \frac{1}{\Sigma }\frac{{{\Phi _1}}}{\gamma }\frac{{{\sigma _{zz}}}}{{\sigma _{RR}^{ - 2}}}\int_{{\mathbb{R}^3}}^{} {\frac{{{\Sigma _0}}}{{{{\left( {2\pi } \right)}^{3/2}}{\sigma _{RR}}{\sigma _{\phi \phi }}{\sigma _{zz}}}} \times }  \hfill \\
   \times {e^{ - \frac{{{u^2} + {v^2} + {w^2}}}{2}}}\left( \begin{gathered}
  1 - \frac{\rm{sinc}^{-1}(\nu \pi)}{{2\pi }} \times \hfill \\
  \int_{ - \pi }^\pi  {{e^{\iota \left( {\nu \tau  + X\left( {u\sin \tau  + v\left( {1 + \cos \tau } \right)} \right)} \right)}}d\tau }  \hfill \\ 
\end{gathered}  \right){u^2}dudvdw \hfill \\
   = \frac{{{\Sigma _0}}}{\Sigma }\sigma _{RR}^2 - \frac{{{\Sigma _0}}}{\Sigma }{\Phi _1}\left( {1 - \rm{sinc}^{-1}(\nu \pi)\left( {_{\left( {\frac{1}{2},1} \right)}{{\hat F}_{\left( {1 - \nu ,1 + \nu } \right)}} - } \right.} \right. \hfill \\
  \left. {\left. { - 2{X^2}\left( {_{\left( {\frac{3}{2},2} \right)}{{\hat F}_{\left( {2 - \nu ,2 + \nu } \right)}} - 3{{\kern 1pt} _{\left( {\frac{5}{2},3} \right)}}{{\hat F}_{\left( {3 - \nu ,3 + \nu } \right)}}} \right)} \right)} \right). \hfill \\ 
\end{gathered} 
\end{equation}
In the same way, we obtain for the azimuthal term (the computation is tedious but straightforward):
\begin{equation}\label{Eq67}
  {\overline {v_\phi ^2}  } = \frac{{{\Sigma _0}}}{\Sigma }v_c^2\left( {1 - \frac{{{\Phi _1}}}{{\sigma _{RR}^2}}\left( {1 - \frac{{\nu \pi}}{\sin({\nu \pi})}{\,_{ ({\frac{1}{2},1}) }}{{\hat F}_{\left\{ {1 - \nu ,1 + \nu } \right\}}}} \right)} \right),
\end{equation}
and for the mixed term:
\begin{equation}\label{Eq68}
\begin{gathered}
  {\overline {{v_R}{v_\phi }}  } = \frac{{{\Phi _1}}}{{{\sigma _{RR}}}}\frac{{{\Sigma _0}}}{\Sigma }\frac{{\rm{sinc}^{-1}\left( {\nu \pi } \right)}}{2}\frac{X}{\gamma } \times  \hfill \\
  \left( {\gamma {v_c}\left( {{{\mkern 1mu} _{\left( {\frac{1}{2},1} \right)}}{{\hat F}_{\left( { - \nu ,\nu  + 2} \right)}} - {{\mkern 1mu} _{\left( {\frac{1}{2},1} \right)}}{{\hat F}_{\left( {\nu ,2 - \nu } \right)}}} \right) + } \right. \hfill \\
  \left. { + \iota X{\sigma _{RR}}\left( {{{\mkern 1mu} _{\left( {\frac{3}{2},2} \right)}}{{\hat F}_{\left( {1 - \nu ,\nu  + 3} \right)}} - {{\mkern 1mu} _{\left( {\frac{3}{2},2} \right)}}{{\hat F}_{\left( {\nu  + 1,3 - \nu } \right)}}} \right)} \right). \hfill \\ 
\end{gathered} 
\end{equation}
The computing of these terms results from a simple application of the Hypergeometric formalism introduced above. The last step to achieve the dispersion velocity terms for the velocity ellipsoid perturbed by spiral arms comes as a  simple collection of the previous results as:
\begin{equation}\label{Eq69}
	\begin{aligned}
		\sigma _{RR}  &= {\overline {v_R^2}  } - \bar v_R^{  2}, \\ 
		\sigma _{\phi \phi }  &= {\overline {v_\phi ^2}  } - \bar v_\phi ^{  2}, \\ 
		\sigma _{R\phi }  &= {\overline {{v_R}{v_\phi }}  } - \bar v_\phi  \bar v_R , \\ 
	\end{aligned}
\end{equation}
and with Eq.\eqref{Eq68}, \eqref{Eq67}, \eqref{Eq66}, \eqref{Eq64}, and \eqref{Eq63} we conclude the computation of the second order moments.

\begin{figure}
\includegraphics[width=\columnwidth]{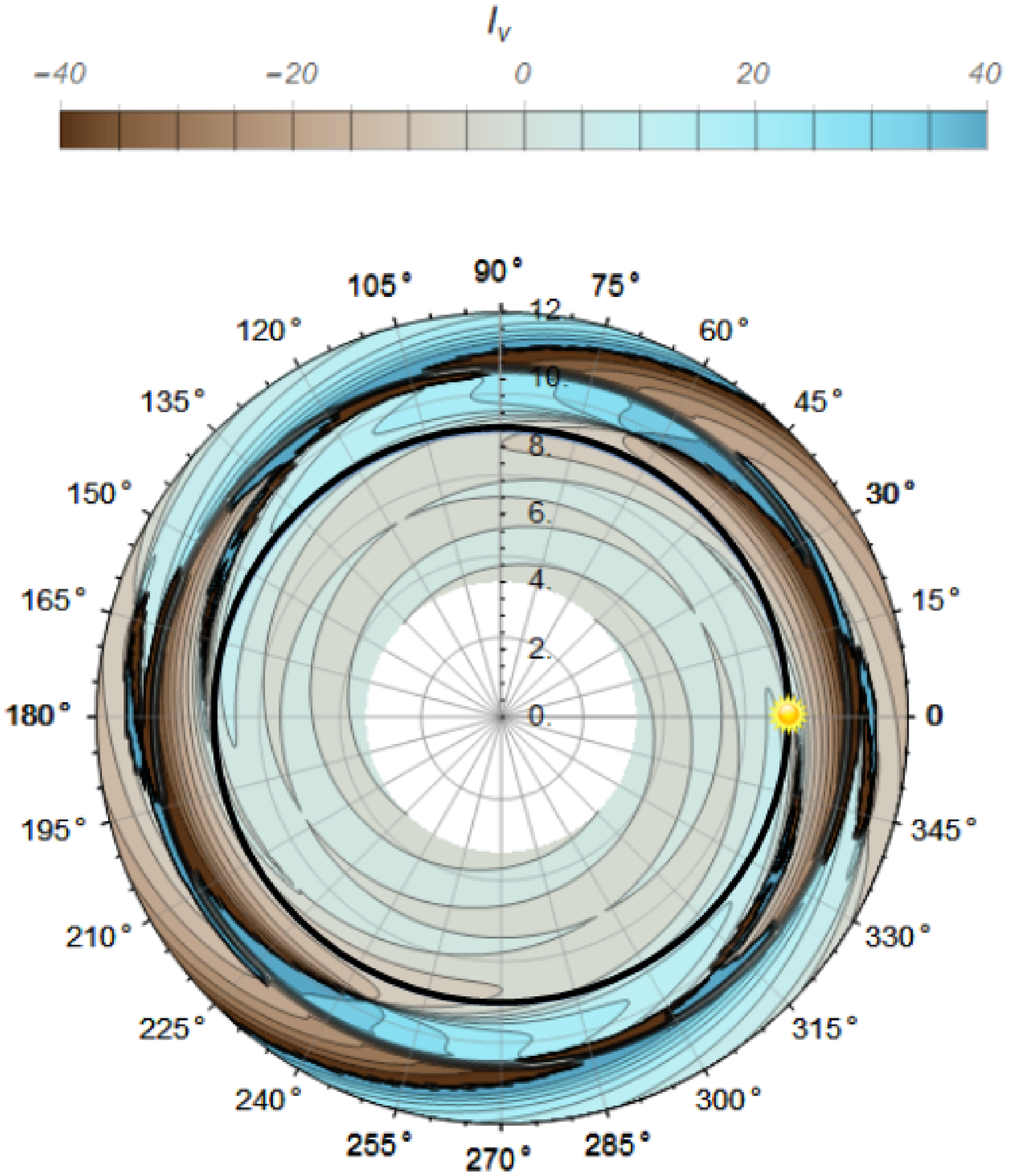}
\caption{Vertex deviation of the velocity ellipsoid at the plane. No graphical smoothing is applied in the contour plot to evidence the grid resolution adopted.
\label{Fig13} }
\end{figure}

\begin{figure}
\includegraphics[width=\columnwidth]{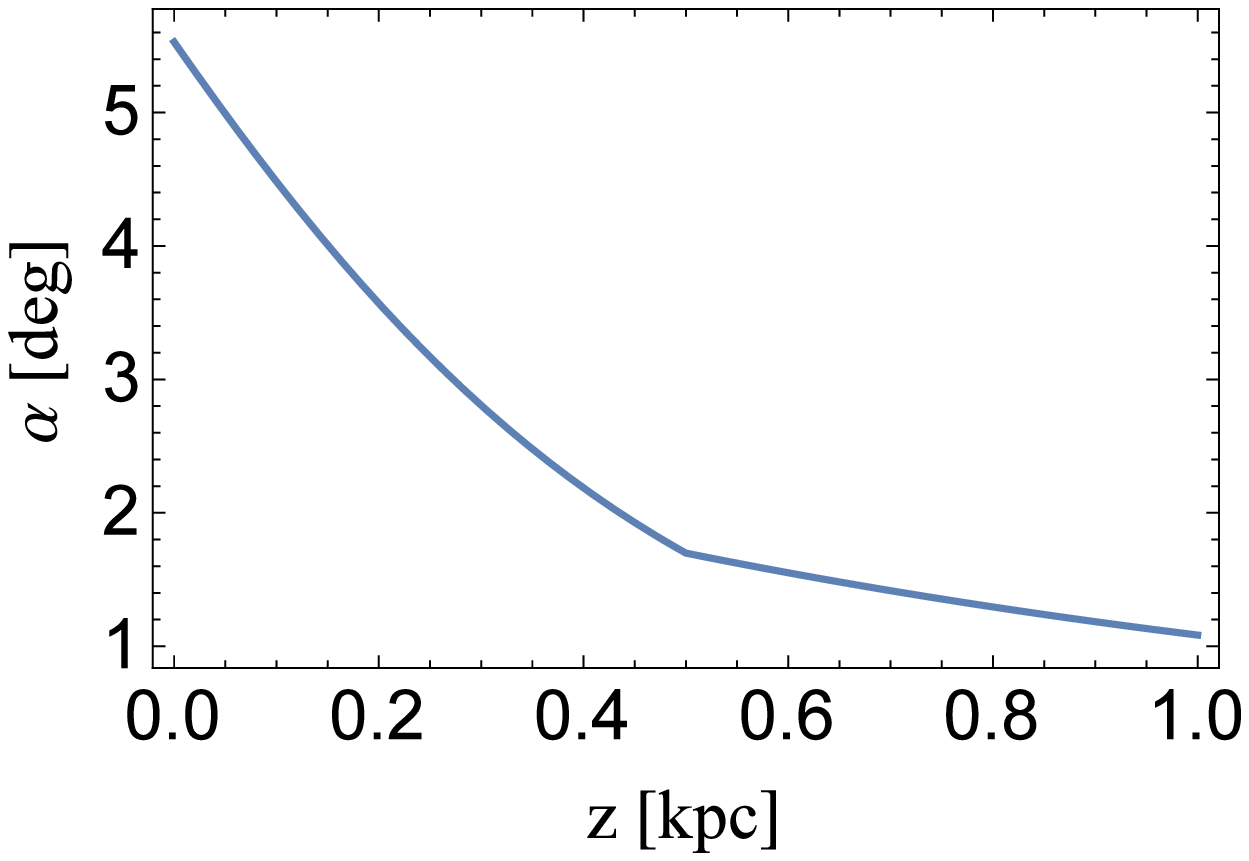}
\caption{Vertex deviation outside the plane on the solar radial position ${l_v} = {l_v}\left( {{R_ \odot },0,z} \right)$.
\label{Fig14} }
\end{figure}

Of particular interest is for example the plot of the mixed term $\sigma _{R\phi } $  because of its connection with the azimuthal tilt (of an angle ${l_v}$ ) with  respect the configuration space cylindrical coordinates of the velocity ellipsoid on the plane:
\begin{equation}\label{Eq70}
	{l_v}\left( {R,\phi ,z} \right) = \frac{1}{2}\arctan \left( {\frac{{2\sigma _{R\phi }^2}}{{\sigma _{RR}^2 - \sigma _{\phi \phi }^2}}} \right),
\end{equation}
i.e. the vertex deviation that we plot in Fig. \ref{Fig13}.
There is much observational evidence for the dependence of the tilting of the velocity ellipsoid on the plane \citep[since][]{1958RA......5..467L, 1970IAUS...38..423W} and recently it has been studied in \citet{2012A&A...547A..71P}. In the latter work, a detailed plot of this trend has been shown not only on the plane but also above and below the plane. These data based on the RAVE survey highlight a decrease of the vertex deviation above and below the plane. This observational trend is important to validate the vertical treatment of the vertex deviation outside the plane. Because the theory is not self-consistently validated outside the plane, we point out that the dependence on ``z'' of Eq.\eqref{Eq70} comes from the vertical dependence of ${\sigma _{R\phi }}$, ${\sigma _{RR}}$ and ${\sigma _{\phi \phi }}$. While the behaviour of  ${\sigma _{RR}}$ and ${\sigma _{\phi \phi }}$ outside the plane are given by Eq.\eqref{Eq44}, which find observational constraints in the values in Table \ref{Table2}, the trend of ${\sigma _{R\phi }} = {\sigma _{R\phi }}\left( z \right)$ is entirely a simplified assumption we adopted in Eq.\eqref{Eq69}. The results of the convolution of just two SSPs of pop 1 and pop 2 of Table \ref{Table1} (Fig. \ref{Fig14}) treated with DWT seem to qualitatively reproduce the observational trend of \citet{2012A&A...547A..71P} (their left panel in Fig \ref{Fig12}).
\subsubsection{Limits of the adopted approach}\label{Sec7.1.4}
Even if the tilt of the velocity ellipsoid presents a regular trend, the present theory accounts only for the contribution of the spiral arms which is well known to be incomplete. The stream motions are expected to have major impact on the velocity distribution in the Solar Neighbourhood as proven by several authors \citep[e.g.,][]{2003A&A...398..141S, 1998AJ....115.2384D, 1999ApJ...524L..35D, 2007MNRAS.380.1348S, 2004MNRAS.350..627D, 1986RMxAA..13....9B, 1982ApJ...255..217H, 1972A&A....18...97M, 1970IAUS...38..423W}. In order to correctly account for the tilt,  a complete map of the distribution  of the molecular clouds encountered  by the stars along their past orbits is needed. To date, this target   is out of reach thus  weakening any study of the kinematics and dynamics of MW based on direct orbit-integration. This is one of the main reasons spurring us to apply the method of moments  in our model of the Galaxy. A limitation of our approach is surely the lack of a self-consistent treatment of the vertical  DF. The DWT is limited to in-plane stellar motions (the vertical and planar motions are uncoupled  because of the epicyclical approximation). This represents a serious drawback  of the theory that makes it not fully coherent   with the assumptions we are going to male  on the vertical tilt of the velocity ellipsoid. Nonetheless, we will include the vertical tilt in the Galaxy Model because, as shown in Fig. \ref{Fig14}, our approach seems to capture, at least qualitatively, the information hidden in the observational data acquired by  \citet{2012A&A...547A..71P}.

\subsection{Vertical velocity field}\label{Sec7.2}
The study of the local vertical profiles of the MW disks has long tradition and is still pushed mostly under axisymmetric assumption \citep[e.g.,][]{2010MNRAS.402..461J,2011MNRAS.411.2586J,2003A&A...398..141S,2003A&A...399..531S,2006A&A...446..933B,2008A&A...480...91S}. The study of the vertical kinematics of spiral arms is still an open research field \citep[e.g.,][]{2013MNRAS.436..101W,2012ApJ...750L..41W}. The kinematical description of the implemented model outside the plane is formally obtained here for the axisymmetric case. Nevertheless, the tilt of the velocity outside the plane and the non-isothermality for single SSP reduce to zero thus matching the in-plane description presented in the previous section. 

\subsubsection{Vertical tilt of the velocity ellipsoid}\label{Sec7.2.1}
The last non null cross term considered in Eq.(43) is $\sigma _{Rz}^2 = \sigma _{Rz}^2\left( {R,\phi ,z} \right)$. It represents the tilt of the principal axis of the velocity ellipsoid with respect to cylindrical coordinates outside the plane. By symmetry in any axisymmetric model we expect that $\sigma _{Rz}^2\left( {R,0} \right) = 0$, so that the principal axes of the velocity ellipsoid are aligned with the cylindrical coordinates in the plane. This is not generally true in a non-axisymmetric model or in a model with non null radial average velocity \citep[][]{2014A&A...561A.141C, 2014A&A...567A..46C}. Out of the plane the alignment is poorly known. For small z we can write
\begin{equation}\label{Eq71}
	\sigma _{Rz}^2\left( {{R_ \odot },z} \right) \approx \sigma _{Rz}^2\left( {{R_ \odot },0} \right) + z\frac{{\partial \sigma _{Rz}^2\left( {{R_ \odot },0} \right)}}{{\partial z}} + O\left( {{z^2}} \right),
\end{equation}
where $\sigma _{Rz}^2\left( {{R_ \odot },0} \right) = 0$ if and only if the model is axisymmetric. The z-derivative of $\sigma _{Rz}^2$  (the last term of Eq. \eqref{Eq71}) evaluated on the plane gives the orientation of the velocity ellipsoid just above or below the plane. \cite{1915MNRAS..76...37E} and \cite{1988ApJ...333...90D} have shown that the velocity tensor is diagonal in coordinates (if they exist) in which the potential is separable. In the case where the MW potential is separable in cylindrical or spherical coordinates, corresponding to mass distributions which are highly flattened and dominated by the disk or spherical halo, we recover the limiting case of the vertical titling of the velocity ellipsoid \citep[][]{2003MNRAS.342.1056V, 2003MNRAS.340..752F, 2003MNRAS.339..834H, 2002MNRAS.331..959V, 1999A&A...341...86B, 1996A&A...314...25M, 1995MNRAS.276..293A, 1994AstL...20..429O, 1992MNRAS.257..152E, 1991MNRAS.252..606D,1990ApJ...363..367H, 1990ApJ...358..399M, 1990MNRAS.244..111E, 1989ApJ...345..647G}.

These two cases correspond to the upper and lower boundaries of the tilt term. They are usually written as
\begin{equation}\label{Eq72}
	\frac{{\partial \sigma _{Rz}^2\left( {R,0} \right)}}{{\partial z}} = \lambda \left( R \right)\frac{{\sigma _{RR}^2\left( {R,0} \right) - \sigma _{zz}^2\left( {R,0} \right)}}{R},
\end{equation}
where$\lambda \left( R \right) \in \left[ {0,1} \right]$. The factor $\lambda (R)$ can be derived either analytically from orbit integration or is assumed $\lambda  = 0$  for simplicity \citep[e.g.,][]{1986ApJ...303..556V, 1989AJ.....97..139L, 1990ApJ...361..408S}. Numerical simulations \citep[e.g.,][]{ 1987AJ.....94..666C, 2000AGM....16..T10B} are performed to calculate explicitly the moments for a given gravitational potential. The result of the above studies is that, at the Sun's position $\lambda  \simeq 0.5$. In our model for the Galactic kinematics, we prefer to adopt the analytical formulation of $\lambda \left( R \right)$ given by \cite{1991ApJ...368...79A}:
\begin{equation}\label{Eq73}
	\lambda \left( R \right) = {\left. {\frac{{{R^2}{\partial _{R,z,z}}\Phi }}{{3{\partial _R}\Phi  + R{\partial _{R,R}}\Phi  - 4R{\partial _{z,z}}\Phi }}} \right|_{\left( {R,z = 0} \right)}}.
\end{equation}
The expression \eqref{Eq73} is null for a potential separable in cylindrical coordinates because the term $\frac{{\partial {\Phi _{{\rm{tot}}}}}}{{\partial R\partial {z^2}}} = 0$. In spherical coordinates $\lambda  = 1$. The relation (73) will be used in the following to describe the tilt of the velocity ellipsoid, obtaining for the DM halo 
\begin{equation}\label{Eq74}
	{\lambda _{DM}}\left( R \right) =  - \frac{{{R^2}}}{{{R^2}\left( {{q_\Phi }^2 - 2} \right) + 2{h_{r,DM}}^2\left( {{q_\Phi }^2 - 1} \right)}},
\end{equation}
and similarly, a unitary constant value for the bulge and stellar halo components is estimated. Finally, for the important contribution of the disk we simplify Eq.(73) as
\begin{equation}\label{Eq75}
	{\lambda _D}\left( R \right) =  - \int_0^\infty  {\frac{{k{R^2}{J_1}(kR)h_z^{ - 1}dk}}{{R\left( {k - \frac{4}{{{h_z}}}} \right){J_0}(kR) + 2{J_1}(kR)}}} 
\end{equation}
that has to be included in Eq.\eqref{Eq72} with a sum over all the disks components.

Finally, we close this section noting that the vertical tilt in the azimuthal-z direction, i.e. ${\sigma _{\phi z}}$, has no clear global trend in dependence on the configuration space ${\sigma _{\phi z}}={\sigma _{\phi z}}(R,\phi,z)$ \citep[e.g.,][]{2012A&A...547A..71P}, hence we simply assume here globally ${\sigma _{\phi z} = \rm{const.} = 0.0}$. This does not mean that locally $\sigma _{\phi z}$ has to be zero: velocity active regions \citep[e.g.,][]{1986PASJ...38..485F} especially in relation to galactic fountains, minor mergers of dwarf galaxies absorbed by the MW and globular clusters passing throughout the MW disk can easily produce areas of the MW disks where ${\sigma _{\phi z} \neq 0.0}$ as well as ${\sigma _{\phi z} \neq \sigma _{z \phi}}$ if large magnetic fields are present.

\subsubsection{Non-isothermal profile of the Galactic Disks}\label{Sec7.2.2}
It is common to assume that the velocity distribution is isothermal in the vertical direction, or more precisely that $\frac{{\partial \sigma _{zz}^2}}{{\partial z}} = 0$ $\forall {f_{{\text{SSP}}}}$ as we have already done in Sec. \ref{Sec7.1}. This is certainly a reasonable assumption for small z, although there is no reason why the galactic disk should be isothermal at all. \citet{1984ApJ...287..926B} \citep[but see also][]{1984BAAS...16..733B, 1984ApJS...55...67B} treated the problem assuming that non isothermality can be simulated by the superposition of more isothermal components. The observations of \citet{1989MNRAS.239..651K} \citep[][]{1989MNRAS.239..605K, 1989MNRAS.239..571K} show significant departures from isothermality at large $z$. One can prove \citep[][]{1991ApJ...368...79A} that in a cool disk $\sigma _{Rz}^2 =  - \sigma _{zz}^2\frac{{\partial \sigma _{zz}^2}}{{\partial z}}{\left( {\frac{{\partial \sigma _{zz}^2}}{{\partial R}}} \right)^{ - 1}}$. This tells us immediately that if the tilt term of the ellipsoid is zero, then the velocity dispersion is constant in the vertical direction. Therefore the assumption of an isothermal structure for $ {f_{{\text{SSP}}}}$ is true only in the case of a gravitational potential which is separable in cylindrical coordinates. In the case of no strict isothermality, this approximation is valid within 1 kpc from the plane, where the fractional change in $\sigma _{zz}^2$ is expected to be less than 3\% \citep[][]{1991ApJ...368...79A, 1991MNRAS.253..427C}. For all reasonable gravitational potentials, $\sigma _{RR}^2 > \sigma _{zz}^2$ and one can prove that $\sigma _{zz}^2$  has an extremum (minimum) on the plane, i.e. first derivative null $\frac{{\partial \sigma _{zz}^2}}{{\partial z}} = 0$ \citep[][]{1979MNRAS.186..813H, 1989ApJ...337..163W, 1987gala.conf..375F, 1988A&A...192..117V}. For small $z$ and fixed $R = {R_ \odot }$, one can perform the Taylor expansion $\sigma _{zz}^2\left( {{R_ \odot },z} \right) \simeq \sigma _{zz}^2\left( {{R_ \odot },0} \right) + \frac{1}{2}{z^2}\frac{{{\partial ^2}\sigma _{zz}^2\left( {{R_ \odot }} \right)}}{{\partial {z^2}}} + o\left( {{z^3}} \right).$ Assuming now that $\sigma _{RR}^2\left( {{R_ \odot },0} \right) = \alpha  \cdot \sigma _{zz}^2\left( {{R_ \odot },0} \right)$ as for Sec. \ref{Sec7.1},  one obtains $\frac{{\sigma _{zz}^2\left( {{R_ \odot },z} \right)}}{{\sigma _{zz}^2\left( {{R_ \odot },0} \right)}} \simeq 1 + \frac{{\lambda \left( {{R_ \odot }} \right)}}{{2{R_ \odot }}}\left( {\alpha  - 1} \right){\left| {\frac{{\partial \ln \sigma _{zz}^2}}{{\partial R}}} \right|_{\left( {{R_ \odot },0} \right)}}{z^2}$. Supposing that $\sigma _{zz}^2 \propto \rho $ in the plane, as in \citet{1982A&A...110...61V}, and that $\rho $ follows an exponential law with constant scale length ${h_R}$  we have $\frac{{\sigma _{zz}^2\left( {{R_ \odot },z} \right)}}{{\sigma _{zz}^2\left( {{R_ \odot },0} \right)}} \simeq 1 + \frac{{\lambda \left( {{R_ \odot }} \right)}}{{2{R_ \odot }}}\left( {\frac{{\sigma _{RR}^2\left( {{R_ \odot },0} \right)}}{{\sigma _{zz}^2\left( {{R_ \odot },0} \right)}} - 1} \right)\frac{{{z^2}}}{{{h_R}}}$ which describes the non-isothermal case as,
\begin{equation}\label{Eq76}
	\begin{aligned}
		\sigma _{zz}^2\left( {{R_ \odot },z} \right) &\simeq \sigma _{zz}^2\left( {{R_ \odot },0} \right)+ \\ 
		&+ \left( {\frac{{\lambda \left( {{R_ \odot }} \right)}}{{2{R_ \odot }}}\left( {\frac{{\sigma _{RR}^2\left( {{R_ \odot },0} \right)}}{{\sigma _{zz}^2\left( {{R_ \odot },0} \right)}} - 1} \right)\frac{{{z^2}}}{{{h_R}}}} \right)\sigma _{zz}^2\left( {{R_ \odot },0} \right), 
	\end{aligned}
\end{equation}
or
\begin{equation}\label{Eq77}
	\sigma _{zz}^2\left( {R,z} \right) \simeq \sigma _{zz}^2\left( {R,0} \right) + {\left. {\frac{{\lambda \left( \Phi  \right)\left( {\sigma _{RR}^2 - \sigma _{zz}^2} \right)}}{{2 \cdot {h_R} \cdot R}}} \right|_{\left( {R,0} \right)}}{z^2},
\end{equation}
which is clearly not constant. We have implemented this last formulation in our models to take the non-isothermal structure of the thin disks into account. This SSP non-isothermality based on \citet{1991ApJ...368...79A} and \citet{1991MNRAS.253..427C} hydro-dynamical model, applies to single stellar population (SSP) alone. However, it is observationally very difficult to identify a truly homogeneous population, i.e. in chemistry but also in age \citep[e.g.,][]{2012ApJ...753..148B}. 

\begin{figure*}
\centering
\resizebox{\hsize}{!}{\includegraphics{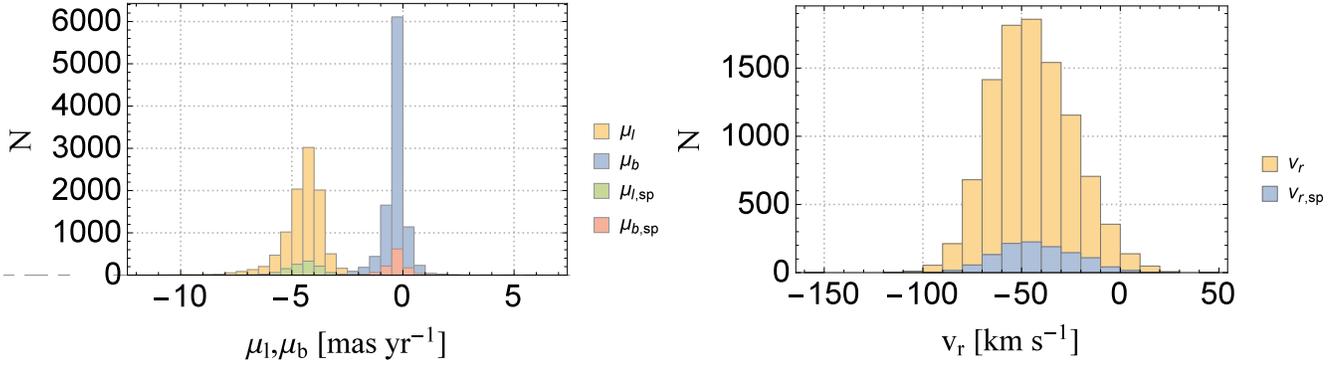}}
\caption{Proper motion distribution (left panel) for a field of about 10000 stars with spiral arms. The contribution of the spiral arms population is evidenced with subscript``sp''. (Right panel) same as left panel but for radial velocity. The CMD of this example is in Fig. \ref{Fig9}.
\label{Fig15} }
\end{figure*}

Finally, we can collect all this kinematic infrastructure to project it on the space of observations by plotting, e.g., the PDF of the proper motion ${\mu _{l,b}}$ and radial velocities ${v_r}$ populated for the same field of Fig. \ref{Fig9} in Fig. \ref{Fig15}.

\section{Comparison with the literature}\label{Sec8}
Star count techniques have a long history and a comparison of all the different flavours of this approach is a complex task. Mainly two kinds of kinematical models and associated star counts are available in the literature. The first ones do not depend on the underlying gravitational potential. They contain a large number of constants treated as free parameters and therefore can quickly fit large samples of data simply because they do not integrate the Poisson equation. A prototype of this modelling approach is \citet{1987BAAS...19..679R} \citep[see e.g.,][]{1989ApJ...339..106R, 1990ApJ...357..435C}, who first applied the kinematical de-convolution of a DF in the phase space. Another model of this kind, with substantially no differences, is by \citet{1984MNRAS.207..223G}. \citet{1996AJ....112..655M} and \citet{2000AJ....119..813M} further refined the star-counts approach to kinematical problems. Their work is based on the epicycle theory of disk kinematics and presents an original treatment of the differential rotation based on the Jeans equations that allow studying of the asymmetric drift for each disk population. The limitation of these models is that the epicycle approximation limits the analysis of the disk kinematics only to regions near the plane and does not consider the vertical tilt of the velocity ellipsoid. Models of this kind, which are not based on a constrained potential may lead to somewhat non-physical solutions. Nevertheless, thanks to their simplification, they are useful to investigate more difficult problems which require formulations that are more sophisticated. For instance, the problem of the vertex deviation requires the axisymmetric hypothesis in the Poisson-solver to be relaxed. These difficulties are the subject of debate and strictly linked with the problems of modelling the bulge too.

The most popular global Galaxy model available in the literature and to which a finer comparison is due, is the Besan\c{c}on model. This model has roots in the works by \citet{1986A&A...157...71R} and \citet{1987A&A...180...94B}, later improved by \citet{2003A&A...409..523R}. This model not only simulates CMDs, taking into account all evolutionary phases of a star down to the white dwarf stages, but makes use of a kinematical description linked to the gravitational potential. 

However, this model is not fully consistent from a dynamical point of view, equipped with much weaker dynamical constraints than what we are presenting here and missing a non-axisymmetric treatment of spiral arms. The method developed by \citet{2003A&A...409..523R} is summarized in \citet{1987A&A...180...94B}. Assuming suitable density profiles for the components of a galaxy, their method calculates the total density profile using the Poisson-solver in axisymmetric stationary conditions. The authors derive the gravitational potential from
\begin{equation}\label{Eq78}
	\Delta \Phi _{tot}^{(I)}\left( {R,z} \right) = 4\pi G{\rho _{tot}}\left( {R,z;\hat R_C^{(I)},\hat \rho _C^{(I)},\hat M_B^{(I)},\hat H_{z,i}^{\left( I \right)}} \right),
\end{equation}
where $\{ \hat R_C^{(I)},\hat \rho _C^{(I)},\hat M_B^{(I)},\hat H_{z,i}^{\left( I \right)}\} $ are respectively the core radius of the halo, the central density of the disk, the total mass of the bulge, and the different scale heights of the disk at their first (I) guess input. This equation yields the first guess of the potential $ \Rightarrow \Phi _{tot}^{(I)}\left( {R,z} \right)$ where $\{ \hat R_C^{(I)},\hat \rho _C^{(I)},\hat M_B^{(I)}\} $ have been considered as parameters. Obviously, some standard constraints are imposed such as the rotation curve $v_c^{\left( I \right)}(R) = {R{\partial _R}\Phi _{tot}^{(I)}(R,z = 0)}^{1/2} $ which can be compared with the observational data. The parameters are varied to match the rotation curve ${v_{{\rm{c}}{\rm{,bestfit}}}}(R) =  {R{\partial _R}\Phi _{tot}^{(I)}(R,z = 0;R_C^{(II)},\rho _C^{(II)},M_B^{(II)},\hat H_{z,i}^{\left( I \right)})}^{1/2} $ and the new parameters $\{ \hat R_C^{(II)},\hat \rho _C^{(II)},\hat M_B^{(II)}\} $ are used to obtain a second guess for the potential $\Delta \Phi _{tot}^{(II)}(R,z) = 4\pi G{\rho _{tot}}(R,z;\hat R_C^{(II)},\hat \rho _C^{(II)},\hat M_B^{(II)},\hat H_{z,i}^{\left( I \right)})$ satisfying the first dynamical constraint set by the rotation curve. The model seeks then to satisfy the Boltzmann equation. Using the Jeans equation their model assumes, in contrast with our model, that ${\sigma _{Rz}} = \overline {\left( {{v_R} - {{\bar v}_R}} \right)\left( {{v_z} - {{\bar v}_z}} \right)}  = \overline {{v_R}{v_z}}  = 0$ i.e. $\frac{{\partial \rho \overline {{v_R}{v_z}} }}{{\partial R}} - \frac{{\rho \overline {{v_R}{v_z}} }}{R} = 0$. With this simplification the Jeans equation reduces to $\frac{{\partial \rho \overline {v_z^2} }}{{\partial z}} + \rho \frac{{\partial {\Phi _{{\rm{tot}}}}}}{{\partial z}} = 0$ and assuming isothermal behaviour $\sigma _{zz}^2\frac{{\partial \rho }}{{\partial z}} + \rho \frac{{\partial {\Phi _{{\rm{tot}}}}}}{{\partial z}} = 0$, where $\sigma _{zz}^2 = \overline {{{\left( {{v_z} - {{\bar v}_z}} \right)}^2}}  = \overline {v_z^2} $. This equation can immediately be solved ${\Phi _{tot}}(R,z) = {c_1} - \sigma _{zz}^2\ln \left( {\rho (R,z)} \right)$. This means that another solution of this equation is ${\Phi _{tot}}(R,z = 0) = {c_1} - \sigma _{zz}^2\ln \left( {\rho (R,z = 0)} \right)$. Subtracting the two solutions we obtain another solution and the implemented equation $\sigma _{zz}^2\ln \left( {\frac{{\rho (R,z)}}{{\rho (R,0)}}} \right) =  - {\Phi _{tot}}(R,z) + {\Phi _{tot}}(R,0)$, that is already present in \citet{1968gaas.book.....M}. The model makes use of this equation to obtain the best fit of the disk scale length with an iterative procedure. The dynamical consistency is clearly poorer than that in our kinematical model, where formulation for the mixed terms  ${\sigma _{R\phi }} = {\sigma _{R\phi }}\left( {{\Phi _{{\rm{tot}}}}} \right)$ and ${\sigma _{Rz}} = {\sigma _{Rz}}\left( {{\Phi _{{\rm{tot}}}}} \right)$  are both obtained with consistency from the potential as well as the non-isothermality of ${\sigma _{zz}}\left( z \right)$. This helps to decrease the large number of parameters, thus strengthening the consistency of our model. Our approximation implicitly assumes a kinematically cold disk. Moreover, in the case of the Galactic potential no analytical, realistic formulation of the third integral of motion (which is likely responsible for the small irregularities present in the Galaxy structure) is available. Therefore, it is not yet analytically possible to derive a correct DF that, thanks to the Jeans’ Theorem, could satisfy the self-consistency requirements. A hypothesis in common to our kinematical model and the one by \citet{2003A&A...409..523R} is the stationary state of the DF. The tri-shifted Gaussian represents the only analytical solution of the Boltzmann equation in a steady state and it is commonly adopted in several models. 

As it is not easy to satisfy the dynamical consistency, many kinematical models are possible and their solution is degenerate. Based on these considerations, instead of the iterative procedure adopted in \citet{2003A&A...409..523R} for which no unique solution is guaranteed, it is perhaps better to let all parameters remain free to converge to the best fit solution with no ad hoc limits. Clearly the best way to proceed is to simultaneously constrain as many parameters as possible with the suitable minimization algorithms that we described in Sec. \ref{Sec5}.

Finally, we point out how in our model, the Poisson solver is exactly the same as in \citet{2003A&A...409..523R}, but our analytical treatment allows us to consider the coupled potential along the vertical-radial direction. This is not possible with the \citet{2003A&A...409..523R}  model  that neither follows the vertical variation of the vertical tilt of the velocity ellipsoid nor the variations of the velocity ellipsoid vertical axis with the stellar populations nor the radial velocity coupling with the presence of spiral arms. Both the \citet{2003A&A...409..523R} and our models allow for a gradient in the vertical component of the temperature profiles, i.e. $\sigma _{zz}^2\left( R \right) = \sum\limits_{i = 1}^n {\sigma _{zz,i}^2} $. However, in the \citet{2003A&A...409..523R} model all the $\sigma _{zz,i}^2$ are considered constant for all stellar populations, i.e. $\sigma _{zz,i}^2 = {\rm{const}}.\forall i$. Therefore, the temperature gradient is a consequence of the different scale height of each population. In contrast each stellar population in our model has its own vertical profile, i.e. $\sigma _{zz,i}^2 = \sigma _{zz,i}^2\left( R \right)\forall i$. This makes it possible to examine separately different types of SSP. The task is nowadays feasible thanks to the wealth of data and even more in the near future with Gaia, whose data will probably give definitive answers to long lasting problems such as the vertical isothermality, presence of dark matter and the origin of the Galactic disks. 

Our model has the significant advantage of reproducing several observational constraints, such as (Sec. \ref{Sec4}) the rotation curve, the outer rotation curve, the Oort functions and constants, the mass inside 100 kpc, the vertical force, the surface density, and the parameter $\lambda $ in the solar neighbourhood. 

Clearly the space of parameter dimensions grows with the square of the number of parameters; however, the number of important parameters is rather small. The interplay between a gravitational potential satisfying all the constraints and parameter adjustment to fit the observational CMDs secures the kinematical consistency of the model, as the kinematics are simultaneously derived from the potential and the properties of the stellar populations generating the potential. One has to remember that the DF has different dispersion axes for each population and that the angular momentum for the orbits is linked to the rotational velocity via the Boltzmann equation moments.

\section{Conclusions}\label{Sec9}
We presented a Galaxy model which can be used to investigate large datasets focused from MW surveys in great detail, of the Milky way with particular attention to the kinematical modelling. This model gathered the heritage of the Padua model that stems from the early studies of   the stellar content of the Palomar-Groeningen survey towards the  Galactic Centre by \citet{1995A&A...295..655N,1995A&A...301..381B}, followed by studies of specific groups of stars and interstellar extinction  by \citet{1995A&A...301..381B,1996A&A...310..115B,1996astro.ph..9134N,1996A&A...315..116N} to mention a few, the studies of the stellar content towards the Galactic Pole \citep{1997A&A...324...65N},  the development of a new minimization technique for the  diagnostics of stellar population synthesis  \citep{1998A&AS..132..133N},
the study of the Galactic Disc Age-Metallicity relation \citep{1998MNRAS.296.1045C},
the possible relationship between the  bulge C-stars and the Sagittarius dwarf galaxy  \citep{1998A&A...338..435N},
 the developments of AMORE (Automatic Observation Rendering)  of a synthetic stellar population's colour-magnitude diagram  based on the genetic algorithm \citep{2002A&A...392.1129N},
 the study of 3-D structure of the Galaxy from star-counts in view of the Gaia mission \citep{2003MmSAI..74..522V}
 and of the kinematics of the Galactic populations towards the North Pole with mock Gaia data  \citep{2004ASPC..317..203V,Pasetto2005,2006A&A...451..125V}.

The building blocks of the Padua Galactic model are a synthetic Hertzsprung-Russell Diagram generator a kinematical model, and a MW gravitational potential model. This tool has now been updated following a novel approach to the theory of population synthesis that borrows and adapts to the present aims a few concepts from Statistical Mechanics:
\begin{enumerate}
	\item The model is grounded  on the concept of the Probability Distribution Function for stellar populations. The system-Galaxy is framed in a theoretical Existence Space   where it is characterized by a number of key relations (mass function, age-metallicity, phase-space, and metallicity/phase-space).
	\item The model is able to analyze and reproduce observable quantities regardless of  size and amounts of the data to a analyse: this is achieved thanks to the use of PDF instead of star-counts.
	\item The distribution of mass and mass density and associated gravitational  potential are thoroughly discussed and formulated for each component of the MW together with a few other important issues such as the rotation curve, the vertical force acting on the plane, the presence of spiral arms and their effects on dynamics and kinematics, the presence of non axisymmetric features etc.
	\item Particular effort is paid to include spiral arms for which we develop a completely new treatment of the mass density, kinematics and extinction. Several treatments of the resonance areas to deal with the star-count technique have been explored and implemented. 
	\item A novel formulation for the extinction  has been implemented to account for the new non-axisymmetric features of the model and in preparation for a forthcoming star-count model of the bulge.
	\item A genetic algorithm has been included to deal simultaneously with photometric information, as well as kinematical and gravitational information.
    \item Particular care is paid to the photometric population synthesis to simulate the photometric properties, magnitudes and colors for samples of stars of unprecedented size taking advantage of the concept of PDF to populate CMDs and luminosity functions bearing in mind the data that will soon be acquired by space observatories  like Gaia.
  \item The model has been  compared with similar models in literature, for instance the  popular Besan\c{c}on model, to highlight  differences and similarities.
\end{enumerate}

 The range of applicability of our Galaxy Model  is very large.  It can already be  applied to existing MW surveys on which the model has already been tested. None of the existing surveys are actually comparable with Gaia for precision and amounts of data, but all of them already investigate different aspects of the Galaxy. To mention a few,  we recall the Radial Velocity Experiment \citep[][]{2006AJ....132.1645S, 2008AJ....136..421Z, 2011AJ....141..187S} of which  the fourth data release \citep[][]{2013AJ....146..134K} has been used to test the model  \citep[][]{2012A&A...547A..70P, 2012A&A...547A..71P}. The Galaxy Model could be applied to the data  coming from the Apache Point Observatory Galactic Evolution Experiment \citep[e.g.][]{2015arXiv150905420M}, a companion program of the Sloan Digital Sky Survey \citep[][]{2014ApJS..211...17A} that makes available  an infrared catalogue of several hundred thousands radial velocities (from high resolution spectra) that are suitable to stellar population studies within the plane. This database is especially interesting when  used in combination with  the data from  the Kepler/K2 mission, because this would allow us  to investigate  stellar populations also with photometry. The ongoing GALAH survey \citep[][]{2015MNRAS.449.2604D}, a large Australian project that will measure the abundances of thirty elements together with HERMES providing spectrographic measurements of  radial velocities  is another example to which the Galaxy Model  could be applied to investigate the galactic archaeology and archeochemistry. 
 
 Finally, our model can be applied also to entirely different astrophysical scales, such as in asteroseismology. When spectroscopic analysis is combined with seismic information, precise constraints on distances, masses, extinction and finally ages can be obtained. The CoRoT red giant field \citep[e.g.][]{2015A&A...576L..12C} analysis is an example, even though the statistical samples are still very small.

A golden age for systematic studies of  MW is imminent. Existing photometric and spectroscopic surveys, as well as future ones such as Gaia, will be crucial to obtain the ultimate model of our own Galaxy, a fundamental local step to interpret the Universe in a cosmological framework. In this context, the Galaxy Model we have developed  is awaiting complete validation by more precise kinematical data, ages and metallicities. In the meantime, we have presented here the first kinematical model that can simultaneously deal with the  age-velocity and dispersion-metallicity relations in a robust dynamical-kinematical framework over the whole space of  variables defined by the proper motions, radial velocities, and multi-band photometric data (magnitudes and colours). Future developments will include an upgrading of the bulge  to include non-axisymmetric descriptions  of the bar and chemical enrichment. 

The Galaxy Model can be access from the internet interface at www.galmod.org and questions addressed at the email: galaxy.model@yahoo.com.

\section*{Acknowledgements}
SP thanks Mark Cropper for the constant support in the realization of this work, and Denija Crnojevic for the hospitality at Texas Tech University of Lubbock where part of this work has been carried out. G.N. would like to acknowledge support from the Leverhulme Trust research project grant RPG-2013-418. We thank the anonymous referee for careful reading of the early versions of the paper.

\begin{small}
\bibliographystyle{mn2e}                    
\bibliography{MNRAS_Biblio}                 
\end{small}

\onecolumn
\appendix
\section{Hypergeometric formulation of the ``Reduction factor''}
There is quite a number of works in the literature, see section 1, that based the interpretation of the reduction factor on the work of \citet{1969ApJ...155..721L} through the computation of the ``q''-factor, i.e. the integral of eq. B8 in Appendix B of the mentioned paper. In this appendix, we express for the first time that integral as a function of the well-known Hypergeometric function. To achieve this result, we make use of the following theorem:

\textbf{Theorem}: from \cite{Erdelyi} (Vol 2, pg. 400, Eq.(8)) we have that for any $\left\{ {z,\alpha ,\beta ,\gamma ,\delta } \right\} \in \mathbb{C}$ with $\operatorname{Re} \gamma  > 0$, $\operatorname{Re} \rho  > 0$ and $\operatorname{Re} \left( {\gamma  + \rho  - \alpha  - \beta } \right) > 0$  the following relation holds:
\begin{equation}\label{Th01}
_2{F_2}\left( {\rho ,\gamma  + \rho  - \alpha  - \beta ;\gamma  + \rho  - \alpha ,\gamma  + \rho  - \beta ;z} \right) = \frac{{\Gamma \left( {\gamma  + \rho  - \alpha } \right)\Gamma \left( {\gamma  + \rho  - \beta } \right)}}{{\Gamma \left( \gamma  \right)\Gamma \left( \rho  \right)\Gamma \left( {\gamma  + \rho  - \alpha  - \beta } \right)}}{e^z}\int_0^1 {{x^{\gamma  - 1}}{{\left( {1 - x} \right)}^{\rho  - 1}}{e^{ - xz}}_2{F_1}\left( {\alpha ,\beta ;\gamma ;x} \right)dx} ,
\end{equation}
with ${_2}{F_2}\left( * \right)$ the hypergeometric function and $\Gamma$ the Eulero-gamma function. 

We want to prove the following corollary to the previous theorem.
 
\textbf{Corollary}: For any $\left\{ {z,\nu } \right\} \in \mathbb{R}$ the following relation holds:
\begin{equation}
		\frac{1}{{2\pi }}\int_{ - \pi }^\pi  {\cos (\nu s){e^{ - z(\cos \left( s \right) + 1)}}} {\mkern 1mu} ds = {\;_2}{\tilde F_2}\left( {\frac{1}{2},1;1 - \nu ,\nu  + 1; - 2z} \right),
\end{equation}
where $_2{\tilde F_2}$ is the Hypergeometric-regularized function. 

\textbf{Proof}. We start relating the Hypergeometric-regularized function, ${_2}{\tilde F_2}$, to the Hypergeometric function, $_2{F_2}$, by writing 
\begin{equation}
		{\;_2}{\tilde F_2}\left( {\frac{1}{2},1;1 - \nu ,\nu  + 1; - 2z} \right) = \frac{{\sin (\pi \nu ){\mkern 1mu} }}{{\pi \nu }}{\,_2}{F_2}\left( {\frac{1}{2},1;1 - \nu ,\nu  + 1; - 2z} \right).
\end{equation}
This equation, because of the Theorem Eq.\eqref{Th01} reduces to:
\begin{equation}
		\begin{aligned}
	\frac{{\sin (\pi \nu ){\mkern 1mu} }}{{\pi \nu }}{\,_2}{F_2}\left( {\frac{1}{2},1;1 - \nu ,\nu  + 1; - 2z} \right)	&= \frac{{\sin (\pi \nu ){\mkern 1mu} }}{{\pi \nu }}\frac{{\Gamma (1 - \nu )\Gamma (\nu  + 1)}}{\pi }{e^{ - 2z}}\int_0^1 {\frac{{{e^{2xz} \,_2}{F_1}\left( {\nu , - \nu ;\frac{1}{2};x} \right)}}{{\sqrt {1 - x} \sqrt x }}dx}  \\ 
		&= \frac{1}{\pi }\frac{{\sin (\pi \nu ){\mkern 1mu} }}{\nu }\frac{\nu }{{\sin \left( {\pi \nu } \right)}}{e^{ - 2z}}\int_0^1 {\frac{{{e^{2xz}}\cos \left( {2\nu \arcsin \sqrt x } \right)}}{{\sqrt {1 - x} \sqrt x }}dx}  \\ 
		&= \frac{{{e^{ - 2z}}}}{\pi }\int_0^1 {\frac{{2{{\rm{e}}^{2{y^2}z}}\cos \left( {2\nu \arcsin y} \right)}}{{\sqrt {1 - {y^2}} }}dy}  \\ 
		&= \frac{{{e^{ - 2z}}}}{\pi }\int_0^{\pi /2} {2{{\rm{e}}^{2z{{\sin }^2}q}}\cos \left( {2q\nu } \right)dq}, 
	\end{aligned} 
	\end{equation}
where in the second line we made use of $_2{F_1}$ and Eulero-gamma function properties, in the third and fourth rows we change variables, $x=y^2$ and $y=\sin q$, accounting for the dominion of integration. The last relation is clearly an even function, thus:
\begin{equation}
		\begin{aligned}
		_2{{\tilde F}_2}\left( {\frac{1}{2},1;1 - \nu ,\nu  + 1; - 2z} \right) &= \frac{1}{\pi }\int_0^\pi  {{e^{- 2z + 2z{{\sin }^2}\frac{s}{2}}}\cos \left( {\nu s} \right)ds}  \\ 
		&= \frac{1}{{2\pi }}\int_{ - \pi }^\pi  {{e^{ - z\left( {1 + \cos s} \right)}}\cos \left( {\nu s} \right)ds,}  \\ 
	\end{aligned} 
\end{equation}
which concludes our proof. The expression for the Reduction factor comes then easily. 

Needless to say that the advantage of having this formulation for the reduction factor stands not only in the compact elegant formalism, but probably more on the rapidity of performing its evaluation. A test on a commercial processor available to date (Intel core-I7, 3.0 GHrz) shows that the integral evaluation against the Hyper-geometrical formulation is $\sim 10^{-4}$ per second. This translates as the possibility (to date) to generate mock catalogues for a small survey of data with about $\sim 10^{4}$ stars each second, against a generation of the same kinematics catalogue by numerical integration (with adaptive Runge-Kutta scheme) in about 3 hrs! This result is even more striking if you distribute the computation on several processors. 

\end{document}